\documentclass[a4paper,12pt,english]{book}
\usepackage[T1]{fontenc}
\usepackage[latin1]{inputenc}
\usepackage{a4}
\usepackage{geometry}
\usepackage{babel}
\usepackage{graphics}

\makeatletter

\providecommand{\LyX}{L\kern-.1667em\lower.25em\hbox{Y}\kern-.125emX\@}

 \newenvironment{lyxlist}[1]
   {\begin{list}{}
     {\settowidth{\labelwidth}{#1}
      \setlength{\leftmargin}{\labelwidth}
      \addtolength{\leftmargin}{\labelsep}
      }}
   {\end{list}}

\usepackage{fancyhdr}
\usepackage{amssymb,amsthm}
\usepackage{exscale}
\usepackage{latexsym}
\usepackage{german}
\fancypagestyle{plain}{\fancyhf{} \fancyhead[LE,RO]{\thepage}}
\newtheorem{Def}{Definition}
\newtheorem{Cl}{Claim}
\newtheorem{Prop}{Proposition}
\newtheorem{Conclusion}{Conclusion}
\newtheorem{Theorem}{Theorem}
\newtheorem{Lemma}{Lemma}
\newtheorem{Remark}{Remark}
\newcommand{\bDef}{\begin{Def}}
\newcommand{\eDef}{\end{Def}}
\newcommand{\bcl}{\begin{Cl}}
\newcommand{\ecl}{\end{Cl}}
\newcommand{\bprop}{\begin{Prop}}
\newcommand{\eprop}{\end{Prop}}
\newcommand{\bpr}{\begin{proof}}
\newcommand{\epr}{\end{proof}}
\newcommand{\bconcl}{\begin{Conclusion}}
\newcommand{\econcl}{\end{Conclusion}}
\newcommand{\btheorem}{\begin{Theorem}}
\newcommand{\etheorem}{\end{Theorem}}
\newcommand{\bla}{\begin{Lemma}}
\newcommand{\ela}{\end{Lemma}}
\newcommand{\brm}{\begin{Remark}}
\newcommand{\erm}{\end{Remark}}
\newcommand{\be}{\begin{equation}}
\newcommand{\ee}{\end{equation}}

\newcommand{\tril}{\triangleleft}
\newcommand{\ar}{\longrightarrow}
\newcommand{\bear}{\begin{array}}
\newcommand{\ear}{\end{array}}
\newcommand{\ten}{\otimes}

\def \ha {\frac{1}{2}}

\makeatother
\begin{document}
\ifx\href\undefined\else\hypersetup{linktocpage=true}\fi
\begin{titlepage}

\begin{center}

\vspace*{1.5cm}

{\Huge\bfseries Models of Gauge Field Theory on} \\ \vspace{0,5cm} {\Huge\bfseries Noncommutative Spaces}

\vspace{1.5cm}

{\large {\it Diplomarbeit der Fakult\"at f\"ur Physik der \\ Ludwig-Maximilians-Universit\"at M\"unchen} \\ \vspace{1cm} {\small vorgelegt von} \\ Frank Meyer \\ {\small aus Dorsten} \\ \vspace{1cm} Februar 2003

\vspace{1.5cm}

{\it Sektion Physik, Lehrstuhl Prof. J. Wess \\ Theresienstr. 37, 80333 M\"unchen }} 

\vspace{1.5cm} 

\resizebox*{4cm}{4cm}{\includegraphics{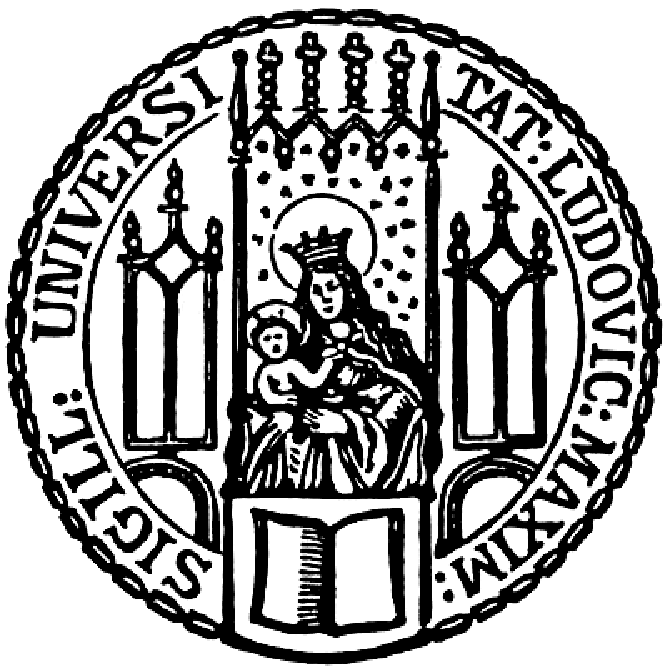}} 

\end{center} 

\end{titlepage}

\section*{~}

~\\
~\\
~\\
~\\
~\\
~\\
~\\
~\\
~\\
~\\
~\\
~\\
~\\
~\\
~\\
~\\
~\\
~\\
~\\
~\\
~\\
~\\
~\\
~\\
~\textbf{}\\
\textbf{~}\\
\textbf{~}\\
\textbf{~}\\
\textbf{~}\\
\textbf{~}\\
\textbf{}\\
\textbf{~}\\
\textbf{~}\\
\textbf{~}\\
\textbf{~}\\
\textbf{~}\\
\textbf{}\\
\textbf{~}\\
\textbf{~}\\
\textbf{~}\\
\textbf{Referent: Prof. Dr. Julius Wess}\\
\textbf{Koreferent: Prof. Dr. Hans-J\"urgen Schneider}

\chapter*{Abstract}

Models of gauge field theories on noncommutative spaces are studied.
In the first chapter we consider a canonically deformed space. Its
noncommutativity is induced by means of a star product with a constant
Poisson tensor. We discuss the dependence of the established gauge
field theory on the choice of the star product itself: The dependence
of the action is derived and it is shown that in general two arbitrary
star products lead to two different actions. A large class of star
products which all correspond to equal actions is determined as well.
The Moyal-Weyl star product and the normal ordered star product are
among this set of star products. 

In the second chapter we discuss noncommutative, abelian gauge field
theory on the \( E_{q}(2)- \)symmetric plane. Two different approaches
are studied.

The first is the generalization of the theory established for a canonically
deformed space. We induce the noncommutativity via a star product
with a non-constant Poisson tensor. To be able to define a gauge invariant
action, we construct an integral with trace property (considerations
made here may be transfered to other noncommutative spaces). The noncommutative
fields are expressed in terms of the ordinary, commutative fields
via the Seiberg-Witten map. This allows us to calculate explicitly
the corrections to the commutative theory predicted by the noncommutative
one in orders of the deformation parameter. We present the results
up to first order, which shows that new interactions appear. The integral
we had to introduce is not invariant with respect to \( E_{q}(2)- \)transformations. 

In the second approach we establish a gauge field theory which is
\( E_{q}(2)-\textrm{covariant}.\textrm{ } \)We construct a \( U_{q}(e(2))-\textrm{invariant} \)
integral. It turns out that this integral is not cyclic but possesses
a {}``deformed'' cyclic property. We establish an \( E_{q}(2)- \)covariant
differential calculus. A frame for one-forms is found as well. A crucial
issue are gauge transformations: In order to obtain a gauge invariant
and \( E_{q}(2)- \)invariant action, we introduce {}``deformed''
gauge transformations. The semi-classical limit \( q\rightarrow 1 \)
of those gauge transformations is examined. 

\tableofcontents{}

\addcontentsline{toc}{chapter}{Introduction}

\newcommand{\rt}{\triangleright }

\newcommand{\lt}{\triangleleft }
 
\newcommand{\te}{\otimes }

\newcommand{\rp}{\rightarrow }

\newcommand{\ra}{\mapsto }

\chapter*{Introduction}

The idea to study noncommutative space-time is very old. It goes back
to Snyder, who first suggested a noncommutative structure at small
length scales to obtain this way an effective ultraviolet cutoff \cite{Snyder:1947}.
This in turn should help to deal with the divergences that already
appeared in the early days of quantum field theory. The problem still
persists and during the last two decades noncommutative spaces have
been intensively studied in the hope to find a natural regularization
of deformed quantum field theories.

There are different ways to introduce a noncommutativity of space-time.
One is inspired by quantum mechanics, where position and momentum
operators obey the Heisenberg commutation relations. Similarly, the
quantized space-time coordinates \( \hat{x}^{i} \) are demanded to
satisfy the so called canonical commutation relations \begin{equation}
\label{intro: canonical commutation relations}
[\hat{x}^{i},\hat{x}^{j}]=i\theta ^{ij}\, \, \, ,
\end{equation}
 where \( \theta ^{ij} \) are real constants. This approach became
especially interesting when in the eighties the study of open strings
in a magnetic background field led to a canonical noncommutativity
in string theory \cite{Seiberg:1999vs}. However, by deforming the
space alone the space will in general loose its classical background
symmetry.

To maintain a background symmetry in the deformed setting, it is therefore
in general necessary to deform both the symmetry group as well as
the space. This leads to so called \( q- \)deformed spaces that admit
a quantum group symmetry. Correctly spoken the algebra of functions
on a manifold and the algebra of functions on a Lie group are \( q- \)deformed
\cite{Madore:Buch}. Such a noncommutative theory is a generalization
of the commutative theory since the deformed space as well as the
deformed symmetry group become in the semi-classical limit \( q\rp 1 \)
the commutative space together with its undeformed background symmetry.
The rich mathematical structure of quantum groups gives rise to the
hope to get a better understanding of physics at short distances and
to solve the above-mentioned problems.

Gauge field theories have been very successful for the understanding
of the fundamental forces of nature. The standard model made a unification
of electromagnetism, the weak and the strong force possible. Therefore
a generalization of the gauge invariance principle to noncommutative
spaces is of particular interest and subject matter of this thesis.
The hope is that a deep understanding of gauge field theories on noncommutative
spaces may make it possible to finally incorporate the last fundamental
force, gravitation, in a unified theory.

This thesis is divided into two chapters. The first chapter deals
with a canonically deformed space. Gauge field theory on this noncommutative
space has been examined intensively in \cite{Madore:2000en,Jurco:2001rq,Jurco:2000ja,Jurco:2000dx,Schupp:2001we}.
A crucial ingredient for the construction of gauge field theory on
a canonically deformed space is the use of a star product that renders
the originally commutative algebra of functions on the space isomorphic
as algebra to the noncommutative algebra of functions. By means of
Seiberg-Witten maps it then is possible to relate ordinary gauge theory
and noncommutative gauge theory and to express noncommutative fields
in terms of their commutative analogs. This allows us to read off
explicitly the corrections to the commutative theory predicted by
the noncommutative one in orders of the deformation parameter \( h \).
In the whole formalism a special star product was used: the Moyal-Weyl
star product that corresponds to the symmetric ordering prescription.
However, there exist infinitely many star products, all corresponding
to a different ordering prescription, that cannot be distinguished
from the algebraic point of view. Thus, the question arises how far
a change of the star product changes the physical theory, i.e. the
action. We will examine in all detail how a change of the ordering
prescription explicitly affects the action. We will see that in general
different star products indeed lead to different actions, but that
nonetheless a large class of star products generates identical actions.
The star products corresponding to normal ordering and symmetric ordering
belong to this class.

The second chapter treats the case of a two-dimensional quantum space
with an \( E_{q}(2)- \)symmetry. By \( E_{q}(2) \) we denote the
\( q- \)deformed algebra of functions on the two-dimensional Euclidean
group. We start introducing the quantum group \( E_{q}(2) \) and
the corresponding \( E_{q}(2)- \)covariant plane. The underlying
commutation relation of the space coordinates is \begin{equation}
\label{intro: commrel for Eq(2)}
z\overline{z}=q^{2}\overline{z}z
\end{equation}
 in the basis \( z=x+iy \), \( \overline{z}=x-iy \). We then proceed
to study two different approaches to abelian gauge field theory on
this noncommutative space. 

The first is a generalization of the formalism developed in the case
of a constant Poisson tensor. If we substitute \( q \) by \( e^{h} \),
we can write down the commutation relations of the coordinates of
the \( E_{q}(2)- \)covariant plane in the form of (\ref{intro: canonical commutation relations}),
where \( \theta ^{ij}=\theta ^{ij}(z,\overline{z}) \) now depends
on the coordinates and is not constant. A star product for this space
is introduced, as well as a noncommutative gauge field \( A^{i} \)
by the concept of covariant coordinates. Moreover, we can again express
the noncommutative quantities in terms of the commutative physical
fields via a solution for the Seiberg-Witten maps. To get a gauge
field which approaches in the semi-classical limit \( h\rp 1 \) the
commutative gauge field, we must introduce vector fields with lower
indices. Since \( \theta  \) is not constant, we cannot use \( \theta  \)
to lower indices as this would spoil gauge covariance of the gauge
field \( A^{i} \). We have to do it covariantly using the {}``covariantizer''
\( \mathcal{D} \) \cite{Jurco:2001my}. At the end of the section
we calculate explicitly the field strength, the Lagrangian and the
action expanded in \( h \) up to first order. As we will see, new
interactions appear. Unfortunately, we will learn that the theory
possesses some freedom in defining the field strength and the action.
Furthermore, we will see that it is not \( E_{q}(2)- \)covariant:
We are forced to introduce an integral with trace property in order
to obtain a gauge invariant action. This action is not \( E_{q}(2)- \)invariant.

In the third section we try to establish an \( E_{q}(2)- \)covariant
gauge field theory. Considering \( E_{q}(2)- \)covariance as a fundamental
concept underlying all considerations, we introduce an \( E_{q}(2)- \)invariant
(or equivalently a \( U_{q}(e(2))-\textrm{invariant} \)) integral.
We will see that this integral is not \emph{}cyclic but possesses
a {}``deformed'' cyclic property. This is a crucial result, telling
us that we cannot introduce gauge transformations of gauge fields
as conjugation with a unitary element as it was done in the previous
section. In a second step, we establish an \( E_{q}(2)- \)covariant
differential calculus. Furthermore, we find a frame, a basis of one-forms
that commutes with all functions, and a generator \( \Theta  \) of
the exterior differential \( d \). This simplifies calculations in
the \( q- \)deformed space. A noncommutative gauge field \( A \)
and a noncommutative field strength \( F \) that approaches the commutative
field strength in the semi-classical limit \( q\rp 1 \) are introduced.
Finally, {}``deformed'' gauge transformations are defined. This
allows us to obtain an action that is both gauge invariant and \( E_{q}(2)-\textrm{invariant} \)
and thereby we get an \( E_{q}(2)- \)covariant gauge field theory.
In the end, we examine the semi-classical limit for the deformed gauge
transformations.

\chapter{Influence of the Ordering Prescription in the Case \protect\( \theta =\mathrm{const}.\protect \)\label{ch: Constant theta Case}}

Noncommutative spaces, especially in the case of a canonical noncommutativity,
have been intensively studied in recent years. In \cite{Jurco:2001rq}
and \cite{Madore:2000en}, for instance, a gauge theory has been developed
on such a noncommutative space. The star product formalism plays a
crucial role in this theory because it gives us, together with the
Seiberg-Witten map, a possibility to express the noncommutative fields
entering the theory in terms of the well known commutative fields.
This makes it possible to read off explicitly the corrections to the
commutative theory predicted by the noncommutative one. The whole
formalism was developed for a special star product, the Moyal-Weyl
star product, and it is necessary to ask the question whether a different
star product leads to a different physical theory. Since a star product
corresponds to an ordering prescription, we are led directly to the
question of how far a change of the ordering prescription affects
the physical theory.

In fact, we will see that for a large class of ordering prescriptions
the theories are physically equivalent in the sense that all those
theories lead to the same action. In particular this is the case for
the usually used star products, the Moyal-Weyl product corresponding
to the symmetric ordering prescription and the normal ordered star
product. Nonetheless, arbitrary star products will in general lead
to different actions and thereby to different physical theories. 

We do not want to repeat in all detail the formalism constructed in
\cite{Jurco:2001rq,Madore:2000en}, presuming that the general ideas
are known. Hence,  we will only present the essential ideas. Nonetheless,
we want to explain in detail the concepts that will be important within
the framework of this thesis starting with the notions of ordering
and star product.

\section{Ordering and Star Products\label{1.1}}

In this section we want to explain what we mean by an ordering prescription,
what a star product is and how a certain ordering prescription corresponds
to a star product. It will be a mixture of well known definitions
and facts (see for example \cite{Waldmann} or \cite{Bayen:1978ha})
with some new results in the application on the case of a constant
Poisson tensor.

\subsection{About Ordering\label{se: ordering}}

Underlying all our considerations is the following noncommutative
algebra of space time generated by the elements \( \hat{x}^{\mu } \),
to which we want to refer as coordinates, satisfying the relations
\begin{equation}
\label{nkomm}
[\hat{x}^{\mu },\hat{x}^{\nu }]=ih\theta ^{\mu \nu },\, \, \, \textrm{ }h,\theta ^{\mu \nu }\in \mathbb {R}\, \, \, ,
\end{equation}
i.e. the algebra \begin{equation}
\label{ncalg}
\mathcal{A}_{\mathrm{nc}}\equiv \mathbb {C}\langle \langle \hat{x}^{0},\hat{x}^{1},\hat{x}^{2},\hat{x}^{3}\rangle \rangle /(\{[\hat{x}^{\mu },\hat{x}^{\nu }]-ih\theta ^{\mu \nu }\})\, \, \, ,
\end{equation}
where Greek indices are to be understood in the whole chapter as elements
of the set \( \{0,1,2,3\} \)%
\footnote{Considerations made in this chapter are true for higher dimensions
as well.
}, if nothing else is indicated. We additionally want to assume that
\( \theta ^{\mu \nu } \) is \emph{non-degenerated}. Furthermore,
we have the well known four dimensional space of commutative coordinates
\( x^{\mu } \),\begin{equation}
\label{kommut}
[x^{\mu },x^{\nu }]=0\, \, \, ,
\end{equation}
generating the algebra \begin{equation}
\label{calg}
\mathcal{A}_{\mathrm{c}}\equiv \mathbb {C}[[x^{0},x^{1},x^{2},x^{3}]]\, \, \, .
\end{equation}
 We want to call a map \( \rho  \) from \( \mathcal{A}_{\mathrm{c}} \)
to \( \mathcal{A}_{\mathrm{nc}} \) an \emph{ordering prescription}%
\footnote{We do not want to understand an ordering prescription as simple {}``ordering''
the coordinates in a special way but admit more general isomorphisms
as well. One could have said \emph{quantization} instead of ordering
prescription, too.
} if

\begin{itemize}
\item \( \rho :\, \mathcal{A}_{\mathrm{c}}\, \rightarrow \, \mathcal{A}_{\mathrm{nc}} \)
is a vector space isomorphism,
\item \( \rho (1)=1 \) and
\item \( \rho  \) approaches the identity for \( h\rp 0 \), i.e. \( \rho =\mathrm{id}+\mathcal{O}(h) \). 
\end{itemize}
Obviously, we have a big freedom in constructing ordering prescriptions.
Let us illustrate that by giving two examples: the normal ordering
and the symmetric ordering.

\subsubsection*{(a) The normal ordering}

We define a vector space isomorphism \[
\rho _{n}:\, \mathcal{A}_{\mathrm{c}}\, \rightarrow \, \mathcal{A}_{\mathrm{nc}}\]
 on the basis of monomials by \begin{equation}
\label{stanord}
(x^{0})^{i}(x^{1})^{j}(x^{2})^{k}(x^{3})^{l}\mapsto (\hat{x}^{0})^{i}(\hat{x}^{1})^{j}(\hat{x}^{2})^{k}(\hat{x}^{3})^{l}\, \, \, \textrm{for }i,j,k,l\in \mathbb {N}\, \, \, .
\end{equation}
This ordering prescription we want to call \emph{normal ordering.}

\subsubsection*{(b) The symmetric ordering}

We can define another vector space isomorphism \[
\rho _{s}:\, \mathcal{A}_{\mathrm{c}}\, \rightarrow \, \mathcal{A}_{\mathrm{nc}}\]
by assigning any monomial of the variables \( x^{0},x^{1},x^{2},x^{3} \)
to its totally symmetric counterpart, i.e. for example \begin{eqnarray}
1 & \mapsto  & 1\label{symord} \\
(x^{\mu })^{k} & \mapsto  & (\hat{x}^{\mu })^{k}\nonumber \\
x^{\mu }x^{\nu } & \mapsto  & \frac{\hat{x}^{\mu }\hat{x}^{\nu }+\hat{x}^{\nu }\hat{x}^{\mu }}{2}\nonumber \label{sord} \\
(x^{\mu })^{2}x^{\nu } & \mapsto  & \frac{(\hat{x}^{\mu })^{2}\hat{x}^{\nu }+\hat{x}^{\mu }\hat{x}^{\nu }\hat{x}^{\mu }+\hat{x}^{\nu }(\hat{x}^{\mu })^{2}}{3}\nonumber \\
 & \dots \nonumber \label{ord} 
\end{eqnarray}
This ordering prescription we want to call \emph{symmetric ordering.} 

Of course it is possible to construct arbitrarily many other ordering
prescriptions. But in the following we will often come back to these
important examples to illustrate our ideas and results. 

While the noncommutative algebra \( \mathcal{A}_{\mathrm{nc}} \)
and the commutative algebra \( \mathcal{A}_{\mathrm{c}} \) can impossibly
be isomorphic as algebras, we use a vector space isomorphism \( \rho  \)
to transfer the multiplicative structure of \( \mathcal{A}_{\mathrm{nc}} \)
onto \( \mathcal{A}_{\mathrm{c}} \). We then obtain a new product,
called \emph{star product} (or \emph{\( \star  \)-product). }

\subsection{Star Products\label{se: star products}}

Star products are usually defined on Poisson manifolds \cite{Bayen:1978ha,Waldmann}.
Taking the smooth functions on a manifold we get an algebra that we
can deform by means of a star product. Here we want to take an analogous
way but do not want to abandon the concept of formal power series
in the coordinates and do not want to speak about smooth functions
on a manifold. This would direct our attention to problems that are
of secondary interest for the general physical statement and intention
of this thesis. Therefore we want to introduce star products for \emph{Poisson
algebras}, where a Poisson algebra is an associative algebra \( \mathcal{A} \)
together with a \emph{Poisson-bracket.} A Poisson bracket on an algebra
\( \mathcal{A} \) in turn is defined as a bilinear map \[
\{\cdot ,\cdot \}:\mathcal{A}\times \mathcal{A}\rp \mathcal{A}\]
 satisfying

\begin{lyxlist}{00.00.0000}
\item [(i)]\( \{f,g\}=-\{g,f\} \) for all \( f,g\in \mathcal{A}_{c} \)
\hfill{}(Anti-symmetry)
\item [(ii)]\( \{\{f,g\},h\}+\{\{h,f\},g\}+\{\{g,h\},f\}=0 \) \hfill{}
(Jacobi-identity)
\item [(iii)]\( \{f,gh\}=\{f,g\}h+g\{f,h\} \) \hfill{}(Leibniz-rule).
\end{lyxlist}
If we define \[
\{f,g\}:=\theta ^{\mu \nu }(\partial _{\mu }f)(\partial _{\nu }g)\, \, \, ,\]
it is easy to see that \( \{\cdot ,\cdot \} \) is a Poisson bracket
for \( \mathcal{A}_{\mathrm{c}} \) in the sense of the above definition.
The tensor \( \theta  \) is called a \emph{Poisson tensor} for \( \mathcal{A}_{\mathrm{c}} \).
It is an antisymmetric tensor because of (i) and satisfies (ii) and
(iii).

\subsubsection{(a) Definition and Remarks}

The deformation theory of associative algebras was first studied by
Gerstenhaber \cite{Gerstenhaber}. A formal deformation of a Poisson
algebra can be defined as follows:

\bDef\label{de:star product} 

Let \( (\mathcal{A},\{\cdot ,\cdot \}) \) be a Poisson algebra. We
call a \( \mathbb {C}[[h]] \)-bilinear map\[
\star :\mathcal{A}[[h]]\times \mathcal{A}[[h]]\rightarrow \mathcal{A}[[h]]\]
that we can write as a formal power series in \( h \)%
\footnote{This is always possible.
}, \[
\star =\sum ^{\infty }_{r=0}h^{r}M_{r}\, ,\]
 with \( \mathbb {C} \)-bilinear maps \( M_{r}:\mathcal{A}\times \mathcal{A}\rightarrow \mathcal{A} \)
a star product (or \( \star  \)-product) for \( (\mathcal{A},\{\cdot ,\cdot \}) \)
if for all \( f,g,h\in \mathcal{A} \) the following holds:

\begin{lyxlist}{00.00.0000}
\item [(i)]\( f\star (g\star h)=(f\star g)\star h \) \hfill{}(associativity)
\item [(ii)]\( M_{0}(f,g)=fg \) \hfill{}(deformation of the commutative
product)
\item [(iii)]\( M_{1}(f,g)-M_{1}(g,f)=i\{f,g\} \)\hfill{} (deformation
in direction of \( \{\cdot ,\cdot \} \))
\item [(iv)]\( f\star 1=f=1\star f \) . 
\end{lyxlist}
A star product that satisfies for all \( f,g\in \mathcal{A}[[h]] \)
the condition\[
\overline{f\star g}=\overline{g}\star \overline{f},\]
is called a hermitian star product (here the bar denotes the usual
complex conjugation on \( \mathbb {C} \)).

\eDef

Similar to this definition, star products are defined on a Poisson
manifold \( M \) taking as algebra the algebra of smooth functions
\( C^{\infty }(M) \) \cite{Bayen:1978ha,Waldmann}.

As we see, a star product is a deformation of the commutative product
assuring that the noncommutative theory we establish approaches in
the semi-classical limit \( h\rp 0 \) the usual commutative physical
theory. The third condition gives the analog to the quantum mechanical
principle of correspondence: The Poisson bracket \( \{U,V\} \) of
two observables in classical mechanics corresponds to the quantum
mechanical commutator \( \frac{-i}{\hbar }[\hat{U},\hat{V}] \) in
the classical limit. In our setting, the noncommutative algebra of
quantum mechanical observables is replaced by the noncommutative algebra
of functions on noncommutative space time. 

Having specified what we want to demand from a star product to assure
physical interpretation, we discuss how to obtain \( \star  \)-products
satisfying these properties. Using an ordering prescription \( \rho :\, \mathcal{A}_{\mathrm{c}}\, \rp \, \mathcal{A}_{\mathrm{nc}} \)
with \( \rho (1)=1 \) (for example \( \rho _{n} \) or \( \rho _{s} \)
given above), we can define a \emph{}product for \( f,g\in \mathcal{A}_{\mathrm{c}} \)
by pulling back the noncommutative product of the algebra \( \mathcal{A}_{\mathrm{nc}} \)
onto the algebra \( \mathcal{A}_{\mathrm{c}} \): \begin{equation}
\label{de: starproduct via ordering prescription}
f\star g:=\rho ^{-1}(\rho (f)\cdot \rho (g))\, \, \, .
\end{equation}
 We want to interpret \( f\star g \) as a formal power series in
the deformation parameter \( h \) \[
f\star g\in \mathcal{A}_{\mathrm{c}}[[h]]\, \, \, ,\]
 such that we obtain by (\ref{de: starproduct via ordering prescription})
\begin{equation}
\label{eq:AcequivAnc}
(\mathcal{A}_{\mathrm{c}}[[h]],\star )\cong (\mathcal{A}_{\mathrm{nc}}[[h]],\cdot )
\end{equation}
as algebras.

We claim that the product \( \star  \) defined in (\ref{de: starproduct via ordering prescription})
actually is a star product on \( \mathcal{A}_{\mathrm{c}}[[h]] \)
in the sense of Definition \ref{de:star product}. To see this, we
have to check some properties:

\begin{lyxlist}{00.00.0000}
\item [(i)]Associativity follows since multiplication in \( \mathcal{A}_{\mathrm{nc}} \)
is associative:\[
\begin{array}{ccccc}
f\star (g\star h) & = & \rho ^{-1}(\rho (f)\cdot \rho (\rho ^{-1}(\rho (g)\cdot \rho (h)))) &  & \\
 & = & \rho ^{-1}(\rho (f)\cdot (\rho (g)\cdot \rho (h)) &  & \\
 & = & \rho ^{-1}((\rho (f)\cdot \rho (g))\cdot \rho (h)) & = & (f\star g)\star h\, \, \, .
\end{array}\]

\item [(ii)]By definition an ordering prescription \( \rho  \) approaches
for \( h\rp 0 \) the identity such that we have \( \rho =\textrm{id}+\mathcal{O}(h) \).
This yields with (\ref{de: starproduct via ordering prescription})
that \( f\star g=fg+\mathcal{O}(h) \).
\item [(iii)]Write \( f\star g=\sum h^{r}M_{r}(f,g) \) with \( M_{r}:\, \mathcal{A}_{\mathrm{c}}\times \mathcal{A}_{\mathrm{c}}\, \rightarrow \, \mathcal{A}_{\mathrm{c}} \)
and define \( \{f,g\}^{'}:=-i(M_{1}(f,g)-M_{1}(g,f)) \). We claim
that \( \{\cdot ,\cdot \}' \) is a Poisson bracket: It is obviously
antisymmetric and to see that the Leibniz rule and the Jacobi identity
are satisfied we have to use a property the maps \( M_{r} \) possess
derived from the fact that \( \star  \) is associative as shown in
(i): We have\[
\begin{array}{cl}
 & f\star (g\star k)=(f\star g)\star k\\
\Leftrightarrow  & f(gk)+h(M_{1}(f,gk)+fM_{1}(g,k))\\
 & +h^{2}(M_{1}(f,M_{1}(g,k))+fM_{2}(g,k)-M_{2}(f,gk))+\mathcal{O}(h^{3})\\
 & =(fg)k+h(M_{1}(fg,k)+M_{1}(f,g)k)\\
 & +h^{2}(M_{1}(M_{1}(f,g),k)+M_{2}(f,g)k-M_{2}(fg,k))+\mathcal{O}(h^{3})\\
\Rightarrow  & M_{1}(f,gk)+fM_{1}(g,k)=M_{1}(fg,k)+M_{1}(f,g)k\\
\textrm{and} & M_{1}(f,M_{1}(g,k))+fM_{2}(g,k)-M_{2}(f,gk)\\
 & =M_{1}(M_{1}(f,g),k)+M_{2}(f,g)k-M_{2}(fg,k)
\end{array}\]
 Using those two properties for \( M_{1} \) a longer but easy calculation
shows that \( \{\cdot ,\cdot \}^{'} \) satisfies the Leibniz rule
and the Jacobi identity so that \( \{\cdot ,\cdot \}' \) defined
as above is indeed a Poisson bracket. Since all Poisson brackets on
\( \mathcal{A}_{\mathrm{c}} \) are of the form\[
\{f,g\}=\alpha ^{\mu \nu }(x)(\partial _{\mu }f)(\partial _{\nu }g)\]
 with \( \alpha ^{\mu \nu } \) antisymmetric (this follows from the
Leibniz rule and the Jacobi identity), we can conclude that actually
\( \{f,g\}^{'}=\alpha ^{\mu \nu }(x)(\partial _{\mu }f)(\partial _{\nu }g) \)
for functions \( \alpha ^{\mu \nu }(x) \). Now we use that \( \star  \)
is defined by means of an ordering prescription \( \rho  \). As an
ordering prescription, \( \rho  \) goes to the identity for \( h\rp 0 \).
Therefore we have \( \rho (x^{\mu })=\hat{x}^{\mu }+\mathcal{O}(h) \).
Moreover, we denote by \( \pi _{1} \) the projection to the first
order of \( h \). With that we obtain\begin{eqnarray*}
h(M_{1}(x^{\mu },x^{\nu })-M_{1}(x^{\nu },x^{\mu })) & = & \pi _{1}(x^{\mu }\star x^{\nu }-x^{\nu }\star x^{\mu })\\
 & = & \pi _{1}(\rho ^{-1}([\rho (x^{\mu }),\rho (x^{\nu })]))\\
 & = & \pi _{1}(\rho ^{-1}([\hat{x}^{\mu },\hat{x}^{\nu }]))\\
 & = & ih\theta ^{\mu \nu }\, \, \, .
\end{eqnarray*}
This yields\[
\{x^{\mu },x^{\nu }\}^{'}=\theta ^{\mu \nu }\, \, \, .\]
 On the other hand we have\[
\{x^{\mu },x^{\nu }\}^{'}=\alpha ^{\sigma \tau }(x)(\partial _{\sigma }x^{\mu })(\partial _{\tau }x^{\nu })=\alpha ^{\mu \nu }(x)\, \, \, .\]
 Hence, we can conclude that \( \alpha ^{\mu \nu }(x)=\theta ^{\mu \nu } \)
and therefore we get \( (M_{1}(f,g)-M_{1}(g,f))=i\{f,g\}^{'}=i\theta ^{\mu \nu }(\partial _{\mu }f)(\partial _{\nu }g) \),
which is exactly what we wanted to show. 
\item [(iv)]\( f\star 1=\rho ^{-1}(\rho (f)\cdot 1)=f=\rho ^{-1}(1\cdot \rho (f))=1\star f \)~~.
\end{lyxlist}
Let us summarize what we learned so far: Starting with an arbitrary
ordering prescription \( \rho :\mathcal{A}_{\mathrm{c}}\, \rp \, \mathcal{A}_{\mathrm{nc}} \)
we get by (\ref{de: starproduct via ordering prescription}) a star
product \( \star  \) on \( \mathcal{A}_{\mathrm{c}} \) such that
\( (\mathcal{A}_{\mathrm{c}},\star ) \) is isomorphic as algebra
to \( \mathcal{A}_{\mathrm{nc}} \). Then the question arises immediately,
whether all existing star products correspond to an ordering prescription.
To answer this question we first have to be able to classify star
products. That is why we want to continue by introducing the notion
of equivalence.\\
\emph{Remark:} By the above proof we showed in particular the existence
of star products for \( \mathcal{A}_{\mathrm{c}}[[h]] \). In the
setting of a general Poisson manifold the question of existence is
much more difficult. Nevertheless, it is known that there exist star
products for any Poisson manifold \cite{Kontsevich:1997vb}.

\subsubsection{(b) Equivalence of Star Products}

\bDef\label{de:equivalencetransformation}

Two star products \( \star  \) and \( \star ^{'} \) for \( \mathcal{A}_{\mathrm{c}}[[h]] \)
are called equivalent if there exists a formal series \begin{equation}
\label{equitrans}
S=\mathrm{id}+\sum ^{\infty }_{k=1}h^{k}S_{k}
\end{equation}
of linear operators \( S_{k}:\mathcal{A}_{\mathrm{c}}\rightarrow \mathcal{A}_{\mathrm{c}} \)
such that for all \( f,g\in \mathcal{A}_{\mathrm{c}} \) \begin{equation}
\label{newstar}
f\star ^{'}g=S^{-1}(S(f)\star S(g))\, \, \, \textrm{and}\, \, \, S(1)=1\, \, \, .
\end{equation}
S is called equivalence transformation from \( \star  \) to \( \star ' \)
and for hermitian star products we also require that for all \( f\in \mathcal{A}_{\mathrm{c}} \)
\[
\overline{S(f)}=S(\overline{f})\, \, \, .\]

\eDef

Let us remark that this notion of equivalence indeed defines an equivalence
relation. Moreover, we see that we can formulate equivalence in the
following way, too:\\
\emph{~~Two star products \( \star  \) and \( \star ^{'} \) are
equivalent if and only if \( (\mathcal{A}_{\mathrm{c}}[[h]],\star )\cong (\mathcal{A}_{\mathrm{c}}[[h]],\star ^{'}) \).}\\
We additionally note that by giving an equivalence transformation
as in (\ref{equitrans}) we can construct a new star product defining
\( \star ^{'} \) as in (\ref{newstar}), so that to a given star
product there exist arbitrarily many equivalent ones. Then immediately
arises the question how many equivalence classes exist. For a symplectic
Poisson manifold, this is a Poisson manifolds with non-degenerated
Poisson tensor, the answer is indeed known (to get a definition of
equivalence in the case of a Poisson manifold \( M \) just substitute
in the above definition \( \mathcal{A}_{\mathrm{c}}[[h]] \) by \( C^{\infty }(M) \))
\cite{Kontsevich:1997vb,Gutt}:

\emph{Let \( M \) be a symplectic Poisson manifold. Then \( \{[\star ]\}\cong H_{\mathrm{dR}}^{2}(M)[[h]] \)
where \( [\star ] \) denotes the equivalence class of the star product
\( \star  \).}%
\footnote{\( H^{2}_{\mathrm{dR}} \) denotes the second de Rham cohomology.
} \emph{}\\
In the case \( M=\mathbb {R}^{n} \) then follows that \emph{all}
star products are equivalent since \( H_{\mathrm{dR}}^{2}(\mathbb {R}^{n})=0. \)
In our example we do not consider a Poisson manifold but the Poisson
algebra \( (\mathcal{A}_{\mathrm{c}},\theta ^{\mu \nu }) \). Nevertheless,
we can interpret the Poisson tensor \( \theta ^{\mu \nu } \) as a
Poisson tensor on \( C^{\infty }(\mathbb {R}^{4}) \) and star products
for \( (\mathcal{A}_{\mathrm{c}}[[h]],\theta ^{\mu \nu }) \) become
star products for the Poisson manifold \( (\mathbb {R}^{4},\theta ^{\mu \nu }) \).
If \( \theta ^{\mu \nu } \) is non-degenerated, we have the case
of a symplectic manifold and since on \( \mathbb {R}^{4} \) all star
products are equivalent this leads us to the following conclusion:

\bconcl\emph{\label{th:all*equivalent}}

For \( (\mathcal{A}_{\mathrm{c}}[[h]],\theta ^{\mu \nu }) \) all
star products are equivalent if \( \theta ^{\mu \nu } \) is non-degenerated.

\econcl

Starting with a certain ordering prescription \( \rho  \) and constructing
the corresponding star product \( \star  \) by (\ref{de: starproduct via ordering prescription}),
we now know that all star products that exist are in fact equivalent
to this one. Moreover, we notice that for an arbitrary equivalence
transformation \( S \), \( \rho ^{'}:=\rho \circ S \) again is an
ordering prescription and obviously we have \( f\star ^{'}g:=S^{-1}(S(f)\star S(g))=S^{-1}\circ \rho ^{-1}(\rho \circ S(f)\cdot \rho \circ S(g))=\rho ^{'^{-1}}(\rho ^{'}(f)\cdot \rho ^{'}(g)) \).
Thus, the answer to the question at the end of (a) is:

\begin{Conclusion}\label{Concl: every star form ordering}

Every star product \( \star  \) for \( (\mathcal{A}_{\mathrm{c}}[[h]],\theta ^{\mu \nu }) \)
can be obtained by \[
f\star g=\rho ^{-1}(\rho (f)\cdot \rho (g))\, \, \, ,\]
 where \( \rho  \) is an ordering prescription.%
\footnote{In particular, every star product renders \( \mathcal{A}_{\mathrm{c}} \)
isomorphic to \( \mathcal{A}_{\mathrm{nc}} \). In the general case
of a non-constant Poisson tensor this is not true.
} 

\end{Conclusion}

\subsubsection{(c) Examples for Star Products}

\begin{itemize}
\item The symmetric ordering \( \rho _{s} \) (\ref{symord}) leads to the
star product \begin{equation}
\label{starsym}
f\star _{s}g=\mu \circ e^{\frac{1}{2}ih\theta ^{\mu \nu }(\partial _{\mu }\te \partial _{\nu })}(f\te g)\, \, \, .
\end{equation}
 where \( \mu  \) is to be understood as the multiplication map \( \mu (f\te g)=fg. \)
This star product is called \emph{Moyal-Weyl star product} \cite{Moyal:1949sk,Madore:2000en}.
Moreover, we see that \( \star _{s} \) is a \emph{hermitian} star
product (see Definition \ref{de:star product}): Taking into account
that \( \theta  \) is antisymmetric and real, we obtain: \begin{eqnarray}
\overline{f\star _{s}g} & = & \overline{\mu \circ e^{\frac{1}{2}ih\theta ^{\mu \nu }(\partial _{\mu }\te \partial _{\nu })}(f\te g)}\label{eq:starsymhermitian} \\
 & = & \mu \circ e^{-\frac{1}{2}ih\theta ^{\mu \nu }(\partial _{\mu }\te \partial _{\nu })}(\overline{f}\te \overline{g})\nonumber \\
 & = & \mu \circ e^{\frac{1}{2}ih\theta ^{\nu \mu }(\partial _{\nu }\te \partial _{\mu })}(\overline{g}\te \overline{f})\nonumber \\
 & = & \overline{g}\star _{s}\overline{f}\, \, \, .\nonumber 
\end{eqnarray}

\item The normal ordering \( \rho _{n} \) (\ref{stanord}) leads to the
star product \begin{equation}
\label{*n}
f\star _{n}g=\mu \circ e^{ihm^{\mu \nu }(\partial _{\mu }\te \partial _{\nu })}(f\te g)\, \, \, ,
\end{equation}
 \( \textrm{where }m^{\mu \nu }:=a^{\mu \nu }\theta ^{\mu \nu }\, (\textrm{no summation}!)\, \textrm{with }a^{\mu \nu }:=\left\{ \begin{array}{c}
0,\, \, \, \mu \leq \nu \\
1,\, \, \, \mu >\nu 
\end{array}\right. \, . \) 
\item Let me introduce the following more general star product:\begin{equation}
\label{*k}
f\star _{k}g:=\mu \circ e^{ihm^{\mu \nu }(\partial _{\mu }\te \partial _{\nu })}(f\te g)\, \, \, ,
\end{equation}
 \( \textrm{where }m^{\mu \nu }:=k^{\mu \nu }\theta ^{\mu \nu }\, \textrm{with }k^{\mu \mu }=0\, \textrm{and}\, k^{\mu \nu }+k^{\nu \mu }=1 \)
for \( \mu \neq \nu . \) \\
It is easy to check that \( \star _{k} \) is a star product in the
sense of Definition \ref{de:star product}. As we know from Conclusion
\ref{Concl: every star form ordering}, we can obtain \( \star _{k} \)
by an ordering prescription.%
\footnote{In general this is not unique.
} To get an impression of how such an ordering prescription looks like,
we calculate explicitly for \( \mu \neq \nu  \):\[
x^{\mu }\star _{k}x^{\nu }=x^{\mu }x^{\nu }+ihm^{\mu \nu }\, \, \, \textrm{and}\, \, \, x^{\nu }\star _{k}x^{\mu }=x^{\nu }x^{\mu }+ihm^{\nu \mu }\]
and therefore\begin{eqnarray}
k^{\nu \mu }(x^{\mu }\star _{k}x^{\nu })+k^{\mu \nu }(x^{\nu }\star _{k}x^{\mu }) & = & (k^{\mu \nu }+k^{\nu \mu })x^{\mu }x^{\nu }+ihk^{\mu \nu }k^{\nu \mu }(\theta ^{\mu \nu }+\theta ^{\nu \mu })\nonumber \\
 & = & x^{\mu }x^{\nu }\, \, \, ,\label{cal1} 
\end{eqnarray}
where the last equality follows because \( \theta  \) is antisymmetric
and because \( k^{\mu \nu }+k^{\nu \mu }=1 \). \\
With (\ref{cal1}) we find that an ordering prescription \( \rho _{k}:\mathcal{A}_{\mathrm{c}}\, \rp \, \mathcal{A}_{\mathrm{nc}} \)
which starts with the assignments \begin{eqnarray}
1 & \ra  & 1\label{ordk} \\
x^{\mu } & \ra  & \hat{x}^{\mu }\nonumber \\
x^{\mu }x^{\nu } & \ra  & k^{\nu \mu }\hat{x}^{\mu }\hat{x}^{\nu }+k^{\mu \nu }\hat{x}^{\nu }\hat{x}^{\mu }\nonumber \\
 & \dots  & \nonumber 
\end{eqnarray}
 leads to the star product \( \star _{k} \). Below we will derive
an equivalence transformation \( T_{k} \) from \( \star _{s} \)
to \( \star _{k} \) such that the entire ordering prescription is
given by \( \rho _{k}=\rho _{s}\circ T_{k} \). We see that we must
require \( k^{\mu \nu }+k^{\nu \mu }=1 \) such that \( \rho _{k} \)
equals the identity map for \( h=0 \). The star product \( \star _{k} \)
is obviously a generalization of the first two examples \( \star _{s} \)
and \( \star _{n} \), since we get \( \star _{s} \) for \( k^{\mu \nu }:=\left\{ \begin{array}{c}
\frac{1}{2},\, \, \, \mu \neq \nu \\
0,\, \, \, \mu =\nu 
\end{array}\right.  \)and \( \star _{n} \) for \( k^{\mu \nu }=a^{\mu \nu } \) as in (\ref{*n}). 
\end{itemize}

\subsubsection{(d) Example for Equivalence}

Can we find an explicit equivalence transformation \( T_{k} \) (see
Definition 2) to pass over from the usually used Moyal-Weyl product
\( \star _{s} \) to the more general star product \( \star _{k}? \)
We already know that such a \( T_{k} \) exists because of Conclusion
\ref{th:all*equivalent} and the following Lemma provides it explicitly.

\bla\label{la:transformationTk}

The star products \( \star _{s} \) defined in (\ref{starsym}) and
\( \star _{k} \) defined in (\ref{*k}) are equivalent and the equivalence
transformation from \( \star _{s} \) to \( \star _{k} \) is given
by\begin{equation}
\label{equivtrans}
T_{k}=e^{-\frac{i}{2}hm^{\mu \nu }\partial _{\mu }\partial _{\nu }}\, \, \, ,
\end{equation}
where \( m^{\mu \nu }:=k^{\mu \nu }\theta ^{\mu \nu } \) (no summation!).%
\footnote{Compare with \cite{Waldmann} where quantum mechanics is considered
as an example for deformation quantization.
}

\ela

\bpr

We have to prove that \( T_{k}(f\star _{k}g)=T_{k}(f)\star _{s}T_{k}(g) \)
for all \( f,g \). For this purpose we have to know how to commute
the multiplication map \( \mu  \) with the operator \( T_{k} \).
In this regard the following equation, nothing else than the Leibniz
rule written in tensor notation, turns out to be very useful: \[
m^{\mu \nu }\partial _{\mu }\partial _{\nu }\circ \mu =\mu \circ (m^{\mu \nu }\partial _{\mu }\partial _{\nu }\te \mathrm{id}+\mathrm{id}\te m^{\mu \nu }\partial _{\mu }\partial _{\nu }+m^{\mu \nu }(\partial _{\mu }\te \partial _{\nu })+m^{\mu \nu }(\partial _{\nu }\te \partial _{\mu }))\, \, \, .\]
 Using this commutation relation we can write \begin{eqnarray*}
T_{k}(f\star _{k}g) & = & e^{-\frac{i}{2}hm^{\mu \nu }\partial _{\mu }\partial _{\nu }}\circ \mu \circ e^{ihm^{\mu \nu }(\partial _{\mu }\te \partial _{\nu })}(f\te g)\\
 & = & \mu \circ e^{-\frac{i}{2}h(m^{\mu \nu }\partial _{\mu }\partial _{\nu }\te \mathrm{id}+\mathrm{id}\te m^{\mu \nu }\partial _{\mu }\partial _{\nu }+m^{\mu \nu }(\partial _{\mu }\te \partial _{\nu })+m^{\mu \nu }(\partial _{\nu }\te \partial _{\mu }))}\\
 &  & \circ e^{ihm^{\mu \nu }(\partial _{\mu }\te \partial _{\nu })}(f\te g)\\
 & = & \mu \circ e^{-\frac{i}{2}hm^{\mu \nu }(\partial _{\mu }\te \partial _{\nu }+\partial _{\nu }\te \partial _{\mu })+ihm^{\mu \nu }(\partial _{\mu }\te \partial _{\nu })}(T_{k}(f)\te T_{k}(g))\\
 & = & \mu \circ e^{\frac{i}{2}hm^{\mu \nu }(\partial _{\mu }\te \partial _{\nu })-\frac{i}{2}hm^{\mu \nu }(\partial _{\nu }\te \partial _{\mu })}(T_{k}(f)\te T_{k}(g))\\
 & = & \mu \circ e^{\frac{i}{2}hm^{\mu \nu }(\partial _{\mu }\te \partial _{\nu })-\frac{i}{2}h\theta ^{\mu \nu }(\partial _{\nu }\te \partial _{\mu })+\frac{i}{2}hk^{\nu \mu }\theta ^{\mu \nu }(\partial _{\nu }\te \partial _{\mu })}(T_{k}(f)\te T_{k}(g))\\
 & = & \mu \circ e^{\frac{i}{2}hm^{\mu \nu }(\partial _{\mu }\te \partial _{\nu })+\frac{i}{2}h\theta ^{\mu \nu }(\partial _{\mu }\te \partial _{\nu })-\frac{i}{2}hk^{\nu \mu }\theta ^{\nu \mu }(\partial _{\nu }\te \partial _{\mu })}(T_{k}(f)\te T_{k}(g))\\
 & = & \mu \circ e^{\frac{i}{2}h\theta ^{\mu \nu }(\partial _{\mu }\te \partial _{\nu })}(T_{k}(f)\te T_{k}(g))\\
 & = & T_{k}(f)\star _{s}T_{k}(g)\, \, \, ,
\end{eqnarray*}
 where we used in the third line that partial derivatives commute
with each other, in the fifth that \( m^{\mu \nu }=k^{\mu \nu }\theta ^{\mu \nu }=(1-k^{\nu \mu })\theta ^{\mu \nu } \)
since \( k^{\mu \nu }+k^{\nu \mu }=1 \) if \( \mu \neq \nu  \) and
in the next to last line again that \( m^{\mu \nu }=k^{\mu \nu }\theta ^{\mu \nu } \).
Furthermore, we used that \( \theta  \) is an antisymmetric tensor.
The last equation follows directly from the explicit expression of
\( \star _{s} \) given in (\ref{starsym}). This proves the lemma.

\epr

We have finished our discussion of orderings and star products and
proceed to study how far the gauge theory developed in \cite{Madore:2000en,Jurco:2001rq,Jurco:2000dx}
for the star product \( \star _{s} \) is affected if we take a different
star product.

\section{Gauge Field Theory in the Case of a constant Poisson Tensor \protect\( \theta \protect \)}

A detailed description of gauge field theory in the case of a constant
Poisson structure can be found in \cite{Jurco:2001rq,Madore:2000en,Jurco:2000dx,Behr,Syko}.
We assume the general ideas to be known and present in the first subsection
just a brief summary of results that are important in our context.
Then we will rather concentrate on what has to be modified in the
theory if we change the star product: We will discuss the issue of
involution and study how the Seiberg-Witten maps depend on the choice
of the star product. Finally, in the last subsection we will discuss
how far physical considerations confine the full freedom in choosing
a star product.

\subsection{Noncommutative Gauge Field Theory\label{se:NCGFT} }

Matter fields are functions in \( \mathcal{A}_{\mathrm{c}} \) and
an infinitesimal noncommutative gauge transformations of a matter
field \( \hat{\psi } \) is defined as \cite{Madore:2000en}:%
\footnote{Functions with a hat are indeed to be understood as elements of \( \mathcal{A}_{\mathrm{c}} \)
and not as elements of \( \mathcal{A}_{\mathrm{nc}} \). We add the
hat to distinguish noncommutative fields and ordinary commutative
fields for the noncommutative ones will be expressed by the commutative
ones (\ref{eq: Seiberg-W}). 
} \begin{equation}
\label{de:trafopsi}
\delta \hat{\psi }(x)=i\hat{\Lambda }(x)\star _{s}\hat{\psi }(x)
\end{equation}
 and \[
\delta x^{\mu }=0\, \, \, .\]
Since multiplication of \( \hat{\psi } \) on the left by coordinates
is not a covariant operation, \emph{covariant coordinates} \( \hat{X}^{\mu } \)
are introduced: \begin{equation}
\label{de:cov.cord}
\hat{X}^{\mu }:=x^{\mu }+\hat{A}^{\mu }\, \, \, ,
\end{equation}
 where \( \hat{A}^{\mu } \) has to transform as\begin{equation}
\label{de: trafo von A oben i}
\delta \hat{A}^{\mu }=-i[x^{\mu }\stackrel{\star _{s}}{,}\hat{\Lambda }]+i[\hat{\Lambda }\stackrel{\star _{s}}{,}\hat{A}^{\mu }]
\end{equation}
to render \( \hat{X}^{\mu } \) covariant.%
\footnote{This means \( \delta \hat{X}^{\mu }=i[\hat{\Lambda }\stackrel{\star _{s}}{,}\hat{X}^{\mu }] \).
} If \( \theta  \) is non-degenerated, we can use \( \theta  \) to
upper and lower indices and can define \[
\hat{A}^{\mu }=:h\theta ^{\mu \nu }\hat{A}_{\nu }\, \, \, .\]
The gauge transformation of \( \hat{A}_{\nu } \) is then given by\begin{equation}
\label{eq:trafoA}
\delta \hat{A}_{\nu }=\partial _{\nu }\hat{\Lambda }+i[\hat{\Lambda }\stackrel{\star _{s}}{,}\hat{A}_{\nu }]\, \, \, ,
\end{equation}
 since we have \begin{equation}
\label{eq: commutator equals derivativer for star_s}
-i[x^{\mu }\stackrel{\star _{s}}{,}f]=h\theta ^{\mu \nu }\partial _{\nu }f
\end{equation}
 for all \( f \). Additionally, a \emph{covariant field strength}
and a \emph{covariant derivative} are introduced:\begin{eqnarray}
\hat{F}_{\mu \nu } & = & \partial _{\mu }\hat{A}_{\nu }-\partial _{\nu }\hat{A}_{\mu }-i[\hat{A}_{\mu }\stackrel{\star _{s}}{,}\hat{A}_{\nu }]\label{de:cov.F} \\
\hat{D}_{\mu }\hat{\psi } & = & \partial _{\mu }\hat{\psi }-i\hat{A}_{\mu }\star _{s}\hat{\psi }\label{de:cov.D} 
\end{eqnarray}
with\begin{equation}
\label{eq:gaugetrafoF}
\delta \hat{F}_{\mu \nu }=i[\hat{\Lambda }\stackrel{\star _{s}}{,}\hat{F}_{\mu \nu }]
\end{equation}
 and\begin{equation}
\label{eq:guagetrafoD}
\delta \hat{D}_{\mu }\hat{\psi }=i\hat{\Lambda }\star _{s}\hat{D}_{\mu }\hat{\psi }\, \, \, .
\end{equation}
In the general case we want to describe non-abelian gauge field theories
based on a Lie algebra\[
[T^{a},T^{b}]=if^{ab}{}_{c}T^{c}\, \, \, .\]
In the noncommutative setting, however, \( \hat{\Lambda } \) has
to lie in the enveloping algebra, i.e.\begin{equation}
\label{eq:lambda in env.algebra}
\hat{\Lambda }(x)=\hat{\Lambda }_{a}(x)T^{a}+\hat{\Lambda }^{1}_{ab}(x):T^{a}T^{b}:+\dots +\hat{\Lambda }_{a_{1}\dots a_{n}}^{n-1}:T^{a_{1}}\dots T^{a_{n}}:+\dots \, \, \, ,
\end{equation}
with \begin{eqnarray*}
:T^{a}: & = & T^{a}\\
:T^{a}T^{b}: & = & \frac{1}{2}\{T^{a},T^{b}\}\\
:T^{a_{1}}\dots T^{a_{n}}: & = & \frac{1}{n!}\sum _{\pi \in S_{n}}T^{a_{\pi (1)}}\dots T^{a_{\pi (n)}}\, \, \, .
\end{eqnarray*}
 This is because the star commutator of two Lie algebra valued transformation
parameters doesn't close in the Lie algebra anymore \cite{Jurco:2000ja}.
On first glance this seems to lead to infinitely many parameters \( \hat{\Lambda }_{a_{1}\dots a_{n}}^{n-1} \),
but the gauge transformations can be restricted to those which permit
to express these infinitely many parameters by the finitely many commutative
ones. This is done by means of the so called \emph{Seiberg-Witten
maps} \begin{eqnarray}
\hat{\Lambda } & = & \hat{\Lambda }_{\alpha }[A_{\mu }]\nonumber \\
\hat{A}_{\mu } & = & \hat{A}_{\mu }[A_{\mu }]\label{eq: Seiberg-W} \\
\hat{\psi } & = & \hat{\psi }[\psi ,A_{\mu }]\, \, \, .\nonumber 
\end{eqnarray}
Here, \( \alpha  \) denotes the commutative gauge parameter, \( A_{\mu } \)
the commutative gauge field and \( \psi  \) the commutative matter
field. To calculate the explicit dependence of the noncommutative
fields on the commutative ones we assume that it is possible to expand
\( \hat{\Lambda }_{\alpha }[A_{\mu }],\, \hat{A}_{\mu }[A_{\mu }],\, \hat{\psi }[\psi ,A_{\mu }] \)
in the formal deformation parameter \( h \):\begin{eqnarray}
\hat{\Lambda }_{\alpha }[A_{\mu }] & = & \alpha +h\hat{\Lambda }^{1}_{\alpha }[A_{\mu }]+h^{2}\hat{\Lambda }^{2}_{\alpha }[A_{\mu }]+\dots \nonumber \\
\hat{A}_{\mu }[A_{\mu }] & = & A_{\mu }+h\hat{A}^{1}_{\mu }[A_{\mu }]+h^{2}\hat{A}^{2}_{\mu }[A_{\mu }]+\dots \label{eq: SW expansion in h} \\
\hat{\psi }[\psi ,A_{\mu }] & = & \psi +h\hat{\psi }^{1}[\psi ,A_{\mu }]+h^{2}\hat{\psi }^{2}[\psi ,A_{\mu }]+\dots \, \, \, .\nonumber 
\end{eqnarray}
Finally, we get the explicit dependence on the commutative fields
by requiring the following \emph{consistency condition }

\begin{equation}
\label{eq:consistency}
\begin{array}{cccc}
 & (\delta _{\alpha }\delta _{\beta }-\delta _{\beta }\delta _{\alpha })\hat{\psi } & = & \delta _{-i[\alpha ,\beta ]}\hat{\psi }\\
\Leftrightarrow  & i\delta _{\alpha }\hat{\Lambda }_{\beta }-i\delta _{\beta }\hat{\Lambda }_{\alpha }+[\hat{\Lambda }_{\alpha }\stackrel{\star _{s}}{,}\hat{\Lambda }_{\beta }] & = & i\hat{\Lambda }_{-i[\alpha ,\beta ]}
\end{array}
\end{equation}
and by requiring that the noncommutative gauge transformations are
induced by the commutative gauge transformations of the commutative
fields the noncommutative ones depend on: 

\begin{eqnarray}
\hat{\Lambda }_{\alpha }[A_{\mu }]+\delta \hat{\Lambda }_{\alpha }[A_{\mu }] & = & \hat{\Lambda }_{\alpha }[A_{\mu }+\delta A_{\mu }]\nonumber \\
\hat{A}_{\mu }[A_{\mu }]+\delta \hat{A}_{\mu }[A_{\mu }] & = & \hat{A}_{\mu }[A_{\mu }+\delta A_{\mu }]\label{eq:SWequ} \\
\hat{\psi }[\psi ,A_{\mu }]+\delta \hat{\psi }[\psi ,A_{\mu }] & = & \hat{\psi }[\psi +\delta \psi ,A_{\mu }+\delta A_{\mu }]\, \, \, .\nonumber 
\end{eqnarray}
These equations were explicitly solved in \cite{Jurco:2001rq} up
to second order in \( h \) for the star product \( \star _{s} \).\\
\emph{Remark}: In \cite{Jurco:2000ja} it is shown that for \( \hat{\Lambda } \)
exists a solution where the \( n \)-th order of \( h \) corresponds
to the \( n+1 \)-th order of \( T^{a} \).

\subsection{Noncommutative Gauge Field Theory for an arbitrary star product\label{se: NCGFT for arbitrary star product}}

Instead of the Moyal-Weyl product \( \star _{s} \) we can also use
every other star product that is equivalent to this one. Let \( T \)
be an equivalence transformation from \( \star _{s} \) to \( \star  \),
i.e. we have \[
f\star g=T^{-1}(T(f)\star _{s}T(g))\, \, \, .\]
Matter fields \( \hat{\psi }' \) are still elements of \( \mathcal{A}_{\mathrm{c}} \)
with the following gauge transformation law:\begin{equation}
\label{eq: gauge trafo for Psi' for arbitrary star product}
\delta \hat{\psi }'(x)=i\hat{\Lambda }'(x)\star \hat{\psi }'(x)\, \, \, .
\end{equation}
 We can again introduce covariant coordinates \[
\hat{X}'^{\mu }:=x^{\mu }+\hat{A}'^{\mu }\]
 where \( \hat{A}' \) must transform under gauge transformations
as\[
\delta \hat{A}'^{\mu }=-i[x^{\mu }\stackrel{\star }{,}\hat{\Lambda }']+i[\hat{\Lambda }'\stackrel{\star }{,}\hat{A}'^{\mu }]\, \, \, .\]
 If \( \theta  \) is non-degenerated, we can again lower indices
using \( \theta  \) as we did in the case of the Moyal-Weyl product
and can define\[
\hat{A}'^{\mu }=\theta ^{\mu \nu }\hat{A}'_{\nu }\, \, \, .\]
 The transformation law of \( \hat{A}'_{\nu } \) is given by\[
\theta ^{\mu \nu }\delta \hat{A}'_{\nu }=-i[x^{\mu }\stackrel{\star }{,}\hat{\Lambda }']+i\theta ^{\mu \nu }[\hat{\Lambda }'\stackrel{\star }{,}\hat{A}'_{\nu }]\, \, \, .\]
 New is that in the case of an arbitrary star product we do \emph{not}
have that \\
\( -i[x^{\mu }\stackrel{\star }{,}f]=h\theta ^{\mu \nu }\partial _{\nu }f \)
as it is the case for \( \star _{s} \). The reason is that in general
the usual partial derivative \( \partial _{\nu } \) is \emph{not}
a derivation for \( (\mathcal{A}_{\mathrm{c}},\star ) \), i.e. in
general \( \partial _{\nu }(f\star g)\neq \partial _{\nu }(f)\star g+f\star \partial _{\nu }(g) \),
whereas the commutator \( -i[x^{\mu }\stackrel{\star }{,}f] \) is
a derivation. Nonetheless, we have: 

\begin{Remark}\label{remark: partial derivative for arbitary star}

Let \( \star  \) be an arbitrary star product equivalent to \( \star _{s} \)
and let \( T \) be the equivalence transformation from \( \star _{s} \)
to \( \star  \). Then \begin{equation}
\label{de: derivation for arbitrary star product}
\partial '_{\mu }:=T^{-1}\circ \partial _{\mu }\circ T
\end{equation}
 is a derivation for \( (\mathcal{A}_{\mathrm{c}},\star ) \).

\end{Remark}

\begin{proof}

Since \( T \) is an equivalence transformation from \( \star _{s} \)
to \( \star  \), we have for arbitrary functions \( f,g \)\begin{eqnarray*}
\partial '_{\mu }(f\star g) & = & \partial '_{\mu }(T^{-1}(T(f)\star _{s}T(g)))\\
 & = & T^{-1}\circ \partial _{\mu }(T(f)\star _{s}T(g))\, \, \, .
\end{eqnarray*}
 As \( \partial _{\mu } \) is a derivation for \( (\mathcal{A}_{\mathrm{c}},\star _{s}) \),
we obtain \begin{eqnarray*}
 & = & T^{-1}(\partial _{\mu }(T(f))\star _{s}T(g)+T(f)\star _{s}\partial _{\mu }(T(g)))\\
 & = & T^{-1}(T(\partial '_{\mu }(f)\star _{s}T(g)+T(f)\star _{s}T(\partial '_{\mu }(g))\\
 & = & \partial '_{\mu }(f)\star g+f\star \partial '_{\mu }(g)\, \, \, .
\end{eqnarray*}
 \end{proof}

Furthermore, let us consider only those star products, which correspond
to an ordering prescription that maps all coordinates \( x^{\mu } \)
to their noncommutative analogs \( \hat{x}^{\mu } \).%
\footnote{If we assign to coordinates the dimension of length, this is a physical
requirement which guarantees that the dimension is preserved.
} Until now, we just demanded \( \rho (x^{\mu })=\hat{x}^{\mu }+\mathcal{O}(h) \)
(Subsection \ref{se: ordering}). In this case we have \[
T(x^{\mu })=x^{\mu }\]
 and we indeed obtain \[
-i[x^{\mu }\stackrel{\star }{,}f]=h\theta ^{\mu \nu }\partial '_{\nu }f\, \, \, .\]
 \begin{proof}

We know from Conclusion \ref{th:all*equivalent} and Conclusion \ref{Concl: every star form ordering}
that \( \star  \) corresponds to an ordering prescription \( \rho  \).
One possibility is given by \( \rho =\rho _{s}\circ T \), where \( \rho _{s} \)
is the symmetric ordering prescription (\ref{symord}), since obviously
\[
\rho ^{-1}(\rho (f)\circ \rho (g))=T^{-1}\circ \rho _{s}^{-1}(\rho _{s}(T(f))\circ \rho _{s}(T(g))=T^{-1}(T(f)\star _{s}T(g))=f\star g\, \, \, .\]
 If we require as above that \( \rho _{s}\circ T(x^{\mu })=\rho (x^{\mu })=\hat{x}^{\mu }=\rho _{s}(x^{\mu }) \),
this yields \( T(x^{\mu })=x^{\mu } \). 

In this case we have because of (\ref{eq: commutator equals derivativer for star_s})
and (\ref{de: derivation for arbitrary star product}) \begin{eqnarray*}
-i[x^{\mu }\stackrel{\star }{,}f] & = & -iT^{-1}[T(x^{\mu })\stackrel{\star _{s}}{,}T(f)]=-iT^{-1}[x^{\mu }\stackrel{\star _{s}}{,}T(f)]\\
 & = & T^{-1}(\theta ^{\mu \nu }\partial _{\nu }T(f))=\theta ^{\mu \nu }\partial '_{\nu }f\, \, \, .
\end{eqnarray*}

\end{proof}

Thus, we finally derive the gauge transformation law for \( \hat{A}'_{\nu } \):\begin{equation}
\label{eq: gauge transform for A' corresponding to arbitrary star product}
\delta \hat{A}'_{\nu }=\partial '_{\nu }\hat{\Lambda }'+i[\hat{\Lambda }'\stackrel{\star }{,}\hat{A}'_{\nu }]\, \, \, .
\end{equation}

Analogously, the field strength and the covariant derivative for an
arbitrary star product \( \star  \) read\begin{equation}
\label{eq: covariant derivative and field strength for arbitrary star product}
\begin{array}{ccc}
\hat{F}'_{\mu \nu } & = & \partial '_{\mu }\hat{A}'_{\nu }-\partial '_{\nu }\hat{A}'_{\mu }-i[\hat{A}'_{\mu }\stackrel{\star }{,}\hat{A}'_{\nu }]\\
\hat{D}'_{\mu }\hat{\psi }' & = & \partial '_{\mu }\hat{\psi }'-i\hat{A}'_{\mu }\star \hat{\psi }'\, \, \, ,
\end{array}
\end{equation}
with\begin{equation}
\label{eq: gauge trafo for Fprime}
\delta \hat{F}'_{\mu \nu }=i[\hat{\Lambda }'\stackrel{\star }{,}\hat{F}'_{\mu \nu }]
\end{equation}
and \begin{equation}
\label{eq: gauge trafo for Dprime}
\delta \hat{D}'_{\mu }\hat{\psi }'=i\hat{\Lambda }'\star \hat{D}'_{\mu }\hat{\psi }'\, \, \, .
\end{equation}
Now we can proceed as done for the star product \( \star _{s} \)
in the previous subsection and express the noncommutative fields in
terms of the commutative ones by means of the Seiberg-Witten maps
for the star product \( \star  \). A solution for those Seiberg-Witten
maps is given later.

\subsection{Noncommutative Gauge Theory and Involution\label{se:NCGT and involution}}

\subsubsection{Definition and Examples}

A \emph{{*}-involution} (Don't mix up {*} and \( \star  \) !) of
an algebra \( \mathcal{A} \) is an anti-linear map such that for
all \( a\in \mathcal{A} \) \[
(ab)^{*}=b^{*}a^{*}\, \, \, \textrm{and}\, \, \, (a^{*})^{*}=a\, \, \, .\]
Now we make an important observation:\\
\( \mathcal{A}_{\mathrm{c}} \) together with the symmetric star product
\( \star _{s} \) admits the usual complex conjugation as involution.
We have already given the proof in Subsection \ref{se: star products}
(c), where we checked that \( \star _{s} \) is a hermitian star product
(see (\ref{eq:starsymhermitian})). On the other hand it can be immediately
shown that for \( (\mathcal{A}_{\mathrm{c}},\star _{n}) \) the complex
conjugation is \emph{not} an involution (the complex conjugation is
not even an involution for \( (\mathcal{A}_{\mathrm{c}},\star _{k}) \)
whenever \( \star _{k}\neq \star _{s} \) (cf. (\ref{*k}))).

The question arises how we can find an involution \( i_{\star } \)
for an arbitrary star product \( \star  \). The following proposition
gives the answer.

\begin{Prop}\label{th:involution}

We get an involution for \( (\mathcal{A}_{\mathrm{c}},\star ) \)
by defining\begin{equation}
\label{involutionarbitrary}
i_{\star }:=T^{-1}\circ \overline{\, \cdot \, }\circ T\, \, \, ,
\end{equation}
where \( \overline{\, \cdot \, } \) denotes the ordinary complex
conjugation and \( T \) is the equivalence transformation from \( \star _{s} \)
to \( \star  \) (whose existence is guaranteed by Conclusion \ref{th:all*equivalent}).

\end{Prop}

\bpr

Let be \( f,g\in \mathcal{A}_{\mathrm{c}} \) arbitrary. With the
definition of \( i_{\star } \) (\ref{involutionarbitrary}), by using
that \( T \) is an equivalence transformation and because \( \overline{\, \cdot \, } \)
is an involution for \( \star _{s} \) we obtain \begin{eqnarray*}
i_{\star }(f\star g) & = & T^{-1}\circ \overline{\, \cdot \, }\circ T(T^{-1}(T(f)\star _{s}T(g)))\\
 & = & T^{-1}(\overline{T(f)\star _{s}T(g)})\\
 & = & T^{-1}(\overline{T(g)}\star _{s}\overline{T(f)})\\
 & = & T^{-1}(T(T^{-1}\circ \overline{\, \cdot \, }\circ T(g))\star _{s}T(T^{-1}\circ \overline{\, \cdot \, }\circ T(f)))\\
 & = & i_{\star }(g)\star i_{\star }(f)\, \, \, .
\end{eqnarray*}
Obviously \( i^{2}_{\star }(f)=f \) is satisfied, too and \( i_{\star } \)
is an involution in the sense of the definition given above.

\epr

\subsubsection{The Gauge Invariant Matter Field Term}

We still haven't defined a gauge invariant action. For the Moyal-Weyl
product \( \star _{s} \) this was done in \cite{Schupp:2001we}.
A noncommutative \emph{Yang-Mills action} is defined in this case
as follows:

\begin{equation}
\label{eq:action}
S=\int d^{4}x\, \mathcal{L}=\int d^{4}x\, \left[ -\frac{1}{4}\textrm{tr}\hat{F}_{\mu \nu }\star _{s}\hat{F}^{\mu \nu }+\overline{\hat{\psi }}\star _{s}(i\gamma ^{\mu }\hat{D}_{\mu }-m)\hat{\psi }\right] \, \, \, ,
\end{equation}
where \( \overline{\hat{\psi }}:=\hat{\psi }^{\dagger }\gamma ^{0} \)
denotes the adjoint matter field. Let us look at this action: In analogy
to the commutative case we want to obtain a gauge invariant expression
\begin{equation}
\label{eq:psibar*psi}
\overline{\hat{\psi }}\star _{s}\hat{\psi }\, \, \, .
\end{equation}
Since \begin{equation}
\label{eq:trafopsibar}
\delta \overline{\hat{\psi }}=-i\overline{\hat{\psi }}\star _{s}\hat{\Lambda }^{\dagger }
\end{equation}
 we have\begin{eqnarray*}
\delta (\overline{\hat{\psi }}\star _{s}\hat{\psi }) & = & -i\overline{\hat{\psi }}\star _{s}\hat{\Lambda }^{\dagger }\star _{s}\hat{\psi }+i\overline{\hat{\psi }}\star _{s}\hat{\Lambda }\star _{s}\hat{\psi }\\
 & = & i\overline{\hat{\psi }}\star _{s}(\hat{\Lambda }-\hat{\Lambda }^{\dagger })\star _{s}\hat{\psi }
\end{eqnarray*}
and therefore\[
\begin{array}{crcl}
 & \delta (\overline{\hat{\psi }}\star _{s}\hat{\psi }) & = & 0\\
\Leftrightarrow  & \hat{\Lambda } & = & \hat{\Lambda }^{\dagger }\, \, \, .
\end{array}\]
This may seem to be trivial but in fact (\ref{eq:trafopsibar}) is
only true because \( \star _{s} \) is a hermitian star product (\ref{eq:starsymhermitian})
and thereby \begin{equation}
\label{eq: dagger is involutive}
(\hat{\Lambda }\star _{s}\hat{\psi })^{\dagger }=\hat{\psi }^{\dagger }\star _{s}\hat{\Lambda }^{\dagger }\, \, \, .
\end{equation}
 We want to demand \( \hat{\Lambda } \) to be hermitian, i.e. \begin{equation}
\label{eq:lambdareal}
\hat{\Lambda }=\hat{\Lambda }^{\dagger }\, \, \, ,
\end{equation}
 in analogy to the commutative case such that (\ref{eq:psibar*psi})
is indeed gauge invariant. Let us recall that in contrast to the commutative
case \( \hat{\Lambda } \) does not lie in the Lie algebra but in
the enveloping algebra. If we use a basis of the enveloping algebra
that is totally symmetric with respect to the hermitian Lie-algebra
generators \( T^{a} \) and write \( \hat{\Lambda } \) as in (\ref{eq:lambda in env.algebra}),
we apparently obtain \[
\begin{array}{cccc}
 & \hat{\Lambda } & = & \hat{\Lambda }^{\dagger }\\
\Leftrightarrow  & \overline{\hat{\Lambda }_{a_{1}\dots a_{n}}^{n-1}} & = & \hat{\Lambda }_{a_{1}\dots a_{n}}^{n-1}
\end{array}\]
for all \( n \), where the bar denotes the usual complex conjugation.
Thus, \( \hat{\Lambda } \) is hermitian if all coefficient functions
are real.

If we want to use an arbitrary star product we have to adjust the
hermiticity condition since the complex conjugation in general will
not be an involution and therefore (\ref{eq: dagger is involutive})
will not hold. Let us define the \emph{star-dagger} \begin{equation}
\label{de:dagger*}
\dagger _{\star }:=(i_{\star }(\cdot ))^{tr}\, \, \, ,
\end{equation}
 where \( i_{\star } \) denotes the involution given in Conclusion
\ref{th:involution} that corresponds to an arbitrary star product
\( \star  \). Let us denote by \( \hat{\psi }' \) the matter field
for the case where we use \( \star  \). Then, we want to define the
\emph{adjoint matter field}\begin{equation}
\label{de:adjointpsi}
\textrm{ad}_{\star }(\hat{\psi }'):=(\hat{\psi }')^{\dagger _{\star }}\gamma ^{0}=\left( i_{\star }(\hat{\psi }')\right) ^{tr}\gamma ^{0}\, \, \, .
\end{equation}
Furthermore, \( \mathrm{ad}_{\star }(\hat{\psi }')\star \hat{\psi }' \)
is to be gauge invariant for an arbitrary star product, i.e.\[
\begin{array}{crcl}
 & \delta (\textrm{ad}_{\star }(\hat{\psi }')\star \hat{\psi }') & = & 0\\
\Leftrightarrow  & \delta \textrm{ad}_{\star }(\hat{\psi }')\star \hat{\psi }'+\textrm{ad}_{\star }(\hat{\psi }'\star \delta \hat{\psi }' & = & 0\\
\Leftrightarrow  & -i\, \textrm{ad}_{\star }(\hat{\psi }')\star (\hat{\Lambda }')^{\dagger _{\star }}\star \hat{\psi }'+i\, \textrm{ad}_{\star }(\hat{\psi }')\star \hat{\Lambda }'\star \hat{\psi }' & = & 0\\
\Leftrightarrow  & i\, \textrm{ad}_{\star }(\hat{\psi }')\star (\hat{\Lambda }'-(\hat{\Lambda }')^{\dagger _{\star }})\star \hat{\psi }' & = & 0\\
\Leftrightarrow  & \hat{\Lambda }' & = & (\hat{\Lambda }')^{\dagger _{\star }}\, \, \, .
\end{array}\]
 Hence,  we demand the \emph{hermiticity condition} \begin{equation}
\label{eq:lambda'real}
\hat{\Lambda }'=(\hat{\Lambda }')^{\dagger _{\star }}\, \, \, .
\end{equation}
 Can we find a connection between \( \hat{\Lambda }' \) and \( \hat{\Lambda } \),
the gauge parameter for \( \star _{s} \)? Having \( \hat{\Lambda } \)
with \( \hat{\Lambda }^{\dagger }=\hat{\Lambda } \) (\ref{eq:lambdareal}),
we can write\[
\begin{array}{crcl}
 & \hat{\Lambda }^{\dagger } & = & \hat{\Lambda }\\
\Leftrightarrow  & \left( (T^{-1}\circ \overline{\, \cdot \, }\circ T)T^{-1}\hat{\Lambda }\right) ^{tr} & = & T^{-1}\hat{\Lambda }\\
\Leftrightarrow  & \left( i_{\star }(T^{-1}\hat{\Lambda })\right) ^{tr} & = & T^{-1}\hat{\Lambda }\\
\Leftrightarrow  & (T^{-1}\hat{\Lambda })^{\dagger _{\star }} & = & T^{-1}\hat{\Lambda }\, \, \, ,
\end{array}\]
where we used the definition of the involution \( i_{\star } \) for
an arbitrary star product given in Proposition \ref{th:involution}.
Defining \begin{equation}
\label{de:lambda'}
\hat{\Lambda }':=T^{-1}\hat{\Lambda }\, \, \, ,
\end{equation}
 \( \hat{\Lambda }' \) actually satisfies (\ref{eq:lambda'real})
and therefore \( \mathrm{ad}_{\star }(\hat{\psi }')\star \hat{\psi }' \)
becomes gauge invariant. \\
\emph{Remark:} The gauge parameter \( \hat{\Lambda }' \)can be expressed
in terms of the commutative fields by solving the Seiberg-Witten equations
(\ref{eq:SWequ}), which in fact depend on the choice of the star
product, too. In the next subsection we will show that (\ref{de:lambda'})
is actually consistent with the results we get this way.

\subsection{Seiberg-Witten Map for Arbitrary Star Products\label{se:SW} }

It is convenient to start determining \( \hat{\Lambda }_{\alpha }[A_{\mu }] \)
by means of the consistency condition (\ref{eq:consistency}). We
can put the expansion (\ref{eq: SW expansion in h}) in the consistency
relation (\ref{eq:consistency}) and solve the equation order by order.
We see that the consistency condition (\ref{eq:consistency}) contains
a \( \star  \). It therefore depends on the special choice of a star
product (remember that \( \hat{\Lambda } \) lies in the enveloping
Lie algebra so that \( [\hat{\Lambda }_{\alpha }\stackrel{\star }{,}\hat{\Lambda }_{\beta }] \)
depends on the star product \( \star  \) even to first order in \( h \)).
The solution to zeroth order is the commutative gauge parameter \( \alpha  \)
whereas for higher orders we obtain differential equations that depend
on the choice of the star product.

For the Moyal-Weyl product the equation and the solution for the first
order is given in \cite{Jurco:2001rq}. We are not interested in the
explicit form of equations and solutions but just want to expand the
solution in orders of \( h \) \[
\hat{\Lambda }^{s}_{\alpha }[A_{\mu }]=\alpha +h\hat{\Lambda }^{s,1}_{\alpha }[A_{\mu }]+h^{2}\hat{\Lambda }^{s,2}_{\alpha }[A_{\mu }]+\dots \, \, \, ,\]
where \( s \) stands for {}``symmetric''. We don't want to specify
these equations but treat the problem generally. Of course it is possible
to solve the equations explicitly order by order if one fixes a star
product, but with the knowledge we have gathered so far we can give
a solution a lot more elegantly and in full generality. So let \( \star  \)
in (\ref{eq:consistency}) be an arbitrary star product. As we know,
\( \star  \) is equivalent to the Moyal-Weyl product \( \star _{s} \)
(Conclusion \ref{th:all*equivalent}) and we want to denote the equivalence
transformation from \( \star _{s} \) to \( \star  \) by \( T \).
Thus,\begin{equation}
\label{eq:equitransT}
f\star g=T^{-1}(T(f)\star _{s}T(g))\, \, \, .
\end{equation}
With this input we get:

\begin{Prop}\label{th:lambda s}

If \( \hat{\Lambda }^{s} \)is a solution of the consistency condition
(\ref{eq:consistency}) for \( \star _{s} \) then \( T^{-1}(\hat{\Lambda }^{s}) \)
is a solution of the consistency condition for \( \star . \)

\end{Prop}

\begin{proof}

To show that \( T^{-1}(\hat{\Lambda }^{s}) \) satisfies the consistency
condition for \( \star  \) given in (\ref{eq:consistency}), we will
reduce the problem to the consistency condition for \( \star _{s} \)
such that we can make use of the fact that \( \hat{\Lambda }^{s} \)
is a solution. This can be done using that \( T \) is an equivalence
transformation from \( \star  \) to \( \star _{s} \) which yields
(\ref{eq:equitransT}). This explicitly reads \[
\begin{array}{ll}
 & i\delta _{\alpha }T^{-1}(\hat{\Lambda }^{s}_{\beta })-i\delta _{\beta }T^{-1}(\hat{\Lambda }^{s}_{\alpha })+[T^{-1}(\hat{\Lambda }^{s}_{\alpha })\stackrel{\star }{,}T^{-1}(\hat{\Lambda }^{s}_{\beta })]\\
= & i\delta _{\alpha }T^{-1}(\hat{\Lambda }^{s}_{\beta })-i\delta _{\beta }T^{-1}(\hat{\Lambda }^{s}_{\alpha })+T^{-1}([T(T^{-1}(\hat{\Lambda }^{s}_{\alpha }))\stackrel{\star _{s}}{,}T(T^{-1}(\hat{\Lambda }^{s}_{\beta }))]\\
= & T^{-1}(i\delta _{\alpha }\hat{\Lambda }^{s}_{\beta }-i\delta _{\beta }\hat{\Lambda }^{s}_{\alpha }+[\hat{\Lambda }^{s}_{\alpha }\stackrel{\star _{s}}{,}\hat{\Lambda }^{s}_{\beta }])\\
= & T^{-1}(i\hat{\Lambda }^{s}_{-i[\alpha ,\beta ]})\\
= & iT^{-1}(\hat{\Lambda }^{s}_{-i[\alpha ,\beta ]})
\end{array}\]
which proves the claim. 

\end{proof}

We emphasize that this result is exactly what we got at the end of
the previous subsection to guarantee an invariant matter field term
in the Lagrangian (cf. (\ref{de:lambda'})).

This is a nice result: Obviously we can immediately give a solution
for the consistency condition (\ref{eq:consistency}) for an arbitrary
star product if we know it for \( \star _{s} \) (this is known up
to second order) and if we know \( T \) (whereas this indeed can
be a problem).

Let us continue with searching a solution for \( \hat{\psi } \) and
\( \hat{A}_{\mu } \). If we write out the left hand sides of the
Seiberg-Witten equations introduced in (\ref{eq:SWequ}), we get with
the gauge transformation laws (\ref{eq: gauge trafo for Psi' for arbitrary star product})
and (\ref{eq: gauge transform for A' corresponding to arbitrary star product})
the following star product-dependent equations\begin{equation}
\label{eq:SWeqwith*}
\begin{array}{rcl}
\hat{\psi }[\psi ,A_{\mu }]+\delta \hat{\psi }[\psi ,A_{\mu }] & = & \hat{\psi }[\psi ,A_{\mu }]+i\hat{\Lambda }_{\alpha }[A_{\mu }]\star \hat{\psi }[\psi ,A_{\mu }]\\
\hat{A}_{\mu }[A_{\mu }]+\delta \hat{A}_{\mu }[A_{\mu }] & = & \hat{A}_{\mu }[A_{\mu }]+\partial '_{\mu }\hat{\Lambda }_{\alpha }[A_{\mu }]+i[\hat{\Lambda }_{\alpha }[A_{\mu }]\stackrel{\star }{,}\hat{A}_{\mu }[A_{\mu }]]\, \, \, ,
\end{array}
\end{equation}
 where \( \partial '_{\mu } \) was introduced in Remark \ref{remark: partial derivative for arbitary star}.
Again we can take the expansions of \( \hat{\psi }[\psi ,A_{\mu }] \)
respectively \( \hat{A}_{\mu }[A_{\mu }] \) in orders of \( h \)
(\ref{eq: SW expansion in h}) and solve the Seiberg-Witten equations
order by order for a fixed star product. This was also done up to
second order in \cite{Jurco:2001rq} for \( \star _{s} \). With this
solution we then obtain:

\begin{Prop}\label{Prop: Solution for Psi and A for arbitrary star product}

If \( \hat{\psi }^{s} \)and \( \hat{A}^{s}_{\mu } \) are a solution
of the Seiberg-Witten equations (\ref{eq:SWeqwith*}) for \( \star _{s} \)
then \( T^{-1}(\hat{\psi }^{s}) \) and \( T^{-1}(\hat{A}^{s}_{\mu }) \)
are a solution for the Seiberg-Witten equations for \( \star  \),
where \( T \) denotes the equivalence transformation from \( \star _{s} \)
to \( \star  \).

\end{Prop}

\begin{proof} \emph{}We have \begin{eqnarray*}
T^{-1}(\hat{\psi }^{s}[\psi ,A_{\mu }])+\delta T^{-1}(\hat{\psi }^{s}[\psi ,A_{\mu }]) & = & T^{-1}(\hat{\psi }^{s}[\psi ,A_{\mu }])+T^{-1}(\delta \hat{\psi }^{s}[\psi ,A_{\mu }])\\
 & = & T^{-1}(\hat{\psi }^{s}[\psi ,A_{\mu }])\\
 &  & +T^{-1}(i\hat{\Lambda }^{s}_{\alpha }[A_{\mu }]\star _{s}\hat{\psi }^{s}[\psi ,A_{\mu }])\\
 & = & T^{-1}(\hat{\psi }^{s}[\psi ,A_{\mu }])\\
 &  & +iT^{-1}(\hat{\Lambda }^{s}_{\alpha }[A_{\mu }])\star T^{-1}(\hat{\psi }^{s}[\psi ,A_{\mu }])\, \, \, .
\end{eqnarray*}
 We used as in the previous proof that \( T \) is an equivalence
transformation from \( \star _{s} \) to \( \star  \) and that the
Seiberg-Witten equation is satisfied by \( \hat{\psi }^{s}[\psi ,A_{\mu }] \)
for the star product \( \star _{s} \). With Proposition \ref{th:lambda s}
we see that \( T^{-1}(\hat{\psi }^{s}[\psi ,A_{\mu }]) \) is a solution
for (\ref{eq:SWeqwith*}).

For the vector field we obtain\begin{eqnarray*}
T^{-1}(\hat{A}^{s}_{\mu }[A_{\mu }])+\delta T^{-1}(\hat{A}^{s}_{\mu }[A_{\mu }]) & = & T^{-1}(\hat{A}^{s}_{\mu }[A_{\mu }])\\
 & + & T^{-1}(\hat{A}^{s}_{\mu }[A_{\mu }]+\partial _{\mu }\hat{\Lambda }^{s}_{\alpha }[A_{\mu }]+i[\hat{\Lambda }^{s}_{\alpha }[A_{\mu }]\stackrel{\star _{s}}{,}\hat{A}^{s}_{\mu }[A_{\mu }]])\\
 & = & T^{-1}(\hat{A}^{s}_{\mu }[A_{\mu }])+\partial '_{\mu }T^{-1}(\hat{\Lambda }^{s}_{\alpha }[A_{\mu }])\\
 & + & i[T^{-1}(\hat{\Lambda }^{s}_{\alpha }[A_{\mu }])\stackrel{\star }{,}T^{-1}(\hat{A}^{s}_{\mu }[A_{\mu }])]\, \, \, ,
\end{eqnarray*}
 where we used in the last step that for \( \partial '_{\mu } \)
(\ref{de: derivation for arbitrary star product}) holds that \( T^{-1}\circ \partial _{\mu }=\partial '_{\mu }\circ T^{-1} \).
Thus, if we define \( \hat{A}'^{\mu }=T^{-1}\hat{A}^{\mu }_{s} \)
we obtain with Proposition \ref{th:lambda s} and the gauge transformation
law (\ref{eq: gauge transform for A' corresponding to arbitrary star product})
that \( \hat{A}'_{\mu }[A_{\mu }]+\delta \hat{A}'_{\mu }[A_{\mu }]=\hat{A}'_{\mu }[A_{\mu }+\delta A_{\mu }] \).

\end{proof}

Having studied star products and its properties enabled us to get
easily these general results for the solution of the Seiberg-Witten
map. Knowing that, we can now treat the problem of how far {}``physics''
changes with the change of the star product.

\subsection{Physics and the Choice of the Star Product\label{se: physics and star products}}

\subsubsection{The Action}

The noncommutative gauge field theory permits to write down a gauge-covariant
Lagrangian and an invariant action (\ref{eq:action}) using the Moyal-Weyl
product \( \star _{s} \).%
\footnote{See below for gauge invariance.
} In the case of an arbitrary star product we obtain with (\ref{eq: covariant derivative and field strength for arbitrary star product})
and (\ref{de:adjointpsi}) \begin{equation}
\label{eq: action for arbitrary star}
S'=\int d^{4}x\, \mathcal{L}'=\int d^{4}x\, \left[ -\frac{1}{4}\textrm{tr}\hat{F}'_{\mu \nu }\star \hat{F}'^{\mu \nu }+\mathrm{ad}_{\star }(\hat{\psi }')\star (i\gamma ^{\mu }\hat{D}'_{\mu }-m)\hat{\psi }'\right] \, \, \, .
\end{equation}
The crucial question is whether this action differs from (\ref{eq:action}).
If the action does not change, the equations of motion will not change
and physics will be unaffected by a change of the star product. But
if it changes we will get different physical predictions. The following
conclusion states how the action is affected.

\begin{Conclusion}\label{th:changeofS}

Let \( S^{s} \) and \( \mathcal{L}^{s} \) be a noncommutative Yang-Mills
action and Lagrangian that correspond to a solution of the Seiberg-Witten
maps for the Moyal-Weyl product \( \star _{s} \). Let \( \star  \)
be another star product equivalent to \( \star _{s} \) by means of
the equivalence transformation \( T \) (from \( \star _{s} \) to
\( \star  \)). Then we get a Yang-Mills action and a Lagrangian,
denoted by \( S^{'} \) and \( \mathcal{L}^{'} \) respectively, that
correspond to a solution of the Seiberg-Witten maps for \( \star  \)
by:\begin{equation}
\label{eq:change of action for T that commutes with derivatives}
S'=\int d^{4}x\, \mathcal{L}^{'}=\int d^{4}x\, T^{-1}\mathcal{L}^{s}\, \, \, .
\end{equation}

\end{Conclusion}

\bpr

We want to understand any matter field and field strength written
with a prime as a solution of the Seiberg-Witten maps for the star
product \( \star  \). We obtain because of \( f\star g=T^{-1}(T(f)\star _{s}T(g)) \),
because of Proposition \ref{th:lambda s} and Proposition \ref{Prop: Solution for Psi and A for arbitrary star product}
and with the definitions for the field strength resp. the covariant
derivative (\ref{eq: covariant derivative and field strength for arbitrary star product})
that\[
\hat{F}^{'}_{\mu \nu }=T^{-1}\hat{F}^{s}_{\mu \nu }\]
 and\[
\hat{D}^{'}_{\mu }\hat{\psi }=T^{-1}\hat{D}^{s}_{\mu }\hat{\psi }\, \, \, .\]
The equation (\ref{eq:change of action for T that commutes with derivatives})
then follows from the definition of the action (\ref{eq:action})
respectively (\ref{eq: action for arbitrary star}) using once again
that \( f\star g=T^{-1}(T(f)\star _{s}T(g)) \).

\epr

Let us comment on this result: As an equivalence transformation, \( T \)
(and therefore \( T^{-1} \) as well) starts in zeroth order with
the identity. Therefore we always recover for any star product to
zeroth order the commutative theory. Changes can only occur in higher
orders of \( h \) . To get a better feeling for what happens, let
us consider for instance \( \star _{k} \) (\ref{*k}) that we introduced
in Subsection \ref{se: star products} (c) with the equivalence transformation
\( T_{k} \) from \( \star _{s} \) to \( \star _{k} \) given in
(\ref{equivtrans}). We get with equation (\ref{eq:change of action for T that commutes with derivatives}):\begin{equation}
\label{eq:SkequalsSs}
\begin{array}{rcl}
S^{k} & = & \int d^{4}x\, T^{-1}_{k}(\mathcal{L}^{s})\\
 & = & \int d^{4}x\, e^{-\frac{i}{2}hm^{\mu \nu }\partial _{\mu }\partial _{\nu }}(\mathcal{L}^{s})\\
 & = & \int d^{4}x\, \mathcal{L}^{s}+\sum ^{\infty }_{k=1}\int d^{4}x\, \frac{(-ih)^{k}}{k!}(m^{\mu \nu }\partial _{\mu }\partial _{\nu })^{k}(\mathcal{L}^{s})\\
 & = & \int d^{4}x\, \mathcal{L}^{s}\, \, \, =\, \, \, S^{s}\, \, \, ,
\end{array}
\end{equation}
 since for non-zero orders we integrate partial derivatives of \( \mathcal{L}^{s} \)
over the whole space which, assuming that fields vanish in infinity,
equals zero. 

Usually, an equivalence transformation is defined to be a differential
operator in all orders of \( h \). In this case the two actions \( S' \)
and \( S^{s} \) in Conclusion \ref{th:changeofS} are actually always
equal if the equivalence transformations \( T \) consists in all
orders of \( h \) of purely \( \mathbb {C} \)-linear combinations
of partial derivatives (a non-constant coefficient would spoil this
property). In this case, as a matter of fact, physics does not \emph{}change.
But it is not difficult to find an ordering prescription for which
the corrections of the action to non-zeroth order in \( h \) do not
vanish. We just have to take any ordering prescription that in a non-zeroth
order contains non-constant coefficients. Nevertheless, there is a
physical reason why we shall not allow star products in full arbitrariness
for our theory: the gauge invariance principle.

\subsubsection{Gauge Invariance}

Gauge invariance is a fundamental concept of our theory. Knowing the
covariant transformation laws (\ref{eq:guagetrafoD}) and (\ref{eq:gaugetrafoF})
respectively (\ref{eq: gauge trafo for Fprime}) and (\ref{eq: gauge trafo for Dprime}),
invariance of the Yang-Mills action is only guaranteed if the integral
satisfies the so called \emph{trace property}, i.e. if the integral
is cyclic

\begin{equation}
\label{eq:traceproperty}
\int d^{4}x\, f\star g=\int d^{4}x\, g\star f
\end{equation}
for all \( f,g \). In the usually considered case, where the Moyal-Weyl
product is used, this property is given. We even have \[
\int d^{4}x\, f\star _{s}g=\int d^{4}x\, fg=\int d^{4}x\, g\star _{s}f\, \, \, .\]
This can be seen by means of partial integration. Again by partial
integration it is not difficult to check that all the star products
\( \star _{k} \) satisfy \[
\int d^{4}x\, f\star _{k}g=\int d^{4}x\, g\star _{k}f\, \, \, ,\]
 too. 

Obviously, if we do not want to loose the concept of an invariant
action, we are forced to accept only star products that satisfy the
trace property (\ref{eq:traceproperty}). Those star products we want
to call \emph{star products with trace property. }

Let \( T \) be an equivalence transformation from \( \star _{s} \)
to a star product \( \star  \). Let us summarize which conditions
we need to get equivalent physical theories in the end:

\begin{lyxlist}{00.00.0000}
\item [(i)]\( \star  \) must be a star product with trace property, i.e.:
\( \int d^{4}x\, f\star g=\int d^{4}x\, g\star f \) for all \( f,g. \)
\item [(ii)]The corresponding actions have to be equal, i.e. \( S'=\int d^{4}x\, \mathcal{L}^{'}=\int d^{4}x\, T^{-1}(\mathcal{L}^{s})\stackrel{!}{=}\int d^{4}x\, \mathcal{L}^{s}=S^{s} \)
(cf. Conclusion \ref{th:changeofS}). 
\end{lyxlist}
If the equivalence transformation \( T \) consists in all orders
of purely \( \mathbb {C} \)-linear combinations of partial derivatives,
the property (ii) is satisfied as we saw above. In this case we obtain,
assuming as always that functions vanish at infinity, that also the
first property is satisfied:\begin{eqnarray*}
\int d^{4}x\, f\star g & = & \int d^{4}x\, T^{-1}(T(f)\star _{s}T(g))\\
 & = & \int d^{4}x\, T(f)\star _{s}T(g)\\
 & = & \int d^{4}x\, T(g)\star _{s}T(f)\\
 & = & \int d^{4}x\, T^{-1}(T(g)\star _{s}T(f))\\
 & = & \int d^{4}x\, g\star f\, \, \, ,
\end{eqnarray*}
 where we used in the third line that \( \star _{s} \) is a star
product with trace property. 

In particular, the transformations \( T_{k} \), which we introduced
in Lemma \ref{la:transformationTk}, are of this type. Hence, in particular
the normal ordered star product, that is sometimes used instead of
the symmetric ordered one, satisfies both conditions. 

However, the first condition, which reflects a physical demand on
the set of star products, does not imply the second one. Thus, the
requirement of trace property is not sufficient to guarantee that
all considered star products lead to the same action.

The question remains, whether there exists another {}``physical reason''
that confines the set of arbitrary star products to those that finally
obey (ii) as well. An answer may be found by restricting the set of
allowed ordering prescriptions. Until now, an ordering prescription
was just defined as a vector space isomorphism with \( \rho (1)=1 \)
that approaches the identity for \( h\rp 0 \). We recall that we
already demanded \( \rho (x^{\mu })=\hat{x}^{\mu } \) in Subsection
\ref{se: NCGFT for arbitrary star product}. Nevertheless, in this
generality ordering prescriptions still are not {}``physical''.
Giving the coordinates the dimension of length, we would like to obtain
the same dimensions for the preimages and images of \( \rho  \).
This implies \( \rho  \) to be homogeneous. The presumption is, that
if we demand an ordering prescription \( \rho  \) to be additionally
homogeneous in all coordinates, that means \( \rho  \) is to preserve
the degree of all coordinates, the corresponding equivalence transformation
\( T \) with \( \rho =\rho _{s}\circ T \) then ends up to consist
in all orders of \( h \) of \( \mathbb {C}- \)linear combinations
of partial derivatives.%
\footnote{The idea is to use the fact that for a monomial \( f \) ordering
\( \rho (f) \) such that we get a linear combination of symmetric
ordered monomials is done by commuting coordinates. But commuting
two coordinates \( \hat{x}^{\mu } \) and \( \hat{x}^{\nu } \) in
turn can be achieved by applying \( \theta ^{\mu \nu }\partial _{\mu }\partial _{\nu } \)
such that \( T \) will consist of linear combinations of such contributions
} Ordering prescriptions that we would understand to be {}``physical''
then are actually those which just {}``order'' the coordinates in
a special way. As we discussed above, star products corresponding
to those ordering prescriptions then lead all to the same action as
the symmetric star product does.

\chapter{Gauge Field Theory on the \protect\( E_{q}(2)\protect \)-Symmetric
Plane}

In this chapter we will treat a noncommutative space with \( E_{q}(2)- \)symmetry.
By \( E_{q}(2) \) we denote the \( q- \)deformed Euclidean group
on the two dimensional space (or more precisely the \( q- \)deformed
algebra of functions on the two dimensional Euclidean group \( E(2) \)).
This is a simple example which allows us to study how gauge field
theory can be implemented on \( q- \)deformed spaces, discussing
the general difficulties we meet and giving possible solutions. The
considerations in this chapter are kept general as far as possible
so that this work may be generalized to other \( q- \)deformed spaces
as well. 

This chapter is divided into three sections. In the first section,
we briefly introduce the quantum group \( E_{q}(2) \) and the \( E_{q}(2)- \)symmetric
plane that is underlying the following considerations. In the second
section, we develop gauge field theories on the \( E_{q}(2) \)-symmetric
plane based on a generalization of the theory developed for the case
of a constant Poisson structure \( \theta  \) (cf. Chapter \ref{ch: Constant theta Case}
and references within). We consider infinitesimal gauge transformations
of matter fields in full analogy to the commutative theory, i.e. taking
the transformation law of commutative matter fields and replacing
the ordinary, commutative multiplication by a convenient star product
that reflects the algebraic properties of the noncommutative \( E_{q}(2)- \)symmetric
plane. In a second step, we introduce covariant coordinates and a
gauge field \( A \). The problem arises that we cannot simply lower
indices using \( \theta  \) as in Chapter \ref{ch: Constant theta Case},
since a non-constant \( \theta  \) in general spoils covariance.
A solution is found by introducing the {}``covariantizer'' \( \mathcal{D} \)
\cite{Jurco:2001my}. Furthermore, this approach allows us to apply
the concept of the Seiberg-Witten maps on our noncommutative plane.
Thus, even for this \( q- \)deformed space, the noncommutative functions
can be expressed in terms of the commutative ones. This makes it possible
to calculate explicitly the corrections to the commutative theory
in the first order of the deformation parameter, which this noncommutative
theory implies. This is done at the end of the second section. Unfortunately,
it turns out that the approach we have chosen in this section leads
to a theory that is not covariant with respect to \( E_{q}(2) \)-transformations
of the space. The problem is the choice of the integral: the invariance
principle forces us to introduce an integral with trace property that
is not invariant under \( E_{q}(2) \)-transformations. But, starting
with a \( q- \)deformed plane possessing a quantum group symmetry,
we would like to set up a theory that is covariant with respect to
this symmetry transformations. We try to establish such a theory in
the third section, this time putting priority on the \( E_{q}(2)- \)symmetry.
First we construct an \( E_{q}(2)- \)invariant integral. As we will
see, this integral in turn looses the trace property. In a second
step we introduce an \( E_{q}(2)- \)covariant differential calculus,
making it possible to speak about covariant one- and two-forms. Difficulties
appear when introducing gauge transformations: As the integral is
not cyclic anymore, gauge transformations cannot be defined by usual
conjugation with an unitary element but we have to adjust them somehow.
We have to define {}``\( q- \)deformed'' gauge transformations
that permit to get a gauge invariant action. Regrettably, we loose
in this approach the concept of Seiberg-Witten map, at least at the
moment we do not know how to convey it to this setting. Nevertheless,
future work could go in this direction, establishing a Seiberg-Witten
map for the \( E_{q}(2)-\textrm{covariant } \)theory...

\section{\protect\( E_{q}(2)\protect \) and the \protect\( E_{q}(2)-\protect \)Symmetric
Plane\label{se: Eq(2) and the symmetric plane}}

In classical physics we have a Lie group acting on a vector space.
Since a Lie group cannot be deformed (the set of semi-simple Lie groups
is a discrete set), one has to consider the algebra of functions on
the considered vector space. For the action of a Lie group on a space
is equivalent to considering the algebra of functions on the Lie group
coacting on the algebra of functions on the space. Those algebras
of functions, both on the Lie algebra and on the space, can be deformed.
The resulting objects are Hopf algebras. First non-trivial examples
of Hopf algebras were for instance introduced by Faddeev, Reshetikhin
and Takhtadjan \cite{FaddeevResTak:1987}. Thus, in the noncommutative
realm, the action of a Lie group becomes the coaction of the corresponding
deformed Hopf algebra on the deformed algebra of functions on the
space. 

We start introducing the generators \( n,v,\bar{n},\bar{v} \) of
the quantum group \( E_{q}(2) \) with their defining relations and
structure maps \cite{Schupp:1992ex} 

\be \label{Eq} \begin{array}{cccc}
\qquad v\bar {v}=\bar  {v} v=1 \qquad n\bar {n}=\bar {n} n \qquad vn=qnv \\
\qquad n\bar {v}=q\bar {v} n \qquad v\bar {n}=q\bar {n} v \qquad \bar {n}\bar {v}=q\bar {v}\bar {n} \\
\Delta(n)=n\otimes\bar {v}+v\otimes n \qquad \Delta(v)=v\otimes v \qquad \Delta(\bar {n})=\bar {n}\otimes v+\bar {v}\otimes \bar {n} \\
\Delta(\bar {v})=\bar {v}\otimes \bar {v} \qquad \varepsilon(n)=\varepsilon(\bar {n})=0 \qquad \varepsilon(v)=\varepsilon(\bar {v})=1 \\
S(n)=-q^{-1}n \quad S(v)=\bar {v} \quad S(\bar {n})=-q\bar {n} \quad S(\bar {v})=v\\
\end{array} \ee
where \( q\in \mathbb {R} \).

If we define new operators \( \theta ,\bar{\theta },t,\bar{t} \)
by \cite{Schupp:1992ex} \be
v=e^{\frac{i}{2}\theta} \qquad \bar {\theta}=\theta \qquad t=nv \qquad \bar {t}=\bar {v} \bar {n} \ee
(note that \( v \) is unitary and can therefore be parametrized by
a hermitian element) the coproduct of \( t \) and \( \bar{t} \)
reads \be \label{copro t} \Delta(t)=t \otimes 1+ e^{i\theta} \otimes t
\qquad \Delta(\bar {t})=\bar {t} \otimes 1 + e^{-i\theta} \otimes \bar {t} \qquad.
\ee

A Hopf algebra \( H \) coacting on an algebra \( A \) means that
algebra is a left (or right) \( H \)-comodule algebra: \\

\begin{Def}\label{de: coaction of a Hopfalgebra on a algebra}

A left coaction of a Hopf algebra H on an algebra A is a linear mapping \( \varrho  \) \be
\begin{array}{rcl}
\rho : A &\longrightarrow& H\otimes A
\end{array} \ee
satisfying \be
\begin{array}{lc}
(\mathrm{id} \otimes \rho)\circ \rho = (\Delta \otimes \mathrm{id})\circ \rho \qquad \mbox{and} \qquad (\varepsilon \otimes \mathrm{id})\circ \rho = \mathrm{id} \\
\rho(ab)=\rho(a)\rho(b)\qquad (m:A\otimes A\ar A \mbox{ is an \(A\)-comodule homomorphism) } \\
\rho(1)=1\otimes1 \qquad \qquad (\eta: \mathbb{C}\longrightarrow A \mbox{ is an \(A\)-comodule homomorphism})~~.\
\end{array} \ee In Sweedler notation we write \cite[p. 32]{Klimyk:1997eb}: 
\[
\rho (a)=:a_{(-1)}\te a_{(0)}\, \, \, .\]
An algebra \( A \) with a left coaction of a Hopf algebra \( H \) is called a left H-comodule algebra.\\
\end{Def}This definition implies that every Hopf algebra \( H \)
admits a comodule structure on itself in virtue of its comultiplication
\be
\Delta : H \longrightarrow H \otimes H \quad . \ee Therefore we can
interpret the algebra generated by \( t,\bar{t} \), which we want
to rename by \( z,\bar{z} \) to distinguish them from the elements
in \( E_{q}(2) \), as the \( q- \)deformed algebra of functions
on the \( E_{q}(2)- \)symmetric plane. We want to write \( \mathbb {C}_{q}^{2} \)
for the algebra generated by \( z,\overline{z} \). From (\ref{copro t})
we get the following left \( E_{q}(2) \)-coaction on \( \mathbb {C}_{q}^{2} \) 

\begin{equation}
\label{coac}
\begin{array}{ccc}
\rho (z) & = & e^{i\theta }\te z+t\te 1\\
\rho (\overline{z}) & = & e^{-i\theta }\te \overline{z}+\bar{t}\te 1\, \, \, .
\end{array}
\end{equation}
 We note that it follows from (\ref{Eq}) (resp. by requiring \( m:A\otimes A\longrightarrow A \)
to be a \( E_{q}(2) \)-comodule homomorphism) that \begin{equation}
\label{com z}
z\overline{z}=q^{2}\overline{z}z\, \, \, ,
\end{equation}
so that we can write \( \mathbb {C}_{q}^{2}=\mathbb {C}\langle z,\bar{z}\rangle /(z\bar{z}-q^{2}\bar{z}z) \).
We want to extend the coaction of \( E_{q}(2) \) to the bigger algebra
of formal power series \begin{equation}
\label{de:Cext}
\mathbb {C}_{q}^{2,\mathrm{ext}}:=\mathbb {C}\langle \langle z,\bar{z}\rangle \rangle /(z\bar{z}-q^{2}\bar{z}z)
\end{equation}
which we call the \emph{algebra of functions on the} \emph{\( E_{q}(2)- \)symmetric
plane}. From now on functions are considered to lie in this algebra.
It is covariant under the \( E_{q}(2)- \)coaction as described above. 

We continue by discussing two different approaches to establish gauge
field theory on \( \mathbb {C}_{q}^{2,\mathrm{ext}} \) starting with
a generalization of the concepts we got to know in Chapter \ref{ch: Constant theta Case}.

\section{Generalization of the Case \protect\( \theta =\mathrm{const}.\protect \)\label{se: Generalization of constant case}}

Let us adjust our notation to the notation we used in Chapter \ref{ch: Constant theta Case},
especially in the section about orderings and star products. Henceforth,
we will use the following abbreviations:\begin{equation}
\label{de: A_c and A_nc (formal power series in z,zbar)}
\mathcal{A}_{\mathrm{c}}:=\mathbb {C}[[z,\overline{z}]]\, \, \, \, \textrm{and}\, \, \, \, \mathcal{A}_{\mathrm{nc}}:=\mathbb {C}_{q}^{2,\mathrm{ext}}\, \, \, .
\end{equation}
In this section we generalize the ideas we studied in Chapter \ref{ch: Constant theta Case}.
We start doing so by constructing a star product for the algebra \( \mathcal{A}_{\mathrm{c}}[[h]] \)
such that it becomes isomorphic as algebra to \( \mathcal{A}_{\mathrm{nc}}[[h]] \)
in full analogy to what we have done in the first chapter. Since a
star product is a deformation in direction of a Poisson structure,
we will first deduce the corresponding Poisson structure.

\subsection{The Poisson Structure for \protect\( \mathcal{A}_{\mathrm{c}}\protect \)}

To distinguish commutative coordinates and noncommutative ones we
want to denote in this subsection the former by \( z,\overline{z} \)
and the latter by \( \hat{z},\overline{\hat{z}} \). In the following
subsections we will only need commutative coordinates and therefore
they cannot be confounded with noncommutative ones. 

We recall that we introduced in the first chapter a formal deformation
parameter \( h \). In this case we want to do so, too. We take \begin{equation}
\label{de:q equals e^h}
h:=\ln q
\end{equation}
 as formal deformation parameter and our aim is to study deviations
from the commutative theory in orders of this parameter \( h. \)
Thus, the commutation relations (\ref{com z}) in \( \mathcal{A}_{\mathrm{nc}} \)
become \[
\hat{z}\overline{\hat{z}}=e^{2h}\overline{\hat{z}}\hat{z}=(1+2h+2h^{2}+\dots )\overline{\hat{z}}\hat{z}\, \, \, .\]
 This yields\[
[\hat{z},\overline{\hat{z}}]=2h\overline{\hat{z}}\hat{z}+\mathcal{O}(h^{2})\, \, \, .\]
We want to construct a star product that renders the commutative algebra
\( \mathcal{A}_{\mathrm{c}} \) isomorphic to \( \mathcal{A}_{\mathrm{nc}} \).
A star product is a deformation in direction of a Poisson structure.
Therefore we need to know which Poisson structure we have to attribute
to \( A_{\mathrm{c}} \). 

Let us suppose we had an arbitrary star product \( \star  \) (star
products exist as we learned in Chapter \ref{ch: Constant theta Case})
reflecting the algebraic properties of \( \mathcal{A}_{\mathrm{nc}} \),
i.e. \( (\mathcal{A}_{\mathrm{c}},\star )\cong \mathcal{A}_{\mathrm{nc}} \).%
\footnote{These are the star products that we are interested in. A priory star
products that correspond to a certain Poisson structure do not lead
to isomorphic algebras. This is only the case if the star products
lie in the same equivalence class. Thus, we need star products that
lie in the equivalence class which consists of exactly those star
products that render \( \mathcal{A}_{\mathrm{c}} \) isomorphic as
algebra to \( \mathcal{A}_{\mathrm{nc}} \). 
} Let \( \rho :\, \mathcal{A}_{\mathrm{c}}\rightarrow \mathcal{A}_{\mathrm{nc}} \)
be the corresponding algebra isomorphism. We want to assume additionally
that \( \mathcal{A}_{\mathrm{nc}}\rp \mathcal{A}_{\mathrm{c}} \)
for \( h\rp 0 \) such that \( \rho =\mathrm{id}+\mathcal{O}(h) \)
(cf. Chapter 1). Thus, we have \( \rho (z^{i})=\hat{z}^{i}+\mathcal{O}(h) \)
and since by definition a star product always starts with the identity
in zeroth order, we obtain for the star-commutator of the coordinates
\begin{equation}
\label{eq: Poisson structure for z,bz}
[z\stackrel{\star }{,}\overline{z}]=2hz\overline{z}+\mathcal{O}(h^{2})\, \, \, .
\end{equation}
 As we know, the Poisson bracket on \( \mathcal{A}_{\mathrm{c}}[[h]] \)
corresponds to the first order term in \( h \) of the star commutator
multiplied by \( -i \) (Definition \ref{de:star product}), such
that we can derive from (\ref{eq: Poisson structure for z,bz}) the
Poisson structure for \( \mathcal{A}_{\mathrm{c}}[[h]] \):\begin{equation}
\label{eq: Poisson structure for Eq(2)}
\{f,g\}=-2iz\overline{z}((\partial _{z}f)(\partial _{\overline{z}}g)-(\partial _{\overline{z}}f)(\partial _{z}g))=-2i\varepsilon ^{ij}z\overline{z}(\partial _{i}f)(\partial _{j}g)\, \, \, .
\end{equation}
Thus, the Poisson tensor is given by

\textbf{\begin{equation}
\label{eq: Poisson tensor for Eq(2)}
\theta ^{ij}=-2i\varepsilon ^{ij}z\overline{z}=-2i\left( \begin{array}{cc}
0 & 1\\
-1 & 0
\end{array}\right) z\overline{z}\, \, \, .
\end{equation}
}\emph{Remark:} If we write down the commutation relations in terms
of the basis \( x,y \), where \( z=x+iy\, \textrm{and}\, \overline{z}=x-iy \),
we find with \( h=\ln q \): \[
[\hat{x},\hat{y}]=-i\frac{1-q^{2}}{1+q^{2}}(\hat{x}^{2}+\hat{y}^{2})=ih(\hat{x}^{2}+\hat{y}^{2})+\mathcal{O}(h^{2})\, \, \, .\]
 Hence, we get for a star product \( \star  \) with the above properties:\[
[x\stackrel{\star }{,}y]=ih(x^{2}+y^{2})+\mathcal{O}(h^{2})\, \, \, .\]
 Therefore the Poisson structure in the basis \( x,y \) reads\[
\{\tilde{f},\tilde{g}\}=(x^{2}+y^{2})\varepsilon ^{\alpha \beta }(\partial _{^{_{\alpha }}}\tilde{f})(\partial _{\beta }\tilde{g})\]
 and we get the following Poisson tensor:\begin{equation}
\label{de: Poisson tensor in real coordinates}
\tilde{\theta }(x,y)=(x^{2}+y^{2})\varepsilon ^{\alpha \beta }\, \, \, .
\end{equation}
 \emph{Notation:} We will always write functions depending on \( x,y \)
with tilde and use Greek indices if we consider the basis \( x,y \),
whereas we will use no tilde for functions depending on \( z,\overline{z} \)
and Latin indices, for example \( i\in \{z,\overline{z}\} \), for
the basis \( z,\overline{z} \). We will also often write \( z^{i} \)
meaning \( z^{1}=z \) and \( z^{2}=\overline{z} \).

\subsection{A Star Product for \protect\( \mathcal{A}_{\mathrm{c}}[[h]]\protect \)}

Again we have a big freedom in choosing a special ordering prescription
\( \rho  \) since every ordering prescription leads to a star product
by defining \( f\star g=\rho ^{-1}(\rho (f)\cdot \rho (g)) \) (cf.
Subsection \ref{se: ordering}). Here, we do not want to repeat all
the discussion of orderings and corresponding star products but just
remark that a star product corresponding to the normal ordering (all
\( z \) to the left and all \( \overline{z} \) to the right) is
given by (see for example \cite{Madore:2000en} or also \cite{Schraml}
where the Manin plane is discussed) \[
f\star _{n}g=\mu \circ e^{-2h(\overline{z}\partial _{\overline{z}}\te z\partial _{z})}(f\te g)\, \, \, .\]
For us the following star product will be of primary interest\begin{equation}
\label{de:q-symm star product}
f\star _{q}g:=\mu \circ e^{h(z\partial _{z}\te \overline{z}\partial _{\overline{z}}-\overline{z}\partial _{\overline{z}}\te z\partial _{z})}(f\te g)\, \, \, .
\end{equation}
 First of all it can be easily verified, checking all requirements
in Definition \ref{de:star product}, that \( \star _{q} \) defined
in (\ref{de:q-symm star product}) is indeed a star product for \( \mathcal{A}_{\mathrm{c}}[[h]] \).
Let us think about which ordering prescription \( \rho _{q} \) corresponds
to \( \star _{q} \): We have \[
z\star _{q}\overline{z}=qz\overline{z}\, \, \, \textrm{and}\, \, \, \overline{z}\star _{q}z=q^{-1}z\overline{z}\, \, \, .\]
 Therefore we get \[
\frac{q^{-1}z\star _{q}\overline{z}+q\overline{z}\star _{q}z}{2}=z\overline{z}\, \, \, .\]
 We assume that \( \star _{q} \) is given by an ordering prescription
\( \rho _{q} \), i.e. we assume that \( f\star _{q}g=\rho _{q}^{-1}(\rho _{q}(f)\cdot \rho _{q}(g)) \).
Than we find with the short calculation given above that an ordering
prescription \( \rho _{q}:\mathcal{A}_{\mathrm{c}}\rp \mathcal{A}_{\mathrm{nc}} \)
that starts with the assignments%
\footnote{Of course this is not the entire isomorphism \( \rho _{q} \) but
we get an impression of how it looks like. 
}\begin{eqnarray*}
z & \ra  & \hat{z}\\
\overline{z} & \ra  & \overline{\hat{z}}\\
z\overline{z} & \ra  & \frac{q^{-1}\hat{z}\overline{\hat{z}}+q\overline{\hat{z}}\hat{z}}{2}\\
 & \dots  & 
\end{eqnarray*}
 leads to \( \star _{q} \). We want to call such an ordering prescription
\emph{\( q- \)symmetric ordering}%
\footnote{This notion goes back to Peter Schupp.
}. In particular we see that \( \star _{q} \) neither corresponds
to the symmetric ordering prescription nor to the normal ordering.

We note that \( \star _{q} \) is indeed a \emph{hermitian} star product.
Thus, complex conjugation is an involution for \( (\mathcal{A}_{\mathrm{c}}[[h]],\star _{q}) \).
\\
\emph{Remark:} In the real basis \( x,y \) (\ref{de:q-symm star product})
becomes \begin{equation}
\label{eq: star product for real basis x,y}
\tilde{f}\star _{q}\tilde{g}=\mu \circ e^{\frac{1}{4}h((x+iy)(\partial _{x}-i\partial _{y})\te (x-iy)(\partial _{x}+i\partial _{y})-(x-iy)(\partial _{x}+i\partial _{y})\te (x+iy)(\partial _{x}-i\partial _{y}))}(\tilde{f}\te \tilde{g})\, \, \, .
\end{equation}
 If we want to determine the corresponding ordering prescription in
this basis things are not so easy. We find for example \[
x^{2}+y^{2}\rp \frac{q^{-1}(\hat{x}+i\hat{y})(\hat{x}-i\hat{y})+q(\hat{x}-i\hat{y})(\hat{x}+i\hat{y})}{2}.\]
 Obviously in this basis the ordering prescription as well as the
star product become more complicated. Therefore it is reasonable to
take the complex basis. Nevertheless, we want to give all results
for the usually used and more familiar real basis as well.

In the next chapter we will learn about gauge field theory on this
noncommutative space. Later we will see why we want to choose especially
\( \star _{q} \) to establish abelian gauge field theory: Actually,
\( \star _{q} \) together with a modification of the common integral
leads to an integral with trace property (see (\ref{eq:traceproperty})),
which is necessary to obtain an invariant action (cf. Subsection \ref{se: physics and star products}).

\subsection{Noncommutative Abelian Gauge Field Theory}

We treat the case of abelian, noncommutative gauge field theory on
\( (\mathcal{A}_{\mathrm{c}}[[h]],\theta ). \) This may be generalized
to non abelian gauge field theories in a second step. 

Infinitesimal gauge transformations of a matter field \( \hat{\psi }(z,\overline{z})\in \mathcal{A}_{\mathrm{c}} \)
are introduced as in Subsection \ref{se:NCGFT}, i.e. \[
\delta \psi (z,\overline{z})=i\Lambda (z,\overline{z})\star \psi (z,\overline{z})\, \, \, .\]
 The concept of covariant coordinates \( Z^{i}:=z^{i}+A^{i} \) with
\( \delta Z^{i}=i[\Lambda \stackrel{\star }{,}Z^{i}] \) leads again
to a gauge field \( A^{i} \) transforming as\[
\delta A^{i}=-i[z^{i}\stackrel{\star }{,}\Lambda ]+i[\Lambda \stackrel{\star }{,}A^{i}]\, \, \, .\]
 One difference to the case where \( \theta  \) is constant arises
now: we cannot simply use \( \theta  \) to lower indices as we did
in Subsection \ref{se:NCGFT}. Since \( \theta  \) is not constant,
this would not lead to a gauge covariant field. Thus, we will stay
with the gauge fields \( A^{i} \) and will try to analyze and to
solve the emerging problems step by step. Let us continue by giving
a solution for the Seiberg-Witten map (\ref{eq: Seiberg-W}).

\subsubsection{Seiberg-Witten map}

The solution for the Seiberg-Witten map for the gauge field \( A^{i} \)
in the case of abelian gauge field theory is given by:

\begin{equation}
\label{eq: Seiberg-Witten for A(z)}
A^{i}(z,\overline{z})=h\theta ^{ij}a_{j}+h^{2}\frac{1}{2}\theta ^{kl}a_{l}(\partial _{k}(\theta ^{ij}a_{j})-\theta ^{ij}f_{jk})+\dots \, \, \, .
\end{equation}
Compare with the publication \cite{Jurco:2001my}%
\footnote{Mind that we want to contribute to every \( \theta  \) an order in
\( h \). 
}, where a solution for an arbitrary Poisson structure \( \theta  \)
up to second order was derived using the Kontsevich star product.
The solution found there was taken and we checked that it is indeed
a solution for \( \star _{q} \), too.

By \( a_{i}=a_{i}(z,\overline{z}) \) and \( f_{ij}(z,\overline{z}) \)
we denote the commutative gauge field and field strength written in
the basis \( z,\overline{z} \). Expressed in terms of the ordinary,
commutative gauge field \( \tilde{a}_{\alpha }(x,y) \) and field
strength \( \tilde{f}_{\alpha \beta }(x,y)=\partial _{\alpha }\tilde{a}_{\beta }-\partial _{\beta }\tilde{a}_{\alpha } \),
we have (Appendix \ref{appendix: change of basis z,zbar to x,y})

\begin{eqnarray*}
a_{z}(z,\overline{z}) & = & \frac{1}{2}\left\{ \tilde{a}_{1}(\phi (z,\overline{z}))-i\tilde{a}_{2}(\phi (z,\overline{z}))\right\} \\
a_{\overline{z}}(z,\overline{z}) & = & \frac{1}{2}\left\{ \tilde{a}_{1}(\phi (z,\overline{z}))+i\tilde{a}_{2}(\phi (z,\overline{z}))\right\} 
\end{eqnarray*}
 and furthermore the commutative field strength \( \tilde{f}_{\alpha \beta }(x,y) \)
becomes in the basis \( z,\overline{z} \) (Appendix \ref{appendix: change of basis z,zbar to x,y})
\[
\frac{1}{2}i\tilde{f}_{ij}(\phi (z,\overline{z}))=f_{ij}(z,\overline{z})=\partial _{i}a_{j}(z,\overline{z})-\partial _{j}a_{i}(z,\overline{z})\, \, \, .\]

\subsection{The Field Strength}

Let us think about the principles that we want to impose on a field
strength \( F \): 

\begin{enumerate}
\item We want the field strength \( F \) to be gauge covariant, that means
that the transformation law of \( F \) is to be given by\begin{equation}
\label{eq: requirement1 for F: cov transf}
\delta F^{ij}=i[\Lambda \stackrel{\star }{,}F^{ij}]\, \, \, .
\end{equation}

\item In the semi-classical  limit \( q\rp 1 \) resp. \( h\rp 0 \) we
want to obtain the commutative field strength \( f_{ij}=\partial _{i}a_{j}-\partial _{j}a_{i} \)
. 
\end{enumerate}
We start treating the first requirement. For this purpose we write
the commutation relations (\ref{com z}) in the following form introducing
the \( \hat{R} \)-matrix:\begin{equation}
\label{eq: commrel for z with R-Matrix}
P^{ij}{}_{kl}z^{k}\star z^{l}:=(\delta _{k}^{i}\delta ^{j}_{l}-\hat{R}^{ij}{}_{kl})z^{k}\star z^{l}=0
\end{equation}
 where \[
\hat{R}^{ij}{}_{kl}:=\left( \begin{array}{cccc}
1 & 0 & 0 & 0\\
0 & 1-q^{-4} & q^{-2} & 0\\
0 & q^{-2} & 0 & 0\\
0 & 0 & 0 & 1
\end{array}\right) \, \, \, .\]
Here the upper indices number the rows in the sequence \( \{11,12,21,22\} \)
and the lower indices the columns in the same sequence. Defining \begin{equation}
\label{de: convarintes F, erste Version}
F^{ij}:=-iP^{ij}{}_{kl}Z^{k}\star _{q}Z^{l}\, \, \, ,
\end{equation}
we get a covariant expression since \( Z^{i} \) are covariant coordinates.
Thereby (\ref{de: convarintes F, erste Version}) can be considered
a good candidate for a field strength. 

Let us calculate the first nontrivial order of \( F^{ij} \) to study
its semi-classical  limit. To do so, we first have to expand the matrix
\( P \) in \( h \)\begin{eqnarray}
P^{ij}{}_{kl} & = & \left( \begin{array}{cccc}
0 & 0 & 0 & 0\\
0 & 1 & -1 & 0\\
0 & -1 & 1 & 0\\
0 & 0 & 0 & 0
\end{array}\right) +h\left( \begin{array}{cccc}
0 & 0 & 0 & 0\\
0 & -4 & 2 & 0\\
0 & 2 & 0 & 0\\
0 & 0 & 0 & 0
\end{array}\right) +\mathcal{O}(h^{2})\label{eq: matrix P up to first order in h} \\
 & =: & P_{0}+hP_{1}+\mathcal{O}(h^{2})\, \, \, .
\end{eqnarray}
 We introduce the notation \[
f\star g=:fg+h(f\star g)_{1}+h^{2}(f\star g)_{2}+\mathcal{O}(h^{3})\]
such that we can write, using that \( Z^{i}=z^{i}+A^{i} \), the commutation
relations (\ref{eq: commrel for z with R-Matrix}) and the Seiberg-Witten
map up to first order in \( h \) for \( A^{i} \) given in (\ref{eq: Seiberg-Witten for A(z)})
\begin{eqnarray*}
F^{ij} & = & -iP^{ij}{}_{kl}(z^{k}\star _{q}A^{l}+A^{k}\star _{q}z^{l}+A^{k}\star _{q}A^{l}+z^{k}\star _{q}z^{l})\\
 & = & -ih^{2}P_{0}^{ij}{}_{kl}((z^{k}\star _{q}(\theta ^{lm}a_{m}))_{1}+((\theta ^{km}a_{m})\star _{q}z^{l})_{1}+\theta ^{km}a_{m}\theta ^{lm}a_{m})\\
 &  & -ih^{2}P_{1}^{ij}{}_{kl}(z^{k}\theta ^{lm}a_{m}+\theta ^{km}a_{m}z^{l})+\mathcal{O}(h^{3})\, \, \, .
\end{eqnarray*}
 We split the calculation and start with the first two terms on the
right hand side for \( i=z \) and \( j=\overline{z} \). Since \( P_{0}^{z\overline{z}}{}_{kl}=\varepsilon _{kl} \)
(see above), we get immediately\[
P_{0}^{z\overline{z}}{}_{kl}((z^{k}\star _{q}(\theta ^{lm}a_{m}))_{1}+((\theta ^{km}a_{m})\star _{q}z^{l})_{1})=([z\stackrel{\star _{q}}{,}\theta ^{\overline{z}m}a_{m}])_{1}+([\theta ^{zm}a_{m}\stackrel{\star _{q}}{,}\overline{z}])_{1}\, \, .\]
As the star-commutator always starts to first order in \( h \) by
\( i \) times the Poisson structure (cf. Definition \ref{de:star product}),
we find\[
\begin{array}{ll}
= & i\theta ^{z\overline{z}}\partial _{\overline{z}}(\theta ^{\overline{z}z}a_{z})+i\theta ^{z\overline{z}}\partial _{z}(\theta ^{z\overline{z}}_{nc}a_{\overline{z}})\\
= & i\theta ^{z\overline{z}}\theta ^{z\overline{z}}(\partial _{\overline{z}}a_{z}-\partial _{z}a_{\overline{z}})+i\theta ^{z\overline{z}}(\partial _{z}(\theta ^{z\overline{z}})a_{\overline{z}}-\partial _{\overline{z}}(\theta ^{z\overline{z}})a_{z})\\
= & i\theta ^{z\overline{z}}\theta ^{z\overline{z}}f_{\overline{z}z}+i\theta ^{z\overline{z}}(\partial _{z}(\theta ^{z\overline{z}})a_{\overline{z}}-\partial _{\overline{z}}(\theta ^{z\overline{z}})a_{z})\, \, \, ,
\end{array}\]
where we used that \( \theta  \) is antisymmetric, too. 

We continue with the following term: \begin{eqnarray*}
P_{0}^{ij}{}_{kl}\theta ^{km}a_{m}\theta ^{lm}a_{m} & \, \, \, .
\end{eqnarray*}
 It is easy to see that this equals zero noting that \( P_{0}^{ij}{}_{kl} \)
is antisymmetric in \( k,l \) whereas \( \theta ^{km}a_{m}\theta ^{lm}a_{m} \)
is symmetric in \( k,l \). The last contribution reads\begin{eqnarray*}
P_{1}^{z\overline{z}}{}_{kl}(z^{k}\theta ^{lm}a_{m}+\theta ^{km}a_{m}z^{l}) & = & -4(z\theta ^{\overline{z}z}a_{z}+\theta ^{z\overline{z}}a_{\overline{z}}\overline{z})+2(\overline{z}\theta ^{z\overline{z}}a_{\overline{z}}+\theta ^{\overline{z}z}a_{z}z)\\
 & = & -2\theta ^{z\overline{z}}(\overline{z}a_{\overline{z}}-za_{z})
\end{eqnarray*}
 such that we find, putting all contributions together multiplied
by \( -i \) as in the definition of \( F \) and taking into account
that \( \theta ^{z\overline{z}}=-2iz\overline{z} \),\begin{eqnarray*}
F^{z\overline{z}} & = & h^{2}\theta ^{z\overline{z}}\theta ^{z\overline{z}}(\partial _{z}a_{\overline{z}}-\partial _{\overline{z}}a_{z})+h^{2}(\partial _{z}\theta ^{z\overline{z}})a_{\overline{z}}-(\partial _{\overline{z}}\theta ^{z\overline{z}})a_{z}\\
 &  & +2ih^{2}\theta ^{z\overline{z}}(\overline{z}a_{\overline{z}}-za_{z})+\mathcal{O}(h^{3})\\
 & = & h^{2}\theta ^{z\overline{z}}\theta ^{z\overline{z}}(\partial _{z}a_{\overline{z}}-\partial _{\overline{z}}a_{z})+\mathcal{O}(h^{3})=h^{2}\theta ^{z\overline{z}}f_{\overline{z}z}\theta ^{z\overline{z}}+\mathcal{O}(h^{3})\, \, \, .
\end{eqnarray*}
It can be easily checked that \( F^{\overline{z}z}=-h^{2}\theta ^{zi}f_{ij}\theta ^{j\overline{z}}+\mathcal{O}(h^{3}) \)
and we can write\begin{equation}
\label{eq:classical limit for F}
F^{kl}=h^{2}\theta ^{ki}\theta ^{jl}f_{ij}+\mathcal{O}(h^{3})\, \, \, .
\end{equation}
 Let us remark that the calculation made so far for the semi-classical
 limit of \( F \) is \emph{independent} of the choice of a star product,
since until now only the star-commutator up to first order in \( h \)
entered into the calculation. 

While \( F^{ij} \) defined above transforms covariantly, the semi-classical
limit does not exactly lead to the commutative field strength \( f_{ij} \).
Therefore \( F^{ij} \) cannot be taken like this as a noncommutative
field strength. Nevertheless, we can modify \( F^{ij} \): We want
to get rid of the two \( \theta  \)'s without loosing covariance.
This can be done by multiplying \( F^{ij} \) by a covariant function
with the property that in the semi-classical limit we actually obtain
\( f_{ij}. \) In \cite{Jurco:2001my} we get a hint how we can do
so: Here, covariant functions are introduced generated by applying
a mapping \( \mathcal{D} \), called {}``covariantizer'', to an
arbitrary function \( f \) \[
\mathcal{D}:f\ra \mathcal{D}f=f+f_{\mathcal{A}}\]
(applying \( \mathcal{D} \) to coordinates leads to covariant coordinates).
The covariantizer is defined by requiring \begin{equation}
\label{eq: conditional equation for covariantizer}
\delta \mathcal{D}(f)=i[\Lambda \stackrel{\star }{,}\mathcal{D}(f)]\, \, \, .
\end{equation}
We can determine \( \mathcal{D} \) up to first order in \( h \).
With \( \Lambda =\alpha +\mathcal{O}(h) \), where \( \alpha  \)
is the commutative gauge parameter, we see that \( \mathcal{D}=\mathrm{id}+h\theta ^{ij}a_{j}\partial _{i}+\mathcal{O}(h^{2}) \)
obeys the conditional equation (\ref{eq: conditional equation for covariantizer}).
Here \( a_{j} \) denotes the commutative gauge field with the gauge
transformation law \( \delta a_{j}=\partial _{j}\alpha  \). Thus,
in our two dimensional case we obtain%
\footnote{The existence of \( \mathcal{D} \) to all orders in \( h \) was
derived in \cite{Jurco:2001my}. 
}\begin{equation}
\label{de: covariantizer D}
\mathcal{D}=\mathrm{id}+h\theta ^{z\overline{z}}(a_{\overline{z}}\partial _{z}-a_{z}\partial _{\overline{z}})+\dots 
\end{equation}
 We define \begin{equation}
\label{de: cov field strength with cl limit}
F^{'}_{ij}:=-\frac{1}{h^{2}}\mathcal{D}(\theta ^{-1})_{ik}\star _{q}F^{kl}\star _{q}\mathcal{D}(\theta ^{-1})_{lj}
\end{equation}
where \begin{equation}
\label{de: theta^-1}
(\theta ^{-1})_{kl}:=-i\frac{1}{2z\overline{z}}\varepsilon _{kl}=-i\frac{1}{2z\overline{z}}\left( \begin{array}{cc}
0 & 1\\
-1 & 0
\end{array}\right) 
\end{equation}
 and \begin{eqnarray}
\mathcal{D}(\theta ^{-1})_{kl} & =- & \frac{1}{2}i\mathcal{D}(\frac{1}{z\overline{z}})\varepsilon _{kl}\nonumber \\
 & =- & \frac{1}{2}i\varepsilon _{kl}(\frac{1}{z\overline{z}}+h\theta ^{12}(a_{\overline{z}}\partial _{z}-a_{z}\partial _{\overline{z}})(\frac{1}{z\overline{z}})+\mathcal{O}(h^{2})\nonumber \\
 & =- & \frac{1}{2}i\varepsilon _{kl}\frac{1}{z\overline{z}}(1-h\frac{\theta ^{12}}{z\overline{z}}(a_{\overline{z}}\overline{z}-a_{z}z))+\mathcal{O}(h^{2})\nonumber \\
 & =- & \frac{1}{2}i\varepsilon _{kl}\frac{1}{z\overline{z}}+h\frac{(a_{\overline{z}}\overline{z}-a_{z}z)}{z\overline{z}}\varepsilon _{kl}+\mathcal{O}(h^{2})\label{eq: D(theta) bis 1. Ordnung in h} 
\end{eqnarray}
 (the explicit contribution to first order in \( h \) will be used
later when we expand the action in \( h \)). We see that because
of (\ref{eq:classical limit for F})\[
F^{'}_{ij}=f_{ij}+\mathcal{O}(h)\]
 and therefore \( F' \) admits the right semi-classical  limit. Moreover,
\( \mathcal{D}(\theta ^{-1}) \) transforms covariantly (\( \mathcal{D} \)
is the {}``covariantizer''). Hence, we can conclude that

\begin{itemize}
\item \( F_{ij}^{'} \) is gauge covariant and that
\item \( F_{ij}^{'}\rp f_{ij} \) for \( h\rp 0 \) 
\end{itemize}
such that \( F_{ij}^{'} \) satisfies all we demanded above from a
field strength. Hence, we consider \( F_{ij}^{'} \) to be the \emph{noncommutative
abelian field strength. }

Now we can write down a Lagrangian for a noncommutative, free photon
field \[
\mathcal{L}=\frac{1}{4}\delta ^{ik}\delta ^{lj}F_{kl}^{'}\star _{q}F_{ij}^{'}\, \, \, .\]
 \emph{Remark:} In the basis of real coordinates we want to define
in analogy to what we studied for the basis \( z,\overline{z} \):%
\footnote{See the remark below to get an explanation for the factor \( \frac{1}{2} \).
}\begin{equation}
\label{de: Ftilde with theta in cl. limit}
\tilde{F}^{\alpha \beta }:=\frac{1}{2}P^{\alpha \beta }{}_{12}(X+iY)\star _{q}(X-iY)+\frac{1}{2}P^{\alpha \beta }{}_{21}(X-iY)\star _{q}(X+iY)
\end{equation}
where \( P \) is given in (\ref{eq: commrel for z with R-Matrix})
and \( X:=x+A^{1} \) resp. \( Y:=y+A^{2} \) are the covariant coordinates
(cf. (\ref{de:cov.cord})). Then \( \tilde{F}^{\alpha \beta } \)
is gauge-covariant. If we undertake a similar calculation as done
above for determining the semi-classical  limit of \( F \), we get
for the first non vanishing order\[
\tilde{F}^{\alpha \beta }=h^{2}\tilde{\theta }^{\alpha \gamma }\tilde{\theta }^{\delta \beta }\tilde{f}_{\gamma \delta }+\mathcal{O}(h^{3})\]
 with \( \tilde{\theta }^{\alpha \beta }=(x^{2}+y^{2})\varepsilon ^{\alpha \beta } \).
Again we want to get rid of the two \( \tilde{\theta } \) that appear
without loosing gauge covariance so that we multiply by \( \mathcal{D}(\tilde{\theta }^{-1}) \).
Since

\begin{equation}
\label{de: theta^-1}
(\tilde{\theta }^{-1})_{\alpha \beta }:=-\frac{1}{x^{2}+y^{2}}\varepsilon _{\alpha \beta }
\end{equation}
 we get with the definition of \( \mathcal{D} \) given in (\ref{de: covariantizer D})
up to first order in \( h \) that \begin{eqnarray}
\mathcal{D}(\tilde{\theta }^{-1})_{\alpha \beta } & = & -\mathcal{D}(\frac{1}{x^{2}+y^{2}})\varepsilon _{\alpha \beta }\nonumber \\
 & = & -\varepsilon _{\alpha \beta }(\frac{1}{x^{2}+y^{2}}+h\tilde{\theta }^{\gamma \delta }\tilde{a}_{\delta }\partial _{\gamma }(\frac{1}{x^{2}+y^{2}})+\mathcal{O}(h^{2})\nonumber \\
 & = & -\varepsilon _{\alpha \beta }\frac{1}{x^{2}+y^{2}}(1-2h\frac{\tilde{\theta }^{\gamma \delta }}{x^{2}+y^{2}}\tilde{a}_{\delta }x_{\gamma })\nonumber \\
 & = & -\varepsilon _{\alpha \beta }\frac{1}{x^{2}+y^{2}}(1-2h\varepsilon ^{\gamma \delta }x_{\gamma }\tilde{a}_{\delta })\, \, \, .\label{eq: Dtilde(theta) bis 1. Ordnung h} 
\end{eqnarray}
 Finally, we define the field strength in real coordinates by\begin{equation}
\label{de: Ftilde' with right classical limit}
\tilde{F}_{\alpha \beta }^{'}:=-\frac{1}{h^{2}}\mathcal{D}(\tilde{\theta }^{-1})_{\alpha \gamma }\star _{q}\tilde{F}^{\gamma \delta }\star _{q}\mathcal{D}(\tilde{\theta }^{-1})_{\delta \beta }=\tilde{f}_{\alpha \beta }+\mathcal{O}(h)
\end{equation}
 and the Lagrangian as \begin{equation}
\label{de: Ltilde}
\tilde{\mathcal{L}}:=\frac{1}{4}\delta ^{\alpha \gamma }\delta ^{\beta \delta }\tilde{F}^{'}_{\gamma \delta }\star _{q}\tilde{F}_{\alpha \beta }^{'}\, \, \, .
\end{equation}
 \emph{Remark:} We found a field strength and a Lagrangian both, in
the real and the complex basis. In both cases we got the right semi-classical
 limit and of course it is possible to come from one to the other
case by basis transformation. But we want to underline the following:
If we compare the definitions (\ref{de: convarintes F, erste Version})
of \( F^{ij} \) resp. (\ref{de: Ftilde with theta in cl. limit})
of \( \tilde{F}^{\alpha \beta } \) made above, we see that \( F^{ij}(\phi ^{-1}(x,y))=-2i\tilde{F}^{ij}(x,y) \)
(for the definition of the basis transformation \( \phi (z,\overline{z}) \)
see Appendix \ref{appendix: change of basis z,zbar to x,y}). Both
definitions obviously differ by the factor \( -2i \). Where does
it come from? We need the factor to guarantee in both cases the right
semi-classical  limit. \( F^{z\overline{z}} \) starts in the first
non-vanishing order with \( h^{2}\theta ^{z\overline{z}}f_{\overline{z}z}\theta ^{z\overline{z}} \)
(see (\ref{eq:classical limit for F})). If we write this expression
in terms of the real basis \( x,y \) we have to use that \( \theta ^{z\overline{z}}(\phi ^{-1}(x,y))=-2i\tilde{\theta }^{12}(x,y) \)
(cf. (\ref{eq: Poisson tensor for Eq(2)}) and (\ref{de: Poisson tensor in real coordinates}))
as well as the fact that \( f_{z\overline{z}}=\frac{1}{2}i\tilde{f}_{12}(\phi (z,\overline{z})) \)
(\ref{app: trafo of antisymm twoform for x,y to z,zbar}). Together
this gives \( F^{ij}(\phi ^{-1}(x,y))=(-2i)(\frac{1}{2}i)(-2i)\tilde{F}^{ij}(x,y)=-2i\tilde{F}^{ij}(x,y) \).
This factor must not be forgotten if translating results from the
case with complex basis to the case where we use the real basis. Therefore
definitions made here are reasonable (best proven by the fact that
they lead to the right semi-classical  limit as they should).

\subsection{The Integral with Trace Property\label{se: integral with trace property}}

In this section we treat the problem of how to find an integral with
trace property for non-constant Poisson structures (for a constant
Poisson structure the usual integral with an appropriate star product
admits the trace property as we saw in Chapter 1). We will treat the
special example \( \theta ^{ij}=-2iz\overline{z}\varepsilon ^{ij} \),
the Poisson tensor for \( \mathcal{A}_{\mathrm{c}} \) (\ref{eq: Poisson tensor for Eq(2)}),
but considerations made here can be transfered to other, more complicated
Poison structures as well. Thus, the aim is not just to give a solution
for our example but also to discuss two approaches how it may be possible
to find solutions for more general Poisson structures.

\subsubsection{General Remarks}

We want to find a measure \( \mu (z,\overline{z}) \) for the usual
integral such that for a given star product \( \star  \) the integral
admits the trace property\begin{equation}
\label{eq: requirement for cyclic property of the integral}
\int \mu (z,\overline{z})dzd\overline{z}\, f\star g=\int \mu (z,\overline{z})dzd\overline{z}\, g\star f
\end{equation}
 for all functions \( f,g \). Let us emphasize that this is a combination
of \emph{two} problems:

\begin{itemize}
\item find a star product and
\item find a measure corresponding to \emph{this} star product.
\end{itemize}
In general it surely is not possible to start with an arbitrary star
product for the considered quantum space in the hope to find a measure
giving the integral the above trace property to all orders in \( h \).

\subsubsection{1. Calculation}

Of course it is possible to start with a special star product that
seems to be {}``good'' in order to try to calculate a corresponding
measure. The problems that arise are that 

\begin{itemize}
\item in general it is difficult to find a solution valid to all orders
in \( h \) 
\item we have to start with a special choice of a star product without knowing
if this concrete star product permits a solution, i.e. if there exists
a measure corresponding to this star product that renders the usual
integral cyclic in the above sense.
\end{itemize}
In our example things can actually be done by calculation: First we
have to fix a star product. The star product \( \star _{q} \) (\ref{de:q-symm star product})
seems to be a good choice since the symmetry of \( \star _{q} \)
tells us that \( f\star _{q}g \) and \( g\star _{q}f \) are equal
for all even orders of \( h \) such that we do not have to bother
about even orders anymore. If one starts first to consider the condition
(\ref{eq: requirement for cyclic property of the integral}) up to
first order in \( h \), a short calculation leads to the result \( \frac{1}{z\overline{z}} \)
for the measure. We can even show that \( \frac{1}{z\overline{z}} \)
is a measure that satisfies (\ref{eq: requirement for cyclic property of the integral})
to all orders in \( h \): 

\bprop\label{prop: measure function is z,zbar}

For all functions \( f,g \) we have\[
\int \frac{dzd\overline{z}}{z\overline{z}}\, f\star _{q}g=\int \frac{dzd\overline{z}}{z\overline{z}}\, fg=\int \frac{dzd\overline{z}}{z\overline{z}}\, g\star _{q}f\, \, \, .\]
 \eprop 

\begin{proof}

We have\begin{eqnarray*}
\int \frac{dzd\overline{z}}{z\overline{z}}\, f\star _{q}g=\int \frac{dzd\overline{z}}{z\overline{z}}\, fg+\int \frac{dzd\overline{z}}{z\overline{z}}\, \mu \circ \sum ^{\infty }_{n=1}\frac{h^{n}}{n!}(\sum ^{2}_{i_{1},j_{1}=1}\varepsilon ^{i_{1}j_{1}}z^{i_{1}}\frac{\partial }{\partial z^{i_{1}}}\te z^{j_{1}}\frac{\partial }{\partial z^{j_{1}}}) & \\
(\sum ^{2}_{i_{2},j_{2}=1}\varepsilon ^{i_{2}j_{2}}z^{i_{2}}\frac{\partial }{\partial z^{i_{2}}}\te z^{j_{2}}\frac{\partial }{\partial z^{j_{2}}})\dots (\sum ^{2}_{i_{n},j_{n}=1}\varepsilon ^{i_{n}j_{n}}z^{i_{n}}\frac{\partial }{\partial z^{i_{n}}}\te z^{j_{n}}\frac{\partial }{\partial z^{j_{n}}})(f\te g)\, \, \, . & 
\end{eqnarray*}
 Let us consider the \( n \) -th term of the sum on the right hand
side\begin{eqnarray*}
\int \frac{dzd\overline{z}}{z\overline{z}}\, \frac{h^{n}}{n!}\mu \circ (\sum ^{2}_{i_{1},j_{1}=1}\varepsilon ^{i_{1}j_{1}}z^{i_{1}}\frac{\partial }{\partial z^{i_{1}}}\te z^{j_{1}}\frac{\partial }{\partial z^{j_{1}}})(\sum ^{2}_{i_{2},j_{2}=1}\varepsilon ^{i_{2}j_{2}}z^{i_{2}}\frac{\partial }{\partial z^{i_{2}}}\te z^{j_{2}}\frac{\partial }{\partial z^{j_{2}}}) &  & \\
\dots (\sum ^{2}_{i_{n},j_{n}=1}\varepsilon ^{i_{n}j_{n}}z^{i_{n}}\frac{\partial }{\partial z^{i_{n}}}\te z^{j_{n}}\frac{\partial }{\partial z^{j_{n}}})(f\te g)\, \, \, . &  & 
\end{eqnarray*}
 We introduce the short hand notation \[
f^{'}\te g^{'}:=(\sum ^{2}_{i_{2},j_{2}=1}\varepsilon ^{i_{2}j_{2}}z^{i_{2}}\frac{\partial }{\partial z^{i_{2}}}\te z^{j_{2}}\frac{\partial }{\partial z^{j_{2}}})\dots (\sum ^{2}_{i_{n},j_{n}=1}\varepsilon ^{i_{n}j_{n}}z^{i_{n}}\frac{\partial }{\partial z^{i_{n}}}\te z^{j_{n}}\frac{\partial }{\partial z^{j_{n}}})(f\te g)\]
 and with that the \( n- \)th term of the sum can be written as\[
\int \frac{dzd\overline{z}}{z\overline{z}}\, \frac{h^{n}}{n!}\mu \circ (\sum ^{2}_{i_{1},j_{1}=1}\varepsilon ^{i_{1}j_{1}}z^{i_{1}}\frac{\partial }{\partial z^{i_{1}}}\te z^{j_{1}}\frac{\partial }{\partial z^{j_{1}}})(f^{'}\te g^{'})\]
\[
=\frac{h^{n}}{n!}\int dzd\overline{z}\, \sum ^{2}_{i_{1},j_{1}=1}\varepsilon ^{i_{1}j_{1}}\frac{\partial }{\partial z^{i_{1}}}(f^{'})\frac{\partial }{\partial z^{j_{1}}}(g^{'})\, \, \, .\]
For \( n>0 \) this leads after partial integration (we assume always
that functions considered here vanish at infinity) to \[
-\frac{h^{n}}{n!}\int dzd\overline{z}\, \sum ^{2}_{i_{1},j_{1}=1}\varepsilon ^{i_{1}j_{1}}f^{'}\frac{\partial }{\partial z^{i_{1}}}\frac{\partial }{\partial z^{j_{1}}}(g^{'})=0\, \, \, .\]
 This is valid for any summand corresponding to \( n>0 \) and only
the zeroth order term does not vanish such that at the end we find
what we claimed: \[
\int \frac{dzd\overline{z}}{z\overline{z}}\, f\star _{q}g=\int \frac{dzd\overline{z}}{z\overline{z}}\, fg\, \, \, .\]

\end{proof}

\subsubsection{2. Using a Theorem from Felder and Shoikhet}

We will use the following theorem from Felder and Shoikhet published
in \cite{Felder}:

\begin{Theorem}

Let \( M \) be a Poisson manifold with the bivector field \( \pi  \)
and let \( \Omega  \) be any volume form on \( M \) such that \( \mathrm{div}_{\Omega }\pi =0 \).
Then there exists a star product on \( C^{\infty }(M) \) such that
for any two functions \( f,g \) one has\[
\int _{M}\Omega \, (f\star g)=\int _{M}\Omega \, fg\, \, \, .\]

\end{Theorem}Let me remark on how we have to understand \( \mathrm{div}_{\Omega }\pi . \)
In \cite{Felder} it is defined as follows:\[
\mathrm{div}\textrm{ }:\, T^{k}_{\mathrm{poly}}(M)\stackrel{\Omega }{\longrightarrow }\Omega ^{d-k-1}(M)\stackrel{\mathrm{d}}{\longrightarrow }\Omega ^{d-k}(M)\stackrel{\Omega ^{-1}}{\longrightarrow }T^{k-1}_{\mathrm{poly}}(M)\]
 where \( d=\mathrm{dim}M \), \( T^{k}_{\mathrm{poly}}(M) \) is
the set of \( k+1 \) - polyvector fields and \( \mathrm{d} \) denotes
the exterior differential. In have the following setting: \( \pi =-2iz\overline{z}\varepsilon ^{ij}\partial _{i}\wedge \partial _{j} \)
and we write for a general volume form \( \Omega (z,\overline{z})=\omega (z,\overline{z})dz\wedge d\overline{z} \).
Then we have\[
\Omega (\pi )=\omega (z,\overline{z})(-2i)z\overline{z}.\]
 Now we see that if we choose \( \omega (z,\overline{z})=\frac{1}{z\overline{z}} \)
we get \( \Omega (\pi )=-2i=\mathrm{const}. \) such that in this
case surely \( \mathrm{d}\circ \Omega (\pi )=0 \). With that\[
\mathrm{div}_{\Omega }\pi =0\, \, \, .\]

What did we achieve this way? We easily could find a volume form \( \Omega  \)
such that \( \mathrm{div}_{\Omega }\pi =0 \). The above theorem then
assures the existence of a star product such that the integral over
that volume form is cyclic (even more: The integral of two functions'
star multiplication is equal to the integral of the ordinarily multiplied
functions). Of course we still do not know which star product has
together with that volume form this nice property. Nonetheless, the
\emph{existence} at least is assured and for usual physical applications
this can be sufficient: usually we do only need to know the star product
explicitly up to second order and these orders can be, if the volume
form is known, determined by calculation. In our example the above
calculation shows that \( \star _{q} \) is the right star product
corresponding to the volume form \( \frac{1}{z\overline{z}}dz\wedge d\overline{z} \).

We want to remark that for other, more complicated poisson structures
where a solution cannot be found easily by calculation, this approach
could be the more systematic one. It should also be possible for more
complicated Poisson structures to determine a volume form \( \Omega  \)
satisfying the condition \( \mathrm{div}_{\Omega }\pi =0 \) guaranteeing
then by Felder's and Shoikhet's theorem the existence of a star product
leading to the trace property. This star product then hopefully can
be determined explicitly in a second step (but in any case at least
the first orders of it).

\subsection{The Action and the Integral}

Having found an integral with trace property for the star product
\( \star _{q} \) as well as a gauge covariant field strength with
right semi-classical  limit, we can write down an invariant action:\[
S=\int \frac{dzd\overline{z}}{z\overline{z}}\, \mathcal{L}=\int \frac{dzd\overline{z}}{z\overline{z}}\, \frac{1}{4}\delta ^{ik}\delta ^{lj}F_{kl}^{'}\star _{q}F_{ij}^{'}\, \, \, .\]
A new problem arises immediately: Let us determine the semi-classical
limit for \( S \). We know that \( \frac{1}{4}F^{'}_{ij}\star _{q}F^{'ij}\rp \frac{1}{4}f_{ij}f^{ij} \)
for \( h\rp 0 \) because of (\ref{de: Ftilde' with right classical limit})
such that we obtain \[
S=\int \frac{dzd\overline{z}}{z\overline{z}}\, \frac{1}{4}f_{ij}f^{ij}+\mathcal{O}(h)\, \, \, .\]
 This is not exactly what we want. In the semi-classical  limit we
do not find the free action for the flat plane, since the measure
we had to introduce to guarantee the trace property of the integral
does not disappear. Again we have to make a modification, similar
to what we already did for the field strength in (\ref{de: cov field strength with cl limit}):
we have to multiply by a covariant term that in the semi-classical
limit reduces to \( z\overline{z} \).%
\footnote{I want to thank Peter Schupp who put forward this idea.
} Then, for \( h\rp 0 \) we get the usual integral over \( z,\overline{z} \).
Let me remark that we do not loose the gauge invariance of the action
if the additional term we introduce is gauge covariant. We want to
multiply by the covariant coordinates and finally define \[
S^{'}:=\int \frac{dzd\overline{z}}{z\overline{z}}\, Z\star _{q}\overline{Z}\star _{q}\mathcal{L}=\int \frac{dzd\overline{z}}{z\overline{z}}\, \frac{1}{4}Z\star _{q}\overline{Z}\star _{q}\frac{1}{4}\delta ^{ik}\delta ^{lj}F_{kl}^{'}\star _{q}F_{ij}^{'}\]
 as the free action. By this definition we get that

\begin{itemize}
\item \( S^{'} \) is gauge invariant as well as that 
\item \( S^{'}=\int dzd\overline{z}\, \frac{1}{4}f_{ij}f^{ij}+\mathcal{O}(h) \),
i.e. it admits the right semi-classical  limit. 
\end{itemize}
(The reader might note that we have some freedom in choosing this
additional, covariant term and also in choosing the position where
to place it under the integral. We come back to this topic later.)

\subsubsection{Expanded Field Strength and Action}

We want to expand this action in \( h \) to be able to read off directly
the corrections we get to the commutative theory in first order. We
start calculating the first order term in \( h \) for \( F_{z\overline{z}}^{'}=-\frac{1}{h^{2}}\mathcal{D}(\theta ^{-1})_{z\overline{z}}\star _{q}F^{\overline{z}z}\star _{q}\mathcal{D}(\theta ^{-1})_{z\overline{z}} \)
using the expression for \( \mathcal{D}(\theta ^{-1})_{z\overline{z}} \)
up to first order in \( h \) (see (\ref{eq: Dtilde(theta) bis 1. Ordnung h}))
and the second order term of \( F^{\overline{z}z} \) given in (\ref{eq:classical limit for F}).
Additionally we have to determine the third order term of \( F^{\overline{z}z} \)
and must put all the results together. This was done making use of
{}``Mathematica''%
\footnote{Special thanks to Fabian Bachmaier who made this possible.
} leading to the following results:\begin{eqnarray}
F^{'}_{z\overline{z}} & = & f_{z\overline{z}}+h\left\{ -f_{z\overline{z}}-2iz\overline{z}f^{2}_{z\overline{z}}+2iz\overline{z}(a_{z}\partial _{\overline{z}}f_{z\overline{z}}-a_{\overline{z}}\partial _{z}f_{z\overline{z}})\right\} +\mathcal{O}(h^{2})\nonumber \\
 & = & f_{z\overline{z}}+h\left\{ -f_{z\overline{z}}+\theta ^{z\overline{z}}f^{2}_{z\overline{z}}-\theta ^{z\overline{z}}(a_{z}\partial _{\overline{z}}f_{z\overline{z}}-a_{\overline{z}}\partial _{z}f_{z\overline{z}})\right\} +\mathcal{O}(h^{2})\label{F' expanded in theta first component} 
\end{eqnarray}
and \begin{eqnarray}
F^{'}_{\overline{z}z} & = & f_{\overline{z}z}+h\left\{ -3f_{\overline{z}z}+2iz\overline{z}f^{2}_{z\overline{z}}+2iz\overline{z}(a_{z}\partial _{\overline{z}}f_{\overline{z}z}-a_{\overline{z}}\partial _{z}f_{\overline{z}z})\right\} +\mathcal{O}(h^{2})\nonumber \\
 & = & f_{\overline{z}z}+h\left\{ -3f_{\overline{z}z}-\theta ^{z\overline{z}}f^{2}_{z\overline{z}}-\theta ^{z\overline{z}}(a_{z}\partial _{\overline{z}}f_{\overline{z}z}-a_{\overline{z}}\partial _{z}f_{\overline{z}z})\right\} +\mathcal{O}(h^{2})\, \, \, .\label{F' expanded in theta second component} 
\end{eqnarray}
Expansion of the Lagrangian \( \mathcal{L}:=\frac{1}{4}\delta ^{ik}\delta ^{lj}F_{kl}^{'}\star _{q}F_{ij}^{'} \)
yields:\begin{eqnarray}
\mathcal{L} & = & \frac{1}{4}f_{ij}f^{ij}+h\left\{ -2f^{2}_{z\overline{z}}-2iz\overline{z}f_{z\overline{z}}(a_{\overline{z}}\partial _{z}f_{z\overline{z}}-a_{z}\partial _{\overline{z}}f_{z\overline{z}}+f_{z\overline{z}}^{2})\right\} +\mathcal{O}(h^{2})\nonumber \\
 & = & \frac{1}{4}f_{ij}f^{ij}\label{eq: L expanded in theta} \\
 &  & +h\left\{ -2f^{2}_{z\overline{z}}+\theta ^{z\overline{z}}f_{z\overline{z}}(a_{\overline{z}}\partial _{z}f_{z\overline{z}}-a_{z}\partial _{\overline{z}}f_{z\overline{z}}+f_{z\overline{z}}^{2})\right\} +\mathcal{O}(h^{2})\, \, \, .\nonumber 
\end{eqnarray}
Finally, multiplying the action with the covariant term \( Z\star _{q}\overline{Z} \)
to guarantee the right semi-classical  limit (see above) we find for
the action:\begin{eqnarray}
S^{'} & := & \int \frac{dzd\overline{z}}{z\overline{z}}\, Z\star _{q}\overline{Z}\star _{q}\mathcal{L}_{free}\nonumber \\
 & = & \int dzd\overline{z}\, \frac{1}{4}f_{ij}f^{ij}\nonumber \\
 &  & +h\left\{ -\frac{3}{2}f^{2}_{z\overline{z}}+(iza_{z}-i\overline{z}a_{\overline{z}})f_{z\overline{z}}^{2}+f_{z\overline{z}}(z\partial _{z}f_{z\overline{z}}-\overline{z}\partial _{\overline{z}}f_{z\overline{z}})\right. \nonumber \\
 &  & -z\overline{z}(2if_{z\overline{z}}^{3}-2if_{z\overline{z}}(a_{z}\partial _{\overline{z}}f_{z\overline{z}}-a_{\overline{z}}\partial _{z}f_{z\overline{z}})\bigg \}+\mathcal{O}(h^{2})\label{eq: S' expanded in theta} \\
 & = & \int dzd\overline{z}\, \frac{1}{4}f_{ij}f^{ij}\nonumber \\
 &  & +h\left\{ -\frac{3}{2}f^{2}_{z\overline{z}}-\frac{1}{2}(\partial _{\overline{z}}(\theta ^{z\overline{z}})a_{z}-\partial _{z}(\theta ^{z\overline{z}})a_{\overline{z}})f_{z\overline{z}}^{2}+\frac{i}{2}f_{z\overline{z}}(\partial _{\overline{z}}(\theta ^{z\overline{z}})\partial _{z}f_{z\overline{z}}\right. \nonumber \\
 &  & -\partial _{z}(\theta ^{z\overline{z}})\partial _{\overline{z}}f_{z\overline{z}})+\theta ^{z\overline{z}}f_{z\overline{z}}^{3}-\theta ^{z\overline{z}}f_{z\overline{z}}(a_{z}\partial _{\overline{z}}f_{z\overline{z}}-a_{\overline{z}}\partial _{z}f_{z\overline{z}})\bigg \}+\mathcal{O}(h^{2})\, \, \, .\nonumber 
\end{eqnarray}

\subsubsection{Expanded Field Strength and Action in the Real Basis \protect\( x,y\protect \) }

The field strength \( \tilde{F}' \) in the basis \( x,y \) was already
introduced in (\ref{de: Ftilde' with right classical limit}). If
we now use the star product \( \star _{q} \) in the basis \( x,y \)
given in (\ref{eq: star product for real basis x,y}), we obtain the
following results up to first order in \( h \):\[
\tilde{F}^{'}_{\alpha \beta }=-\frac{1}{h^{2}}\mathcal{D}(\theta ^{-1})_{\alpha \gamma }\star _{q}\tilde{F}^{\gamma \delta }\star _{q}\mathcal{D}(\theta ^{-1})_{\delta \beta }\]
 gives \begin{eqnarray}
\tilde{F}'_{12} & = & \tilde{f}_{12}+h\left\{ -\tilde{f}_{12}+(x^{2}+y^{2})\tilde{f}_{12}^{2}-(x^{2}+y^{2})(\tilde{a}_{1}\partial _{y}\tilde{f}_{12}-\tilde{a}_{2}\partial _{x}\tilde{f}_{12})\right\} +\mathcal{O}(h^{2})\nonumber \\
 & = & \tilde{f}_{12}+h\left\{ -\tilde{f}_{12}+\tilde{\theta }^{12}\tilde{f}_{12}^{2}-\tilde{\theta }^{12}(\tilde{a}_{1}\partial _{y}\tilde{f}_{12}-\tilde{a}_{2}\partial _{x}\tilde{f}_{12})\right\} +\mathcal{O}(h^{2})\label{eq: expansion of Ftilde' first component} 
\end{eqnarray}
 and \begin{eqnarray}
\tilde{F}'_{21} & = & \tilde{f}_{21}+h\left\{ -3\tilde{f}_{21}-(x^{2}+y^{2})\tilde{f}_{12}^{2}-(x^{2}+y^{2})(\tilde{a}_{1}\partial _{y}\tilde{f}_{21}-\tilde{a}_{2}\partial _{x}\tilde{f}_{21})\right\} +\mathcal{O}(h^{2})\nonumber \\
 & = & \tilde{f}_{21}+h\left\{ -3\tilde{f}_{21}-\tilde{\theta }^{12}\tilde{f}_{12}^{2}-\tilde{\theta }^{12}(\tilde{a}_{1}\partial _{y}\tilde{f}_{21}-\tilde{a}_{2}\partial _{x}\tilde{f}_{21})\right\} +\mathcal{O}(h^{2})\, \, \, .\label{eq: expansion of Ftilde' second component} 
\end{eqnarray}
 The Lagrangian introduced in (\ref{de: Ltilde}) has the following
form up to first order in \( h \): \begin{eqnarray}
\tilde{\mathcal{L}} & = & \frac{1}{4}\tilde{f}_{\alpha \beta }\tilde{f}^{\alpha \beta }\nonumber \\
 &  & +h\left\{ -2\tilde{f}^{2}_{12}+(x^{2}+y^{2})(\tilde{f}_{12}(\tilde{a}_{2}\partial _{x}\tilde{f}_{12}-\tilde{a}_{1}\partial _{y}\tilde{f}_{12})+\tilde{f}_{12}^{3})\right\} +\mathcal{O}(h^{2})\nonumber \\
 & = & \frac{1}{4}\tilde{f}_{\alpha \beta }\tilde{f}^{\alpha \beta }\label{eq: expansion of Ltilde} \\
 &  & +h\left\{ -2\tilde{f}^{2}_{12}+\tilde{\theta }^{12}(\tilde{f}_{12}(\tilde{a}_{2}\partial _{x}\tilde{f}_{12}-\tilde{a}_{1}\partial _{y}\tilde{f}_{12})+\tilde{f}_{12}^{3})\right\} +\mathcal{O}(h^{2})\, \, \, .\nonumber 
\end{eqnarray}
 To get the right classical limit for the integral we must again multiply
by a covariant term. We take \( X\star _{q}X+Y\star _{q}Y \) (note
that this is not equal to \( (X+iY)\star _{q}(X-iY) \)!) and finally
find for the modified field strength 

\begin{eqnarray}
\tilde{S}^{'} & = & \int dxdy\, \frac{1}{4}\tilde{f}_{\alpha \beta }\tilde{f}^{\alpha \beta }\nonumber \\
 &  & +h\left\{ 2\tilde{f}^{2}_{12}-(x^{2}+y^{2})\tilde{f}_{12}^{3}-i(x\tilde{a}_{2}\partial _{y}\tilde{f}_{12}-y\tilde{a}_{1}\partial _{x}\tilde{f}_{12})\tilde{f}_{12}\right. \nonumber \\
 &  & \left. -(x^{2}+y^{2})\tilde{f}_{12}(\tilde{a}_{2}\partial _{x}\tilde{f}_{12}-\tilde{a}_{1}\partial _{y}\tilde{f}_{12})+(y\tilde{a}_{1}-x\tilde{a}_{2})\tilde{f}^{2}_{12}\right\} +\mathcal{O}(h^{2})\nonumber \\
 & = & \int dxdy\, \frac{1}{4}\tilde{f}_{\alpha \beta }\tilde{f}^{\alpha \beta }\label{eq: expansion of Stilde'} \\
 &  & +h\left\{ 2\tilde{f}^{2}_{12}-\tilde{\theta }^{12}\tilde{f}_{12}^{3}-\frac{i}{2}(\partial _{x}(\tilde{\theta }^{12})\tilde{a}_{2}\partial _{y}\tilde{f}_{12}-\partial _{y}(\tilde{\theta }^{12})\tilde{a}_{1}\partial _{x}\tilde{f}_{12})\tilde{f}_{12}\right. \nonumber \\
 &  & \left. -\tilde{\theta }^{12}\tilde{f}_{12}(\tilde{a}_{2}\partial _{x}\tilde{f}_{12}-\tilde{a}_{1}\partial _{y}\tilde{f}_{12})+\frac{1}{2}(\partial _{y}(\tilde{\theta }^{12})\tilde{a}_{1}-\partial _{x}(\tilde{\theta }^{12})\tilde{a}_{2})\tilde{f}^{2}_{12}\right\} +\mathcal{O}(h^{2})\, \, \, .\nonumber 
\end{eqnarray}

Again we find the right semi-classical  limit, i.e. the theory we
established approaches for both cases, complex and real basis, the
commutative theory in the limit \( h\rp 0. \) Moreover, we can now
read off the correction to the commutative theory this noncommutative
theory leads to up to first order in \( h \). \\
Let us comment on the results: 

\begin{itemize}
\item First we see that we get after transformation in both bases the same
results for the field strength \( F \) respectively \( \tilde{F} \)
and the Lagrangian \( \mathcal{L} \) respectively \( \tilde{\mathcal{L}} \)
providing us a correctness check.
\item The two results for \( S' \) respectively \( \tilde{S} \) are not
identically equal. The reason is that we multiplied in the case of
the real basis by the additional covariant term \( X\star _{q}X+Y\star _{q}Y \)
to compensate the measure of the integral in the limit \( q\rp 1 \),
whereas we took in the case of the complex basis the covariant term
\( Z\star _{q}\overline{Z}=(X+iY)\star _{q}(X-iY)\neq X\star _{q}X+Y\star _{q}Y \).
This was done on purpose to show that the results we get here depend
on the special choice of the additional term (we just required gauge-covariance
and a special semi-classical  limit but these requirements do not
uniquely determine such an additional term). We will comment on this
in the following subsection. 
\item The results we get lead us to the following interpretation: In first
order of \( h \) we obtain a correction to the common photon propagator
(quadratic term in \( f \)). Additionally, we get, if we interpret
\( \theta  \) as a background field, some interaction term between
three photons and the \( \theta - \)field (the rest of the action's
first-order term). 
\item Matter fields could easily be introduced, too, leading to more new
interactions in the first order of the deformation parameter \( h \).
\end{itemize}

\subsection{Freedom of the Theory}

We established in all detail a gauge field theory on \( \mathcal{A}_{\mathrm{nc}} \),
the \( E_{q}(2)- \)symmetric space. We defined a field strength,
introduced a cyclic integral and got finally an invariant action with
right semi-classical  limit. At the end the expansion of field strength,
Lagrangian and action in the deformation parameter \( h \) was calculated
up to first order. We can read off explicitly how far the theory differs
from the commutative theory. 

Nevertheless, it is our obligation to look back and to examine whether
all the assumptions and the definitions we made are dictated by principles
we want to require or how far our considerations actually admit freedom
in defining the introduced physical quantities such as field strength,
Lagrangian and action. The reader might already have noticed that
in the definition of \( F' \) and of \( \mathcal{L} \) (cf. (\ref{de: cov field strength with cl limit})
and (\ref{de: Ltilde})) at least the position where to put the appearing
\( \mathcal{D}(\theta ^{-1}) \) terms is arbitrary. There is no conceptual
criterion that distinguishes one special choice of position. All possible
choices of position lead to the same semi-classical  limit and give
covariant expressions and this is all we required. The hope was, that
nevertheless physics might be unaffected of the choice of position:
We hoped that at least the first order term of the expanded quantities
would not depend on the choice of position. Unfortunately, calculation
shows that for example \( \frac{1}{h^{2}}\mathcal{D}(\theta ^{-1})_{z\overline{z}}\star _{q}F^{\overline{z}z}\star _{q}\mathcal{D}(\theta ^{-1})_{z\overline{z}} \)
and \( \frac{1}{h^{2}}\mathcal{D}(\theta ^{-1})_{z\overline{z}}\star _{q}\mathcal{D}(\theta ^{-1})_{z\overline{z}}\star _{q}F^{\overline{z}z} \),
two equivalent, possible ways to define \( F' \), differ in first
order of \( h \) by \[
-2h(\overline{z}\partial _{\overline{z}}f_{z\overline{z}}-z\partial _{z}f_{z\overline{z}})\]
 such that the choice of position of the intervening \( \mathcal{D}(\theta ^{-1}) \)
terms influences the physical results. 

Moreover, when we defined the action we had to introduce a covariant
expression that compensates in the semi-classical limit the measure
\( \frac{1}{z\overline{z}} \), which in turn we had to introduce
to guarantee the trace property of the integral (see (\ref{eq: requirement1 for F: cov transf})
and Proposition \ref{prop: measure function is z,zbar}). At least
if we want to treat gauge field theory on the flat space we cannot
do without this additional term. We added \( Z\star _{q}\overline{Z} \)
but a lot of other terms could have been chosen. For example \( \mathcal{D}(z\overline{z}) \)
admits the same semi-classical  limit as \( Z\star _{q}\overline{Z} \)
and is also covariant. None of both can be regarded the better one.
In the way we established the theory, no conceptual criterion exists
to get rid of this freedom. 

The other general problem we have with this approach is the following:
Our considerations did not include in all points the background symmetry
of the space, the \( q-\textrm{deformed} \) two-dimensional Euclidean
group. We see that the integral with the measure we had to choose
to get the trace property is certainly not \emph{\( E_{q}(2) \)}
- invariant. Nevertheless, having a background symmetry we should
try to keep it and try to set up an \( E_{q}(2) \) - covariant theory. 

That is why we study in the next section an alternative approach to
the problem. We will establish a theory where the \( E_{q}(2)- \)
symmetry stands in the foreground of all considerations.

\section{\protect\( E_{q}(2)\protect \) - Covariant Abelian Gauge Field Theory}

A covariant, abelian gauge theory was already established on another
noncommutative space: the \( q- \)deformed fuzzy sphere that is covariant
with respect to the coaction of \( SU_{q}(2) \) \cite{Grosse:2000Fuzzy1,Grosse:2001pr,Steinacker:2001fj}.
Guided by these publications, we set up a gauge theory based on \( E_{q}(2)- \)covariance.
At the beginning, we construct an \( E_{q}(2)- \)invariant integral.
As we will see, it is not cyclic. In a second step an \( E_{q}(2)- \)covariant
differential calculus is introduced. We define an exterior differential,
\( q- \)one-forms, \( q- \)two-forms and \( q- \)deformed derivatives.
Moreover, a generator of the exterior differential as well as a frame,
a basis of one-forms that commutes with arbitrary functions, are derived.
We additionally introduce a gauge field \( A \) and a field strength
\( F \) that in the semi-classical limit \( q\rp 1 \) becomes the
commutative field strength. Problems arise when trying to define gauge
transformations: Since the integral does not possess the trace property,
we cannot introduce gauge transformations of gauge fields as we did
in the previous section. By means of an algebra homomorphism \( \alpha :U_{q}(e(2))\rp \mathcal{A}_{\mathrm{nc}} \),
we define gauge transformations of one forms in a way that allows
us to speak about a gauge invariant action. 

For a detailed discussion of quantum groups the reader is refered
to \cite{Kassel:1995xr,Klimyk:1997eb,Chaichian:1996ah}.

\subsection{An Invariant Integral on the \protect\( E_{q}(2)\protect \) - Symmetric
Plane\label{se: q-integra}}

We need an \( E_{q}(2) \) - invariant integral to define an action.
Considerations made in this subsection are guided by \cite{Koe},
where a Haar-functional on \( E_{q}(2) \) extended to formal power
series is derived. Nevertheless, we use another, physically more intuitive
basis of the dual of \( E_{q}(2) \) and present the necessary calculations.
Whereas we introduced in Section \ref{se: Eq(2) and the symmetric plane}
the quantum group \( E_{q}(2) \) and the \( E_{q}(2)- \)symmetric
plane, we proceed to introduce at this point the quantum dual of \( E_{q}(2) \).

\subsubsection{The Quantum Universal Enveloping Algebra \protect\( U_{q}(e(2))\protect \)
and Duality}

In many cases it is convenient to consider the action of the quantum
groups dual instead of the coaction of the quantum group itself. Therefore
we introduce \( U_{q}(e(2)) \), the dual of \( E_{q}(2) \). 

\begin{Def}\label{de: dual pairing}

\textit{Let \( (H,m,\eta ,\Delta ,\varepsilon ,S) \) and \( (H^{'},m^{'},\eta ^{'},\Delta ^{'},\varepsilon ^{'},S^{'}) \)
be two Hopf algebras. We say that \( H \) and \( H^{'} \) are in
duality if there exists a bilinear form, called dual pairing,}\textit{\emph{\[
\langle \cdot ,\cdot \rangle :\, H\te H'\rp \mathbb {C}\]
}} \textit{satisfying\begin{equation}
\label{defduality}
\begin{array}{rclcrcl}
\langle gh,x\rangle  & = & \langle g\te h,\Delta '(x)\rangle  &  & \langle h,xy\rangle  & = & \langle \Delta (h),x\te y\rangle \\
\langle 1,x\rangle  & = & \varepsilon '(x) &  & \langle h,1\rangle  & = & \varepsilon (h)
\end{array}
\end{equation}
} \textit{\emph{\[
\langle S(h),x\rangle =\langle h,S'(x)\rangle \]
}} \textit{for all \( g,h\in H \) and \( x,y\in H^{'} \), where
\( \langle g\ten h,x\ten y\rangle :=\langle g,x\rangle \langle h,y\rangle  \)
is the extension of \( \langle \cdot ,\cdot \rangle  \) on tensor
products.}

\end{Def}In \cite{Schupp:1992ex} the dual to \( E_{q}(2) \), called
\( U_{q}(e(2)) \), is constructed. With the following identifications
for the generators \( \mu ,\nu ,\xi  \) introduced in this publication,
\[
\begin{array}{ccccccccccc}
\mu  & \equiv  & T &  & -q^{2}\nu  & \equiv  & \overline{T} &  & \xi  & \equiv  & J
\end{array}\, \, \, ,\]
 we then get that \( U_{q}(e(2)) \) is generated by \( T,\overline{T},J \)
with the following commutation relations and structure maps\[
\begin{array}{ccccccccccc}
T\overline{T} & = & q^{2}\overline{T}T &  & [J,T] & = & iT &  & [J,\overline{T}] & = & -i\overline{T}
\end{array}\]
\begin{equation}
\label{eq: commrel and structure maps for Uq(e(2))}
\begin{array}{ccccccc}
\Delta (T) & = & T\te q^{2iJ}+1\te T &  & \Delta (\overline{T}) & = & \overline{T}\te q^{2iJ}+1\te \overline{T}
\end{array}
\end{equation}

\[
\begin{array}{ccccccccccc}
\Delta (J) & = & J\te 1+1\te J &  & \varepsilon (T) & = & \varepsilon (\overline{T}) & = & \varepsilon (J) & = & 0
\end{array}\]
\[
\begin{array}{ccccccccccc}
S(T) & = & -Tq^{-2iJ} &  & S(\overline{T}) & = & -\overline{T}q^{-2iJ} &  & S(J) & = & -J
\end{array}\, \, \, ,\]
 where the dual pairing on the generators is given by\begin{equation}
\label{eq: dual pairing Uq(e(2)) and Eq(2)}
\begin{array}{ccccc}
\langle T,\theta ^{i}t^{j}\overline{t}^{k}\rangle =\delta _{0i}\delta _{1j}\delta _{0k}, & \langle \overline{T},\theta ^{i}t^{j}\overline{t}^{k}\rangle =-q^{2}\delta _{0i}\delta _{0j}\delta _{1k}, & \langle J,\theta ^{i}t^{j}\overline{t}^{k}\rangle =\delta _{1i}\delta _{0j}\delta _{0k} & .
\end{array}
\end{equation}
Moreover, we have \begin{equation}
\label{eq: complex conjugation of J}
\overline{J}=-J\, \, \, .
\end{equation}
The generators \( T,\overline{T} \) as duals of \( t,\overline{t} \)
can be interpreted as the \( q- \)analogs of the generators of translations
whereas \( J \), the dual of \( \theta  \), can be viewed as the
generator of rotations (rotations are not deformed). After having
introduced the dual, we can now use the dual pairing to pass over
from the coaction of \( E_{q}(2) \) to an action of the dual \( U_{q}(e(2)) \)
on \( \mathbb {C}^{2}_{q} \).

\subsubsection{The Action of \protect\( U_{q}(e(2))\protect \) on \protect\( \mathcal{A}_{\mathrm{nc}}\protect \)}

First of all we want to define what me mean by the action of a Hopf
algebra on an algebra. 

\begin{Def}\label{de: def of action on an algebra}

\textit{We say that a Hopf algebra \( H \) is acting on an algebra
\( A \) from the right (left) if \( A \) is a right (left) \( H \)-module
such that \( m:A\ten A\ar A \) and \( \eta :\mathbb {C}\ar A \)
are right (left) \( H \)-module homomorphisms, that means if holds }

\textit{\emph{\begin{equation}
\label{action}
ab\lt h=(a\te b)\lt \Delta (h)=(a\lt h_{(1)})(b\lt h_{(2)})\, \, \, \, \, \, \textrm{and}\, \, \, \, \, \, 1\lt h=\varepsilon (h)1
\end{equation}
}} respectively\emph{\[
h\rt ab=\Delta (h)\rt (a\te b)=(h_{(1)}\rt a)(h_{(2)}\rt b)\, \, \, \, \, \, \textrm{and}\, \, \, \, \, \, h\rt 1=\varepsilon (h)1\, \, \, ,\]
}\textit{for any \( h\in H \) and \( a,b\in A \). An algebra \( A \)
with a Hopf algebra \( H \) acting on it from the right (left) is
called a right (left) \( H- \)module algebra.}

\end{Def}

\begin{Lemma}

\textit{Given a dual pairing between Hopf algebras \( H \) and \( H^{'} \)
and a left coaction \( \rho  \) of \( H^{'} \) on an algebra \( A \),
we get a right action}%
\footnote{Similarly we a get a left action via a dual pairing from a right coaction.
} \textit{of \( H \) on \( A \), \( \tril :A\ten H\ar A \), by defining}

\begin{equation}
\label{de: action from coaction}
f\lt X:=(\langle X,\cdot \rangle \te id)\circ \rho (f)=\langle X,f_{(-1)}\rangle f_{(0)}\, ,\, \, \, \, \, \, \, \, \, X\in H,\, f\in A\, \, \, .
\end{equation}

\end{Lemma}

\begin{proof}

We use Sweedler notation. \\
If we define as in (\ref{de: action from coaction}) we get from the
properties of the dual pairing (Definition \ref{de: dual pairing})
that \( A \) is a right \( H \)-module. For example we have \begin{eqnarray*}
i)\quad (f\tril X)\tril Y & = & (\langle X,f_{(-1)}\rangle f_{(0)})\tril Y\\
 & = & \langle X,f_{(-1)}\rangle \langle Y,f_{(0)_{(-1)}}\rangle f_{(0)_{(0)}}\\
 & = & \langle X,f_{(-2)}\rangle \langle Y,f_{(-1)}\rangle f_{(0)}\\
 & = & \langle X\ten Y,\Delta f_{(-1)}\rangle f_{(0)}\\
 & = & \langle XY,f_{(-1)}\rangle f_{(0)}\\
 & = & f\tril (XY)\qquad .
\end{eqnarray*}
 We prove the remaining requirements again using Definition \ref{de: dual pairing}:\begin{eqnarray*}
ii)\quad fg\tril X & = & \langle X,(fg)_{(-1)}\rangle (fg)_{(0)}\\
 & = & \langle X_{(1)},f_{(-1)}\rangle \langle X_{(2)},g_{(-1)}\rangle f_{(0)}g_{(0)}\\
 & = & (f\tril X_{(1)})(g\tril X_{(2)})\\
 &  & \\
iii)\quad 1\tril X & = & \langle X,1\rangle 1\\
 & = & \varepsilon (X)1\qquad .
\end{eqnarray*}

\end{proof}

Thus, with the dual pairing given in (\ref{eq: dual pairing Uq(e(2)) and Eq(2)})
and with the coaction of \( z,\overline{z} \) given in (\ref{coac})
we can define an action of \( U_{q}(e(2)) \) on \( \mathbb {C}^{2}_{q} \)
in analogy to (\ref{de: action from coaction}). We obtain for the
action of \( J,T \) and \( \overline{T} \) on the generators \( z,\overline{z} \):\begin{equation}
\label{eq: action of Uq on generators z, zbar}
\begin{array}{ccccc}
z\lt T & = & \langle T,e^{i\theta }\rangle z+\langle T,t\rangle  & = & 1\\
z\lt \overline{T} & = & \langle \overline{T},e^{i\theta }\rangle z+\langle \overline{T},t\rangle  & = & 0\\
z\lt J & = & \langle J,e^{i\theta }\rangle z+\langle J,t\rangle  & = & iz\\
\overline{z}\lt T & = & \langle T,e^{i\theta }\rangle \overline{z}+\langle T,\overline{t}\rangle  & = & 0\\
\overline{z}\lt \overline{T} & = & \langle \overline{T},e^{-i\theta }\rangle \overline{z}+\langle \overline{T},\overline{t}\rangle  & = & -q^{2}\\
\overline{z}\lt J & = & \langle J,e^{-i\theta }\rangle \overline{z}+\langle J,\overline{t}\rangle  & = & -i\overline{z}\, \, \, .
\end{array}
\end{equation}
 Now we want to extend the action of \( U_{q}(e(2)) \) on the algebra
of formal power series in \( z,\overline{z} \) respecting the commutation
relation (\ref{com z}). This algebra we denoted by \( \mathcal{A}_{\mathrm{nc}} \)
(\ref{de: A_c and A_nc (formal power series in z,zbar)}). Before
giving explicit formulas, let us remark that any formal power series
\( f(z,\bar{z})\in \mathcal{A}_{\mathrm{nc}} \) can be decomposed
as follows:

\begin{equation}
\label{de}
f(z,\overline{z})=f_{1}(z,z\overline{z})+f_{2}(\overline{z},z\overline{z})
\end{equation}
where \( f_{1},f_{2} \) are formal power series. To see that, we
have to split a formal power series depending of \( z,\bar{z} \)
into two series, the one consisting of all summands where the exponent
of \( z \) is bigger or equal to that of \( \bar{z} \) and the other
consisting of the remaining part. Taking into account that \be 
z^{k}\bar {z} ^{k} = q^{k(k-1)}(z\bar {z})^{k} \qquad, \ee
which can be verified using the commutation relation (\ref{com z})
and that \( \sum _{i=0}^{k-1}i=\ha k(k-1) \), we get a decomposition
of the above type (\ref{de}). \( f_{1} \) and \( f_{2} \) in turn
can be decomposed in sums consisting of summands of the following
type: \be \label{dec} z^{k}f(z\bar {z}) \qquad \mbox{respectively} \qquad \bar {z}^{k}f(z\bar {z})~~.
\ee Moreover, positive powers of \( \bar{z} \) can be rewritten in
terms of negative powers of \( z \) by using \be
\bar {z}^{k}= q^{k(k-1)}z^{-k}(z\bar {z})^{k} \qquad . \ee This leads
to the following remark:

\begin{Remark}

\textit{Any formal power series \( f(z,\bar{z}) \) can be written
in the following form:} 

\emph{\begin{equation}
\label{eq: decomposition of arbitrary f}
f(z,\overline{z})=\sum _{m\in \mathbb {Z}}z^{m}f_{m}(z\overline{z})
\end{equation}
}where \( f_{m} \) are formal power series in \( z\overline{z} \). 

\end{Remark}

To know the action on \( \mathcal{A}_{\mathrm{nc}} \) we now just
have to define it on a summand \be
z^{k}f(z\bar {z}) \qquad, k\in \mathbb{Z} \ee
and extend it linearly to power series (\ref{eq: decomposition of arbitrary f}).
Using the structure maps of \( U_{q}(e(2)) \) given in (\ref{eq: commrel and structure maps for Uq(e(2))})
it is possible to determine the right-action of \( J,T,\overline{T} \)
on such a summand. This is done in Appendix \ref{appendix: Action of Uq on A_nc}
leading to the following results:\begin{eqnarray}
z^{k}f(z\overline{z})\lt T & = & \frac{z^{k-1}}{1-q^{-2}}(f(q^{2}z\overline{z})-q^{-2k}f(z\overline{z}))\nonumber \\
z^{k}f(z\overline{z})\lt \overline{T} & = & \frac{q^{4}}{1-q^{2}}z^{k+1}\frac{f(z\overline{z})-f(q^{-2}z\overline{z})}{z\overline{z}}\label{eq: action of J,T,Tbar on z^kf(zzbar)} \\
z^{k}f(z\overline{z})\lt J & = & i^{k}z^{k}f(z\overline{z})\, \, \, .\nonumber 
\end{eqnarray}
 We see that the action of \( U_{q}(e(2)) \) on a function \( z^{k}f(z\overline{z}) \)
closes in \( \mathcal{A}_{\mathrm{nc}} \), i.e. we have

\begin{equation}
\label{de: Alg_nc_ext}
f\in \mathcal{A}_{\mathrm{nc}}\, \, \Longrightarrow \, \, f\lt X\in \mathcal{A}_{\mathrm{nc}}\, \, \, \textrm{for all }X\in U_{q}(e(2))\, \, \, .
\end{equation}
 \emph{Remark:} A priori it is not clear whether an extension of the
action to formal power series is possible. But the special form of
the action on summands as we got it in (\ref{eq: action of J,T,Tbar on z^kf(zzbar)})
shows, that in our case this indeed can be done: The action on \( z \)
to the power of \( k \) multiplied by a function \( f(z\overline{z}) \)
leads in all cases to the powers \( k-1,k \) or \( k+1 \). Thus,
the action on a formal power series, where also negative powers of
\( z \) are allowed, again leads in all cases to finitely many contributions
in every power of \( z \) and therefore the linear extension is actually
well-defined.

\subsubsection{The \protect\( U_{q}(e(2))\protect \) Invariant Integral}

We will construct a \( U_{q}(e(2))-\textrm{invariant} \) integral
following the methods given in \cite{Koe} starting by defining what
we want to call an invariant integral.

\begin{Def}\label{de: Uq(e(2)) invariant iintegral}

\textit{We call an integral (i.e. a linear functional) invariant with
respect to the right action of \( U_{q}(e(2)) \) if it satisfies
the following invariance condition:\begin{equation}
\label{eq: invariance condition for the q-integral}
\int ^{q}f(z,\overline{z})\lt X=\varepsilon (X)\int ^{q}f(z,\overline{z})
\end{equation}
for all} \( f \) and \( X\in U_{q}(e(2)) \).

\end{Def}

Here the action of \( X \) on \( f(z,\bar{z}) \) is to be understood
as in (\ref{eq: action of J,T,Tbar on z^kf(zzbar)}).\\
Since \( \varepsilon  \) is an algebra homomorphism and \( \tril  \)
an algebra action, it is sufficient to check the condition (\ref{eq: invariance condition for the q-integral})
for \( T,\overline{T} \) and \( J \) (recall that \( \{J^{k}T^{l}\overline{T}^{m}|k\in \mathbb {Z},l,m\in \mathbb {Z}_{+}\} \)
is a basis of \( U^{\mathrm{ext}}_{q}(e(2)) \), the extension of
\( U_{q}(e(2)) \) to formal power series). Let us first consider
functions of the type \be
z^{m}f(z\bar {z}),\qquad m\ne 0. \ee
In this case it is apparently true that\begin{equation}
\label{pr: q-integral for m not equal 0}
\int ^{q}z^{m}f(z\overline{z}):=0\, \, \, \textrm{for}\, \, \, m\neq 0\textrm{ }
\end{equation}
 is invariant if for \( z^{m}f(z\bar{z})\tril X \) holds that the
exponent of \( z \) still is not equal to 0 (where \( X\in \{J,T,\overline{T}\} \)).
The only cases where this is not guaranteed are (cf. (\ref{eq: action of J,T,Tbar on z^kf(zzbar)}))\[
zf(z\overline{z})\lt T=\frac{1}{1-q^{-2}}(f(q^{2}z\overline{z})-q^{-2}f(z\overline{z}))\]
and\[
z^{-1}f(z\overline{z})\lt \overline{T}=\frac{q^{4}}{1-q^{2}}\frac{f(z\overline{z})-f(q^{-2}z\overline{z})}{z\overline{z}}=\frac{q^{4}}{1-q^{2}}\left( \frac{f(z\overline{z})}{z\overline{z}}-q^{-2}\frac{f(q^{-2}z\overline{z})}{q^{-2}z\overline{z}}\right) \, \, \, .\]
 Since \( \varepsilon (T)=\varepsilon (\overline{T})=0 \), we have
to require that \( \int ^{q}zf(z\overline{z})\lt T \) and \( \int ^{q}z^{-1}f(z\overline{z})\lt \overline{T} \)
equal zero if the integral is to satisfy the invariance condition
(\ref{eq: invariance condition for the q-integral}). This implies
\begin{equation}
\label{pr: q-integral 1}
\int ^{q}f(q^{2}z\overline{z})-q^{-2}f(z\overline{z})=0
\end{equation}
 respectively\begin{equation}
\label{pr: q-integral2}
\int ^{q}\tilde{f}(z\overline{z})-q^{-2}\tilde{f}(q^{-2}z\overline{z})=0\, \, \, ,
\end{equation}
 where \( \tilde{f}(z\overline{z})-q^{-2}\tilde{f}(q^{-2}z\overline{z}):=\frac{f(z\overline{z})}{z\overline{z}}-q^{-2}\frac{f(q^{-2}z\overline{z})}{q^{-2}z\overline{z}} \).
If we define

\[
\int ^{q}f(z\overline{z}):=\sum ^{\infty }_{k=-\infty }q^{2k}f(q^{2k}r_{0}^{2})\, \, \, ,\]
where \( r_{0}^{2}q^{2k},k\in \mathbb {Z} \) are the eigenvalues
of \( z\bar{z} \) in a representation of \( \mathbb {C}_{q}^{2} \)
(see for example \cite{Chaichian:1999wy}), we see that this integral
satisfies (\ref{pr: q-integral 1}) and (\ref{pr: q-integral2}).
Together with (\ref{pr: q-integral for m not equal 0}), we can now
define an invariant integral. Let us summarize the result in the following
Lemma: 

\begin{Lemma}

For all functions \( f\in \mathcal{A}_{\mathrm{nc}} \) 

\begin{equation}
\label{de: Eq(2) invariant q-integral}
\int ^{q}z^{m}f(z\overline{z}):=\delta _{m,0}r^{2}_{0}(q^{2}-1)\sum ^{\infty }_{k=-\infty }q^{2k}f(q^{2k}r_{0}^{2})
\end{equation}
is an invariant integral under the action of \( U_{q}(e(2)) \) in
the sense of Definition \ref{de: Uq(e(2)) invariant iintegral}. 

\end{Lemma}

The factor \( r_{0}^{2}(q^{2}-1) \) was added to guarantee the right
semi-classical  limit of the integral.

\subsubsection{Cyclic property of the Invariant Integral}

Let us examine whether the integral (\ref{de: Eq(2) invariant q-integral})
is cyclic as it is in the undeformed case. 

First of all we provide a commutation relation which we will need
later on.

\begin{Remark}

For any formal power series \( f(z\overline{z}) \) we have:

\begin{equation}
\label{eq: comm-rel for f(zzbar) with z^m}
f(z\overline{z})z^{m}=z^{m}f(q^{-2m}z\overline{z})
\end{equation}
and \begin{equation}
\label{eq: comm-rel for f(zzbar) with (zbar)^m}
f(z\overline{z})\overline{z}^{m}=\overline{z}^{m}f(q^{2m}z\overline{z})
\end{equation}
for \emph{\( m\in \mathbb {Z} \)}.

\end{Remark}

\begin{proof}

Using (\ref{com z}) we immediately get \[
(z\bar{z})z=q^{-2}z(z\bar{z})\]
 implying \[
(z\bar{z})z^{m}=q^{-2m}z^{m}(z\bar{z})\qquad .\]
Extending this to formal power series, we get the first equation in
the Remark. The second follows similarly.

\end{proof}

Let us undertake the following calculation to examine the cyclic property
of the \( q- \)integral (\ref{de: Eq(2) invariant q-integral}) using
in the first step the commutation relation (\ref{eq: comm-rel for f(zzbar) with z^m}):

\begin{eqnarray*}
\lefteqn {\int ^{q}z^{m}f_{m}(z\bar{z})z^{n}g_{n}(z\bar{z})} & \\
 & = & \int ^{q}z^{m+n}f_{m}(q^{-2n}z\bar{z})g_{n}(z\bar{z})\\
 & = & \delta _{m+n,0}r_{0}^{2}(q^{2}-1)\sum _{k=-\infty }^{+\infty }q^{2k}f_{m}(q^{2k}q^{-2n}r_{0}^{2})g_{n}(q^{2k}r_{0}^{2})\\
 & = & \left\{ \begin{array}{c}
r_{0}^{2}(q^{2}-1)\sum _{k=-\infty }^{+\infty }q^{2k}f_{m}(q^{2k}q^{2m}r_{0}^{2})g_{-m}(q^{2k}r_{0}^{2})\, \, \, \, \, \textrm{if}\, \, \, n+m=0\\
\, \, \, \, \, \, \, \, \, 0\, \, \, \, \, \, \, \, \, \, \, \, \, \, \, \, \, \, \, \, \, \, \, \, \, \, \, \, \, \, \, \, \, \, \, \, \textrm{if}\, \, \, n+m\neq 0
\end{array}\right. \\
 & \stackrel{(k\rightarrow k+m)}{=} & \left\{ \begin{array}{c}
r_{0}^{2}(q^{2}-1)\sum _{k=-\infty }^{+\infty }q^{-2m}q^{2k}f_{m}(q^{2k}r_{0}^{2})g_{-m}(q^{2k}q^{-2m}r_{0}^{2})\, \, \, \, \textrm{if}\, \, \, n+m=0\\
0\, \, \, \, \, \, \, \, \, \, \, \, \, \, \, \, \, \, \, \, \, \, \, \, \, \, \, \, \, \, \, \, \, \, \, \, \, \textrm{if}\, n+m\neq 0
\end{array}\right. \\
 & = & \delta _{m+n,0}r_{0}^{2}(q^{2}-1)\sum _{k=-\infty }^{+\infty }q^{-2m}q^{2k}g_{-m}(q^{2k}q^{-2m}r_{0}^{2})f_{m}(q^{2k}r_{0}^{2})\\
 & = & q^{-2m}\int ^{q}z^{n}g_{n}(z\bar{z})z^{m}f_{m}(z\bar{z})\qquad .
\end{eqnarray*}
If we now define an operator\begin{equation}
\label{de: operator D (cyclic property of the q-integral)}
\mathcal{D}(z^{m}):=q^{-2m}z^{m}\, \, \, \, \, \, \, \, \, \, \mathcal{D}(\overline{z}^{m}):=q^{2m}\overline{z}^{m}
\end{equation}
 and extend it linearly on the entire algebra \( \mathcal{A}_{\mathrm{nc}} \),
we can write:

\begin{Lemma}\label{la: cyclic property of the q-integral}

\textit{For any two functions \( f,g\in \mathcal{A}_{\mathrm{nc}} \)
we have} \begin{equation}
\label{eq: cyclic property of the q-integral}
\int ^{q}fg=\int ^{q}g\mathcal{D}(f)\, \, \, .
\end{equation}

\end{Lemma}

A cyclic property of this kind for invariant integrals of functions
on \( q- \)deformed spaces was first derived by Harold Steinacker
in \cite{Steinacker:1995jh}, where a \( SO_{q}(N)- \)covariant space
is considered. 

We see that the integral we defined in (\ref{de: Eq(2) invariant q-integral})
is not cyclic. Obviously the demand for \( U_{q}(e(2))- \)invariance
spoils the trace property, whereas the demand for a cyclic integral
implies loosing \( U_{q}(e(2))- \)invariance (as we saw in Subsection
\ref{se: integral with trace property}). We will comment on this
{}``dilemma'' later. The considerations made here are important
if we want to write down an action that is invariant under gauge transformations
later on. If we define the gauge transformation of a field strength
\( f\in \mathcal{A}_{\mathrm{nc}} \) in the usual way, i.e. by \be
f \mapsto \Omega f \Omega^{-1} \qquad, \Omega \mbox{ unitary}, \ee
we will now in the deformed case not get an invariant action because
of (\ref{eq: cyclic property of the q-integral}). The only way out
without abandoning the demand for an invariant integral is to define
new gauge transformations, to {}``deform'' them in a way that compensates
the property (\ref{eq: cyclic property of the q-integral}). How this
can be done will be treated later.

\subsection{Covariant Differential Calculus on the \protect\( E_{q}(2)\protect \)
- Symmetric Plane}

Differential calculus on quantum groups was first studied by Woronowicz
\cite{Woronowicz:1989rt}. For more details the reader is also refered
to \cite{Zumino:1991} and \cite{Jurco:1991}. Calculi on noncommutative
spaces that are covariant with respect to the coaction of a quantum
group have been studied by Wess and Zumino in \cite{Wess:1991vh},
where the quantum hyper plane, which is covariant with respect to
the coaction of the quantum group \( GL_{q}(n) \), is discussed.
We will develop such a differential calculus that is covariant with
respect to the coaction of \( E_{q}(2) \).

\subsubsection{\protect\( q-\protect \)One-Forms and \protect\( q-\protect \)Derivatives}

We start with introducing variables \( dz \) and \( d\overline{z} \),
the \( q- \)differentials of \( z \) and \( \overline{z}. \) These
are noncommutative differentials which do not commute with the space
coordinates \( z,\overline{z} \), either. How can we find the corresponding
commutation relations? Certainly, the commutation relations have to
be covariant with respect to the coaction of \( E_{q}(2) \) on \( \mathcal{A}_{\mathrm{nc}} \)
given in (\ref{coac}). Let me explain how this is to be understood:
We get the coaction of \( E_{q}(2) \) on the \( q- \)differentials
\( dz,d\overline{z} \) by applying \( d \) on the coaction of \( E_{q}(2) \)
on \( z,\overline{z} \), where we want to assume that \( d \) applied
on elements of \( E_{q}(2) \) equals zero. In other words \[
\begin{array}{ccc}
\rho (z) & = & e^{i\theta }\te z+t\te 1\\
\rho (\overline{z}) & = & e^{-i\theta }\te \overline{z}+\bar{t}\te 1
\end{array}\]
 leads to \[
\begin{array}{ccc}
\rho (dz) & = & e^{i\theta }\te dz\\
\rho (d\overline{z}) & = & e^{-i\theta }\te d\overline{z}\, \, \, .
\end{array}\]
 The requirement of covariance implies the commutation relations between
coordinates and their differentials in the following sense: If we
want to extend the algebra \( \mathcal{A}_{\mathrm{nc}} \) by generating
elements \( dz,d\overline{z} \) with the above coaction and if we
want the newly generated algebra to be a left \( E_{q}(2)- \)comodule
algebra, then \( \rho  \) has to be a left coaction in the sense
of Definition \ref{de: coaction of a Hopfalgebra on a algebra} on
the extended algebra. One property of a coaction on an algebra is
that the multiplication in the algebra has to be a comodule homomorphism,
i.e. it has to satisfy \( \rho (ab)=\rho (a)\rho (b) \) for all algebra
elements \( a,b \).%
\footnote{The remaining properties can be easily checked, too. 
} In our setting, it therefore has to be true that \[
\rho (zdz)=\rho (z)\rho (dz)=(e^{i\theta }\te z+t\te 1)(e^{i\theta }\te dz)=e^{2i\theta }\te zdz+te^{i\theta }\te dz\, \, \, .\]
 On the other hand we find with the same argument that \[
\rho (dzz)=\rho (dz)\rho (z)=(e^{i\theta }\te dz)(e^{i\theta }\te z+t\te 1)=e^{2i\theta }\te dzz+e^{i\theta }t\te dz\]
has to be true, too. But in the quantum group \( E_{q}(2) \) we have
\( te^{i\theta }=q^{-2}e^{i\theta }t \) (see commutation relations
(\ref{Eq})). Thus, \( \rho  \) as defined above is only well-defined
if \[
zdz=q^{-2}dzz\, \, \, .\]
 Analogously follow the remaining commutation relations:\begin{equation}
\label{eq: comm-rel dz,dzbar with z,zbar}
\begin{array}{cccccccc}
zdz & = & q^{-2}dzz &  &  & \overline{z}dz & = & q^{-2}dz\overline{z}\\
zd\overline{z} & = & q^{2}d\overline{z}z &  &  & \overline{z}d\overline{z} & = & q^{2}d\overline{z}\, \overline{z}\, \, \, .
\end{array}
\end{equation}
 We see that for \( q\rp 1 \) coordinates and differentials commute.
Let us remark that these considerations do not lead to any information
about the commutation relations of \( dz \) and \( d\overline{z} \)
themselves, since \( e^{-i\theta }e^{i\theta }=e^{i\theta }e^{-i\theta } \).
How to find them anyway will be discussed below. 

Now we can define an \emph{exterior differential} \( d:\mathcal{A}_{\mathrm{nc}}\rp \Omega _{q}^{1} \)
on the entire algebra \( \mathcal{A}_{\mathrm{nc}} \), where by \( \Omega _{q}^{1} \)
we want to denote the \emph{space of \( q- \)deformed one-forms},
by demanding \[
d(\mathrm{const}.)=0\, \, \, ,\]
 as well as the \emph{Leibniz rule} \begin{equation}
\label{de: Leibniz rule for exterior differential}
d(fg)=(df)g+f(dg)
\end{equation}
 for all elements \( f \) and \( g \) in \( \mathcal{A}_{\mathrm{nc}} \).
To see that \( d \) is indeed well-defined we have to check whether
it respects the commutation relations of the algebra, i.e. we need
to examine whether \begin{equation}
\label{eq: differential d well-defined}
d(z\overline{z}-q^{2}\overline{z}z)\stackrel{!}{=}0\, \, \, .
\end{equation}
 This follows immediately from (\ref{eq: comm-rel dz,dzbar with z,zbar})
and (\ref{de: Leibniz rule for exterior differential}).

In the next step we can introduce \emph{\( q- \)deformed partial
derivatives} by defining in analogy to the noncommutative case\begin{equation}
\label{de: exterior differential gives derivatives}
d=:dz^{i}\partial _{i}\, \, \, .
\end{equation}
 In (\ref{eq: differential d well-defined}) we saw that \( d \)
is well-defined such that the partial derivatives \( \partial _{z} \)
and \( \partial _{\overline{z}} \) are well-defined, too. \\
\emph{Remark about notation:} We want to distinguish notationally
on the one hand \emph{applying} partial derivatives to a function
and on the other hand understanding \( \partial _{z},\partial _{\overline{z}} \)
as elements of the algebra and \emph{multiplying} them by functions
in the algebra itself. When applying \( \partial _{z},\partial _{\overline{z}} \)
to a function, we always want to put the function the derivatives
are applied on in brackets, i.e. \[
\partial _{z}(f)\, \, \, \, \, \, \textrm{and}\, \, \, \, \, \, \partial _{\overline{z}}(f)\, \, \, ,\]
 whereas we won't use brackets when we interpret \( \partial _{z},\partial _{\overline{z}} \)
as part of the algebra itself. 

According to definition (\ref{de: exterior differential gives derivatives}),
we find for example for the \( q- \)deformed partial derivatives
applied to the coordinates\begin{equation}
\label{eq: partial drerivatives of z and zbar}
\begin{array}{cccccccc}
\partial _{z}(z) & = & 1 &  &  & \partial _{\overline{z}}(z) & = & 0\\
\partial _{z}(\overline{z}) & = & 0 &  &  & \partial _{\overline{z}}(\overline{z}) & = & 1
\end{array}
\end{equation}
which is just what we expected. Nevertheless, the derivatives \( \partial _{z},\partial _{\overline{z}} \)
do not satisfy the usual Leibniz rule but a modified one that we want
to call the \emph{\( q- \)Leibniz rule}. It can be derived from the
Leibniz rule of the exterior differential together with the commutation
relations of differentials and coordinates (\ref{eq: comm-rel dz,dzbar with z,zbar}):\begin{eqnarray}
d(fg) & = & (df)g+f(dg)\nonumber \\
 & = & dz^{i}\partial _{i}(f)g+fdz^{i}\partial _{i}(g)\nonumber \\
 & = & dz^{i}\partial _{i}(f)g+dzf(q^{-2}z,q^{-2}\overline{z})\partial _{z}(g)+d\overline{z}f(q^{2}z,q^{2}\overline{z})\partial _{\overline{z}}(g)\label{eq: calculating the q-Leibniz rule} 
\end{eqnarray}
where we have used the following commutation relations\[
\begin{array}{ccc}
f(z,\overline{z})dz & = & dzf(q^{-2}z,q^{-2}\overline{z})\\
f(z,\overline{z})d\overline{z} & = & d\overline{z}f(q^{2}z,q^{2}\overline{z})\, \, \, ,
\end{array}\]
 which we want to prove:

\begin{proof}

From (\ref{de: Alg_nc_ext}) we get that \( f(z,\overline{z})=\sum _{m\in \mathbb {Z}}z^{m}f_{m}(z\overline{z}) \),
where the \( f_{m} \) are formal power series for all \( m \). Since
\( z\overline{z}dz=q^{-4}dz(z\overline{z}) \), we deduce that \( f_{m}(z\overline{z})dz=dzf_{m}(q^{-4}z\overline{z}). \)
Moreover, \( z^{m}dz=dz(q^{-2}z)^{m} \), such that together follows
\( f(z,\overline{z})dz=dzf(q^{-2}z,q^{-2}\overline{z}) \), the first
claim. A similar calculation proves the second claim.

\end{proof} 

On the other hand we have by definition \[
d(fg)=dz\partial _{z}(fg)+d\overline{z}\partial _{\overline{z}}(fg)\, \, \, ,\]
 such that with the result in (\ref{eq: calculating the q-Leibniz rule})
we conclude that\begin{equation}
\label{eq: q-Leibniz rule}
\begin{array}{ccc}
\partial _{z}(fg) & = & \partial _{z}(f)g+f(q^{-2}z,q^{-2}\overline{z})\partial _{z}(g)\\
\partial _{\overline{z}}(fg) & = & \partial _{\overline{z}}(f)g+f(q^{2}z,q^{2}\overline{z})\partial _{\overline{z}}(g)\, \, \, .
\end{array}
\end{equation}
 Obviously, we obtain for \( q\rp 1 \) the usual commutative Leibniz
rule just as expected. With the \( q- \)Leibniz rule it is now possible
to derive the commutation relations of partial derivatives and space
coordinates. Applying the \( q- \)Leibniz rule (\ref{eq: q-Leibniz rule})
on the function \( z=zf \) resp. \( \overline{z}=\overline{z}f \)
and using the results in (\ref{eq: partial drerivatives of z and zbar})
we find\[
\begin{array}{cccccccc}
\partial _{z}(zf) & = & 1f+q^{-2}z\partial _{z}(f) &  &  & \partial _{z}(\overline{z}f) & = & q^{-2}\overline{z}\partial _{z}(f)\\
\partial _{\overline{z}}(zf) & = & q^{2}z\partial _{\overline{z}}(f) &  &  & \partial _{\overline{z}}(\overline{z}f) & = & 1f+q^{2}\overline{z}\partial _{\overline{z}}(f)
\end{array}\]
which yields the following commutation relations\begin{equation}
\label{eq: comm rel derivatives with coordinates}
\begin{array}{cccccccc}
\partial _{z}z & = & 1+q^{-2}z\partial _{z} &  &  & \partial _{z}\overline{z} & = & q^{-2}\overline{z}\partial _{z}\\
\partial _{\overline{z}}z & = & q^{2}z\partial _{\overline{z}} &  &  & \partial _{\overline{z}}\overline{z} & = & 1+q^{2}\overline{z}\partial _{\overline{z}}\, \, \, .
\end{array}
\end{equation}
 Furthermore, we find, applying \( \partial _{z}\partial _{\overline{z}} \)
on the function \( z\overline{z} \) using the \( q- \)Leibniz rule,
that\[
\partial _{z}\partial _{\overline{z}}(z\overline{z})=\partial _{z}(q^{2}z)=q^{2}\]
 whereas\[
\partial _{\overline{z}}\partial _{z}(z\overline{z})=\partial _{\overline{z}}(\overline{z})=1\, \, \, ,\]
 such that we derive the following commutation relation for the \( q- \)derivatives:
\begin{equation}
\label{eq: comm rel for q-derivatives}
\partial _{z}\partial _{\overline{z}}=q^{2}\partial _{\overline{z}}\partial _{z}\, \, \, .
\end{equation}
 Again for \( q\rp 1 \) these commutation relations become those
of the commutative theory. We still do not know the commutation relations
for the \( q- \)differentials. To derive them, we need to extend
the exterior differential to one-forms: we want to demand for the
exterior differential \( d \) in addition to the Leibniz rule (\ref{de: Leibniz rule for exterior differential})
that in analogy to the commutative case \[
d^{2}\equiv 0\]
 as well as\begin{equation}
\label{eq: Leibniz rule for n-forms}
d(\alpha \beta )=(d\alpha )\beta +(-1)^{\mathrm{deg}\alpha }\alpha (d\beta )
\end{equation}
 for forms \( \alpha ,\beta  \), where \( \mathrm{deg}\alpha  \)
denotes the degree of \( \alpha  \). Using this and applying \( d \)
on the commutation relation \( dz\overline{z}=q^{2}\overline{z}dz \)
(cf. \ref{eq: comm-rel dz,dzbar with z,zbar}), we find\begin{equation}
\label{eq: comm-rel for differentials}
dzd\overline{z}=-q^{2}d\overline{z}dz\, \, \, ,
\end{equation}
becoming for \( q\rp 1 \) the usual commutation rule of differentials.
Applying \( d \) to \( zdz=q^{-2}dzz \) respectively \( \overline{z}d\overline{z}=q^{2}d\overline{z}\, \overline{z} \)
(\ref{eq: comm-rel dz,dzbar with z,zbar}), we obtain\[
(dz)^{2}=(d\overline{z})^{2}=0\, \, \, .\]
 To see that this extension of \( d \) is indeed well-defined, we
remark that applying \( d \) to the remaining equation of (\ref{eq: comm-rel dz,dzbar with z,zbar}),
namely \( \overline{z}dz=q^{-2}dz\overline{z} \), leads to (\ref{eq: comm-rel for differentials}),
too. 

For completeness we want to derive the commutation relations for \( q- \)differentials
and \( q- \)derivatives. For this purpose we assume that \[
\partial _{z^{i}}dz^{j}=c_{i}^{j}dz^{j}\partial _{z^{i}}\, \, \, \, \, (\textrm{no summation})\]
for some constants \( c_{j}^{i} \). Multiplying for example \( \partial _{z}dz-c_{1}^{1}dz\partial _{z}=0 \)
by \( z \) from the right and commuting \( z \) through to the left
using the above commutation relations (\ref{eq: comm-rel dz,dzbar with z,zbar})
and (\ref{eq: comm rel derivatives with coordinates}) leads to the
following:\begin{eqnarray*}
0 & =(\partial _{z}dz-c_{1}^{1}dz\partial _{z})z=q^{2}\partial _{z}zdz=q^{2}(1+q^{-2}z\partial _{z})dz-c^{1}_{1}dz(1+q^{-2}z\partial z) & \\
 & =(q^{2}-c^{1}_{1})dz+(1-c^{1}_{1}q^{-2})z\partial _{z}dz & 
\end{eqnarray*}
 such that we conclude \[
c_{1}^{1}=q^{2}\, \, \, .\]
 Similar calculations provide the remaining constants \( c^{i}_{j} \)
and we get as result:\begin{equation}
\label{eq: comm rel for derivatives and differentials}
\begin{array}{cccccccc}
\partial _{z}dz & = & q^{2}dz\partial _{z} &  &  & \partial _{z}d\overline{z} & = & q^{-2}d\overline{z}\partial _{z}\\
\partial _{\overline{z}}dz & = & q^{2}dz\partial _{\overline{z}} &  &  & \partial _{\overline{z}}d\overline{z} & = & q^{-2}d\overline{z}\partial _{\overline{z}}\, \, \, .
\end{array}
\end{equation}
As we see, the \( q- \)differentials and \( q- \)derivatives become
the classical differentials resp. derivatives in \emph{}the semi-classical
limit \( q\rp 1 \).\\
\emph{Remark:} Above we derived all possible types of commutation
relations between coordinates, differentials and derivatives. Nevertheless,
we still have to check if all these relations are consistent with
each other. If we extend the algebra by \( dz,d\overline{z} \) with
the relations (\ref{eq: comm-rel dz,dzbar with z,zbar}) and (\ref{eq: comm-rel for differentials}),
these relations have to be consistent with the commutation relation
of the space coordinates \( z\overline{z}=q^{2}\overline{z}z \),
i.e. multiplying them by coordinates from the right and commuting
the coordinates to the left must not lead to any inconsistency. For
example we have:\begin{eqnarray*}
(zdz-q^{-2}dzz)z & = & z(q^{2}dzz-zdz)\\
(zdz-q^{-2}dzz)\overline{z} & = & \overline{z}(q^{4}zdz-q^{2}dzz)
\end{eqnarray*}
 obviously not leading to any inconsistency. Multiplying the remaining
relations of (\ref{eq: comm-rel dz,dzbar with z,zbar}) or (\ref{eq: comm-rel for differentials})
by coordinates from the right and commuting them through to the left,
does not lead to inconsistencies, either. The reason is that the commutation
relations are of the following type: Commuting two arbitrary elements
does again reproduce exactly those two elements multiplied by some
power of \( q \). Then, multiplying any relation from the right by
an element of the algebra and commuting it through to the left, must
lead to the same relation we started with multiplied by a power of
\( q \) and therefore never produces any inconsistency. In a second
step we introduced the partial derivatives \( \partial _{z},\partial _{\overline{z}} \)
with the relations (\ref{eq: comm rel derivatives with coordinates}),(\ref{eq: comm rel for q-derivatives})
and (\ref{eq: comm rel for derivatives and differentials}). A lot
of the commutation relations are again of the type that commuting
an arbitrary derivative with a differential or a coordinate reproduces
exactly the same elements multiplied by a power of \( q \). By the
same consideration as above, at least all cases, in which the commutation
relations \( \partial _{z}z=1+q^{-2}z\partial _{z} \) or \( \partial _{\overline{z}}\overline{z}=1+q^{2}\overline{z}\partial _{\overline{z}} \)
do not enter the calculation, don't lead to inconsistencies. If these
two commutation relations enter in the calculation, an additional
term \( +1 \) appears. Those cases remain to check. We easily verify\begin{eqnarray*}
(\partial _{z}z-1-q^{-2}z\partial _{z})z & = & zq^{-2}(\partial _{z}z-1-q^{-2}z\partial _{z})\\
(\partial _{z}z-1-q^{-2}z\partial _{z})\overline{z} & = & \overline{z}(\partial _{z}z-1-q^{-2}z\partial _{z})\\
(\partial _{z}z-1-q^{-2}z\partial _{z})dz & = & dz(\partial _{z}z-1-q^{-2}z\partial _{z})\\
(\partial _{z}z-1-q^{-2}z\partial _{z})d\overline{z} & = & d\overline{z}(\partial _{z}z-1-q^{-2}z\partial _{z})\, \, \, .
\end{eqnarray*}
 A short calculation shows that we don't get inconsistencies for the
other equation \( \partial _{\overline{z}}\overline{z}=1+q^{2}\overline{z}\partial _{\overline{z}} \),
either. The last equation, \( \partial _{z}\partial _{\overline{z}}=q^{2}\partial _{\overline{z}}\partial _{z} \),
still has to be multiplied by coordinates from the right commuting
them through to the left since again \( +1 \) terms will enter:\begin{eqnarray*}
(\partial _{z}\partial _{\overline{z}}-q^{2}\partial _{\overline{z}}\partial _{z})z & =q^{2}\partial _{\overline{z}}+\partial _{z}\partial _{\overline{z}}-q^{2}\partial _{\overline{z}}-q^{2}z\partial _{\overline{z}}\partial _{z} & =z(\partial _{z}\partial _{\overline{z}}-q^{2}\partial _{\overline{z}}\partial _{z})\\
(\partial _{z}\partial _{\overline{z}}-q^{2}\partial _{\overline{z}}\partial _{z})\overline{z} & =\partial _{z}+\overline{z}\partial _{\overline{z}}\partial _{z}-\partial _{z}-q^{2}\overline{z}\partial _{\overline{z}}\partial _{z} & =\overline{z}(\partial _{z}\partial _{\overline{z}}-q^{2}\partial _{\overline{z}}\partial _{z})\, \, \, .
\end{eqnarray*}
All together we have seen that the calculus established above withstands
all consistency checks. 

The above considerations give in all detail an \( E_{q}(2)- \)covariant
differential calculus on \( \mathcal{A}_{\mathrm{nc}} \). We saw
that differentials and functions do not commute. This makes calculations
complicated and therefore we will try in a next step to find a basis
\( \theta ^{i} \) of the space of one-forms, called \emph{frame,}
that commutes with all functions (Frames exist on other noncommutative
spaces as well (cf. for example \cite{Steinacker:2001fj,Madore:Buch})
and \( \theta ^{i} \) is the usually used notation that is not to
be mixed up with the Poisson structure \( \theta ^{ij} \) !).

\subsubsection{A Nicer Basis of One-Forms: A Frame}

We are looking for a new basis \( \theta =:\theta ^{z},\overline{\theta }=:\theta ^{\overline{z}} \)
of \( \Omega _{q}^{1} \) commuting with all functions. Let us consider
\begin{equation}
\label{de: Frame theta^i}
\begin{array}{ccc}
\theta  & := & z^{-1}\overline{z}dz\\
\overline{\theta } & := & d\overline{z}z\overline{z}^{-1}\, \, \, .
\end{array}
\end{equation}
 Then:

\bla\label{la: commutation relations of theta (small)}

We have\begin{equation}
\label{eq: theta commutes with all functions f}
[\theta ,f]=[\overline{\theta },f]=0
\end{equation}
 for all functions \( f\in \mathcal{A}_{\mathrm{nc}} \) and\[
\theta \overline{\theta }=-q^{2}\overline{\theta }\theta \, \, \, .\]

\ela

\begin{proof}

Using the commutation relations (\ref{com z}),(\ref{eq: comm-rel dz,dzbar with z,zbar}),
we get\[
\theta z=z^{-1}\overline{z}dzz=q^{2}z^{-1}\overline{z}zdz=z\theta \]
 and \[
\theta \overline{z}=z^{-1}\overline{z}dz\overline{z}=q^{2}z^{-1}\overline{z}\bar{z}dz=\overline{z}\theta \, \, \, .\]
 \( \theta \overline{\theta }=-q^{2}\overline{\theta }\theta  \)
follows making use of the same commutation relations. 

\end{proof}

It is even possible to find a one-form \( \Theta  \) that generates
the exterior differential.

\subsubsection{A Generator for the Exterior Differential}

In the following lemma we introduce a one-form \( \Theta  \) that
generates the exterior differential \( d. \) Later on, \( \Theta  \)
will play an important role when considering gauge fields and gauge
transformations.

\bla\label{la: generator of exterior differentil: Ttheta}

Consider \[
\Theta :=\theta ^{i}\lambda _{i}:=\theta \frac{1}{1-q^{-2}}\overline{z}^{-1}-\overline{\theta }\frac{1}{1-q^{-2}}z^{-1}\, \, \, .\]
 Then\[
df=[\Theta ,f]=[\lambda _{i},f]\theta ^{i}\]
 for all functions \( f \) and\[
d\Theta =\Theta ^{2}=0\, \, \, .\]
 \ela

\begin{proof}

We remember that the \( \theta ^{i} \) commute with all functions
and with that we get\[
[\Theta ,f]=\theta [\frac{1}{1-q^{-2}}\overline{z}^{-1},f]-\overline{\theta }[\frac{1}{1-q^{-2}}z^{-1},f]\, \, \, .\]
 Plugging in the explicit expressions (\ref{de: Frame theta^i}) for
\( \theta ^{i} \) we find\begin{eqnarray*}
[\Theta ,f] & = & z^{-1}\overline{z}dz[\frac{1}{1-q^{-2}}\overline{z}^{-1},f]-d\overline{z}z\overline{z}^{-1}[\frac{1}{1-q^{-2}}z^{-1},f]\\
 & = & dzz^{-1}\overline{z}[\frac{1}{1-q^{-2}}\overline{z}^{-1},f]-d\overline{z}z\overline{z}^{-1}[\frac{1}{1-q^{-2}}z^{-1},f]\, \, \, ,
\end{eqnarray*}
 where in the last step we used the commutation relations (\ref{eq: comm-rel dz,dzbar with z,zbar}).
Taking \( f=z \) and \( f=\overline{z} \) we get with the commutation
relations of the space coordinates \begin{equation}
\label{pr: deriv on generator z}
z^{-1}\overline{z}[\frac{1}{1-q^{-2}}\overline{z}^{-1},z]=\frac{1}{1-q^{-2}}-\frac{q^{-2}}{1-q^{-2}}=1
\end{equation}
 and\begin{equation}
\label{pr: deriv on generator zbar}
z\overline{z}^{-1}[\frac{1}{1-q^{-2}}z^{-1},z]=0\, \, \, .
\end{equation}
Thus, \( [\Theta ,z]=dz \) and analogously follows \( [\Theta ,\overline{z}]=d\overline{z} \).
Hence, on the generators of the algebra of functions the claim is
true. Since \( [\Theta ,f] \) is a derivation, we can now conclude
that \[
df=[\Theta ,f]\]
 for all functions \( f \). 

The second claim, \( d\Theta =\Theta ^{2}=0 \), is shown by the following
calculations: First\begin{eqnarray*}
((1-q^{-2})\Theta )^{2} & = & (\theta \overline{z}^{-1}-\overline{\theta }z^{-1})^{2}=(q^{-2}z^{-1}dz-\overline{z}^{-1}d\overline{z})^{2}\\
 & = & -q^{-2}z^{-1}dz\overline{z}^{-1}d\overline{z}-\overline{z}^{-1}d\overline{z}q^{-2}z^{-1}dz\\
 & = & -q^{-4}z^{-1}\overline{z}^{-1}dzd\overline{z}-\overline{z}^{-1}z^{-1}d\overline{z}dz=0\, \, \, ,
\end{eqnarray*}
 where we have used the commutation relations (\ref{coac}), (\ref{eq: comm-rel dz,dzbar with z,zbar})
and (\ref{eq: comm-rel for differentials}), and secondly\[
(1-q^{-2})d\Theta =d(q^{-2}z^{-1}dz-\overline{z}^{-1}d\overline{z})=-q^{-4}z^{-2}dzdz+q^{2}\overline{z}^{-2}d\overline{z}d\overline{z}=0\]
 where we used \( d(z^{-1})=-q^{-2}z^{-2}dz \) and \( d(\overline{z}^{-1})=-q^{2}\overline{z}^{-2}d\overline{z} \)
which follows with the \( q- \)Leibniz rule applied on \( 0=d1=d(zz^{-1})=d(\bar{z}\bar{z}^{-1}). \)
\end{proof} \textbf{}\emph{Remark}\textbf{:} We notice that \[
\Theta =\theta ^{i}\lambda _{i}=q^{-2}z^{-1}dz-\overline{z}^{-1}d\overline{z}\]
 is not \emph{}invariant under translations. It is only invariant
under rotations.

\subsubsection{\protect\( q-\protect \)Two-Forms}

We can also write the extension of the exterior differential \( d \)
to one-forms in a nice form:

\begin{Lemma}\label{la: Theta generates two-forms}

Let \( \Theta  \) be defined as in Lemma \ref{la: generator of exterior differentil: Ttheta}.
Then \[
d\alpha =\{\Theta ,\alpha \}\]

for any one-form \( \alpha  \). Here \( \{\cdot ,\cdot \} \) denotes
the anti-commutator.

\end{Lemma}

\begin{proof}

We have \( \{\Theta ,\alpha f\}=\{\Theta ,\alpha \}f-\alpha [\Theta ,f] \)
and \( \{\Theta ,f\alpha \}=[\Theta ,f]\alpha +f\{\Theta ,\alpha \} \)
for arbitrary functions \( f \) and arbitrary one-forms \( \alpha  \).
Thus, \( \{\Theta ,\cdot \} \) satisfies the Leibniz-rule (\ref{eq: Leibniz rule for n-forms})
which implies that \( d\alpha =\{\Theta ,\alpha \} \).

\end{proof}

This finishes the discussion of the differential calculus and we will
now attend to gauge fields and the action.

\subsection{Gauge Fields and Action}

We consider only the easiest case: noncommutative abelian gauge fields.
They are one-forms \( B\in \Omega ^{1}_{q} \). We can express \( B \)
in the basis \( \theta ^{i} \) (found in the last subsection) that
commutes with arbitrary functions:\[
B=B_{i}\theta ^{i}\, \, \, .\]
Furthermore, we write \cite{Grosse:2000Fuzzy1} \begin{equation}
\label{de: B equal theta plus A}
B=\Theta +A\, \, \, ,
\end{equation}
 where, as we will see, \( A \) is the analog of the commutative
gauge field, and define \begin{equation}
\label{de: Field strength F equals B to the two}
F:=B^{2}\, \, \, ,
\end{equation}
 the noncommutative field strength. With (\ref{de: B equal theta plus A})
and since \( \Theta ^{2}=0 \) (Lemma \ref{la: generator of exterior differentil: Ttheta})
and \( \{\Theta ,A\}=dA \) (Lemma \ref{la: Theta generates two-forms})
we get\begin{equation}
\label{F equals  A to the two plus dA}
F=A^{2}+dA\, \, \, .
\end{equation}
In terms of components with respect to the frame \( \theta ^{i} \)
we have\begin{eqnarray*}
F & = & (A_{i}\theta ^{i})^{2}+d(A_{i}\theta ^{i})\\
 & = & A_{i}A_{j}\theta ^{i}\theta ^{j}+\{\lambda _{i}\theta ^{i},A_{j}\theta ^{j}\}\\
 & = & A_{1}A_{2}\theta \overline{\theta }+A_{2}A_{1}\overline{\theta }\theta +(\lambda _{1}A_{2}+A_{1}\lambda _{2})\theta \overline{\theta }+(A_{2}\lambda _{1}+\lambda _{2}A_{1})\overline{\theta }\theta \\
 & = & (A_{1}A_{2}-q^{-2}A_{2}A_{1})\theta \overline{\theta }+((\lambda _{1}A_{2}-q^{-2}A_{2}\lambda _{1})-(q^{-2}\lambda _{2}A_{1}-A_{1}\lambda _{2}))\theta \overline{\theta }
\end{eqnarray*}
where we used in the last step that \( \theta \overline{\theta }=-q^{2}\overline{\theta }\theta  \)
(Lemma \ref{la: commutation relations of theta (small)}). If we additionally
denote the components of \( A \) in the basis \( dz,d\overline{z} \)
by \( A_{z} \) and \( A_{\overline{z}} \), i.e.\[
A=A_{1}\theta +A_{2}\overline{\theta }=A_{z}dz+A_{\overline{z}}d\overline{z}\]
and therefore \begin{eqnarray*}
A_{1}\theta  & = & A_{z}dz\\
A_{2}\overline{\theta } & = & A_{\overline{z}}d\overline{z}\, \, \, ,
\end{eqnarray*}
 we can write\[
F=(A_{1}A_{2}-q^{-2}A_{2}A_{1})\theta \overline{\theta }+(\theta (\lambda _{1}A_{\overline{z}}d\overline{z}-q^{-2}A_{\overline{z}}d\overline{z}\lambda _{1})-(q^{-2}\lambda _{2}A_{z}dz-A_{z}dz\lambda _{2}))\overline{\theta }\, \, \, .\]
 The explicit expressions of \( \lambda _{i} \) given in Lemma \ref{la: generator of exterior differentil: Ttheta}
and the commutation relations (\ref{eq: comm-rel dz,dzbar with z,zbar})
yield \( \lambda _{1}d\overline{z}=q^{-2}d\overline{z}\lambda _{1} \)
and \( dz\lambda _{2}=q^{-2}\lambda _{2}dz \). Thus, we obtain\[
F=(A_{1}A_{2}-q^{-2}A_{2}A_{1})\theta \overline{\theta }+((\lambda _{1}A_{\overline{z}}-A_{\overline{z}}\lambda _{1})\theta d\overline{z}-q^{-2}(\lambda _{2}A_{z}-A_{z}\lambda _{2})dz\overline{\theta }\]
 By Lemma \ref{la: generator of exterior differentil: Ttheta} we
get that\[
[\lambda _{1},f]\theta =dz\partial _{z}(f)\, \, \, \textrm{and}\]
\[
[\lambda _{2},f]\overline{\theta }=d\overline{z}\partial _{\overline{z}}(f)\, \, \, .\]
Together with \( dz\overline{\theta }=-q^{2}\overline{\theta }dz \)
we finally have \[
F=(A_{1}A_{2}-q^{-2}A_{2}A_{1})\theta \overline{\theta }+dz\partial _{z}(A_{\overline{z}})d\overline{z}+d\overline{z}\partial _{\overline{z}}(A_{z})dz\, \, \, .\]
 Thereby the above expression for \( F \) approaches in the semi-classical
limit \( q\rp 1 \) the following expression:\begin{equation}
\label{eq: limit q to 1 for F equals B to the two}
F\stackrel{q\rp 1}{\longrightarrow }(\partial _{z}A_{\overline{z}}-\partial _{\overline{z}}A_{z})dzd\overline{z}\, \, \, .
\end{equation}
 Hence,  we see that in fact the field strength defined as above becomes
the commutative field strength (cf. Appendix \ref{appendix: change of basis z,zbar to x,y})
in the semi-classical limit and \( A \) admits the interpretation
of a noncommutative gauge field. Therefore (\ref{de: Field strength F equals B to the two})
seems to be in this regard a reasonable definition for a noncommutative
field strength. Of course gauge covariance is another important requirement
for \( F \). This issue will be treated in the after next subsection.

At this point we want to continue introducing the action. For this
purpose we consider the \emph{Hodge dual} \( *_{H}F \) of \( F \):
In our two-dimensional case we can express any two-form \( F \) in
the basis \( \theta \overline{\theta }=z^{-1}\overline{z}dzd\overline{z}z\overline{z}^{-1}=q^{-2}dzd\overline{z} \):
\[
F=f\theta \overline{\theta }=q^{-2}fdzd\overline{z}\]
 for some coefficient function \( f. \) Then we define \( *_{H} \)
on two-forms as\begin{equation}
\label{de: Hodge dual of functions}
*_{H}F:=\frac{1}{2}f\, \, \, .
\end{equation}
 With the integral found in Subsection \ref{se: q-integra} we can
now write down an action:\[
S:=\int ^{q}F(*_{H}F)=\int ^{q}\frac{1}{2}f^{2}\theta \overline{\theta }\, \, \, .\]
It admits the right semi-classical  limit for \( q\rp 1 \): \[
S\stackrel{q\rp 1}{\longrightarrow }\int \frac{1}{2}(\partial _{z}A_{\overline{z}}-\partial _{\overline{z}}A_{z})^{2}dzd\overline{z}\, \, \, .\]
 We can even assign to \( S \) a gauge-invariance property for special
{}``deformed'' gauge transformations, as we will see later, but
first we need the following subsection as preparation.

\subsection{The Algebra Homomorphism \protect\( \alpha :U_{q}(e(2))\rp \mathcal{A}_{\mathrm{nc}}\protect \)\label{se: alg hom Uq(e(2)) to Alg_nonc}}

We construct an algebra homomorphism \( \alpha :U_{q}(e(2))\rp \mathcal{A}_{\mathrm{nc}} \)
which we will need in the following subsection when defining gauge
transformations in analogy to \cite{Grosse:2000Fuzzy1}, where such
a homomorphism is introduced, too (even though in a different way).
Nevertheless, we do not look for an arbitrary algebra homomorphism.
It shall have the additional property that the right-action \( x\lt u \)
of an arbitrary element \( u\in U_{q}(e(2)) \) on an element \( x\in \mathcal{A}_{\mathrm{nc}} \)
is given by\[
\alpha (S(u_{(1)}))x\alpha (u_{(2)})=x\lt u\, \, \, .\]
 This property we will need to obtain a gauge invariant action later
on. For this purpose it is convenient to consider the \emph{cross
product algebra} \( U_{q}(e(2))\ltimes \mathcal{A}_{\mathrm{nc}} \)
(cf. for example \cite{Klimyk:1997eb}) which is as vector space isomorphic
to the tensor space \( U_{q}(e(2))\te \mathcal{A}_{\mathrm{nc}} \)
and where multiplication is defined as follows:\begin{equation}
\label{de: multiplication in cross algebra}
(a\te x)\cdot (b\te y):=ab_{(1)}\te (x\lt b_{(2)})y\, \, \, .
\end{equation}
 It is common to omit the tensor signs and to identify \( 1\te x\equiv x \)
and \( a\te 1\equiv a \), leading with the multiplication as defined
above to the so called commutation relation\[
x\cdot a=a_{(1)}(x\lt a_{(2)})\, \, \, .\]
In our case this reads, if we use the explicit form for coproduct
and right-action of \( U_{q}(e(2)) \) given in (\ref{eq: action of Uq on generators z, zbar}):

\begin{enumerate}
\item \( z\cdot T=T(z\lt q^{2iJ})+z\lt T=q^{-2}Tz+1 \)
\item \( z\cdot \overline{T}=\overline{T}(z\lt q^{2iJ})+z\lt \overline{T}=q^{-2}\overline{T}z \)
\item \( z\cdot J=Jz+z\lt J=Jz+iz \)
\item \( \overline{z}\cdot T=T(\overline{z}\lt q^{2iJ})+\overline{z}\lt T=q^{2}T\overline{z} \)
\item \( \overline{z}\cdot \overline{T}=\overline{T}(\overline{z}\lt q^{2iJ})+\overline{z}\lt \overline{T}=q^{2}\overline{T}\overline{z}-q^{2} \) 
\item \( \overline{z}\cdot J=J\overline{z}+\overline{z}\lt J=J\overline{z}-i\overline{z} \)
\end{enumerate}
Let us try to find an algebra homomorphism \( \alpha ':U_{q}(e(2))\ltimes \mathcal{A}_{\mathrm{nc}}\rp \mathcal{A}_{\mathrm{nc}} \)
that on \( \mathcal{A}_{\mathrm{nc}} \) is the identity (in a second
step such a \( \alpha ' \) can be restricted to \( U_{q}(e(2)) \)
leading to an algebra homomorphism \( \alpha  \) with all the demanded
properties). Knowing the above commutation relations we can explicitly
construct \( \alpha ' \). Being an algebra homomorphism, for example
implies because of 1. above: \[
z\alpha '(T)=\alpha '(z)\alpha '(T)\stackrel{!}{=}\alpha '(z\cdot T)=q^{-2}\alpha '(T)z+1\]
 which is equivalent to \begin{equation}
\label{eq: on the way to Hom: Uq to Alg-noncom}
z\alpha '(T)-q^{-2}\alpha '(T)z=1\, \, \, .
\end{equation}
 Defining \[
\alpha '(T):=\frac{z^{-1}}{1-q^{-2}}\]
 we see that (\ref{eq: on the way to Hom: Uq to Alg-noncom}) is indeed
satisfied. Similarly we find how to define \( a'(\overline{T}) \):\[
a'(\overline{T}):=\frac{\overline{z}^{-1}}{1-q^{-2}}\, \, \, .\]
 Finally, 3. above yields, if \( \alpha ' \) is to be an algebra
homomorphism with \( \alpha '|_{\mathcal{A}_{\mathrm{nc}}}=\mathrm{id}|_{\mathcal{A}_{\mathrm{nc}}} \),\[
z(\alpha '(J)-i)=\alpha '(J)z\]
 implying \[
\alpha '(q^{-2iJ})z=q^{-2}z\alpha '(q^{-2iJ})\]
 which is satisfied for \[
\alpha '(q^{-2iJ}):=z\overline{z}\, \, \, .\]
 To show that \( \alpha ' \) defined on the generators of \( U_{q}(e(2)) \)
as suggested above really gives an algebra homomorphism \( U_{q}(e(2))\ltimes \mathcal{A}_{\mathrm{nc}}\rp \mathcal{A}_{\mathrm{nc}} \),
we have to prove that the defining commutation relations of the generators
in \( U_{q}(e(2)) \), 

\[
T\overline{T}=q^{2}\overline{T}T\]
 \[
[J,T]=iT\, \, \, \Leftrightarrow \, \, \, q^{-2iJ}T=q^{2}Tq^{-2iJ}\]
 \[
[J,\overline{T}]=-i\overline{T}\, \, \, \Leftrightarrow \, \, \, q^{-2iJ}\overline{T}=q^{-2}\overline{T}q^{-2iJ}\, \, \, ,\]
are respected, too. This can easily be checked using the commutation
relation of the coordinates \( z\overline{z}=q^{2}\overline{z}z. \)
For example \[
\alpha (T)\alpha (\overline{T}):=\frac{z^{-1}\overline{z}^{-1}}{1-q^{-2}}=q^{2}\frac{\overline{z}^{-1}z^{-1}}{1-q^{-2}}=q^{2}\alpha (\overline{T})\alpha (T)\]
 and the rest follows similarly. Anyhow, as explained above, our aim
is to find a homomorphism \( \alpha :U_{q}(e(2))\rp \mathcal{A}_{\mathrm{nc}} \).
By restricting \( \alpha ' \) to \( U_{q}(e(2)) \) we obtain such
an \( \alpha  \):

\bla\label{la: the homomorphism alpha}

The assignments\begin{eqnarray}
T & \ra  & \frac{z^{-1}}{1-q^{-2}}\nonumber \\
\overline{T} & \ra  & \frac{\overline{z}^{-1}}{1-q^{-2}}\label{de: HOM alpha} \\
q^{-2iJ} & \ra  & z\overline{z}\nonumber 
\end{eqnarray}
define an algebra homomorphism \[
\alpha :U_{q}(e(2))\rp \mathcal{A}_{\mathrm{nc}}\]
 with the property\begin{equation}
\label{eq: property of alpha}
\alpha (S(u_{(1)}))x\alpha (u_{(2)})=x\lt u
\end{equation}
 for all \( u\in U_{q}(e(2)) \) and \( x\in \mathcal{A}_{\mathrm{nc}}. \)

\ela

\begin{proof}

As restriction of the algebra homomorphism \( \alpha ' \) we get
that \( \alpha  \) is an algebra homomorphism. Moreover, the property
\( \alpha (S(u_{(1)}))x\alpha (u_{(2)})=x\lt u \) follows directly
because in the cross algebra \( U_{q}(e(2))\ltimes \mathcal{A}_{\mathrm{nc}} \)
the right-action is given by \[
x\lt u=S(u_{(1)})\cdot x\cdot u_{(2)}\]
 (just use the definition of the multiplication in the cross algebra
(\ref{de: multiplication in cross algebra}) to verify this). Taking
into account that \( \alpha ' \) is an algebra homomorphism with
\( \alpha '|_{\mathcal{A}_{\mathrm{nc}}}=\mathrm{id}|_{\mathcal{A}_{\mathrm{nc}}} \)
we then have \[
x\lt u=\alpha '(x\lt u)=\alpha '(S(u_{(1)})\cdot x\cdot u_{(2)})=\alpha '(S(u_{(1)}))x\alpha '(u_{(2)})=\alpha (S(u_{(1)}))x\alpha (u_{(2)})\, \, \, .\]

\end{proof}\emph{}\\
\emph{Remark:} The element \[
T\overline{T}q^{-2iJ}\]
is a Casimir operator of the quantum group \( U_{q}(e(2)) \). This
may give another hint how to define the algebra homomorphism \( \alpha  \)
mapping it on a constant in \( \mathcal{A}_{\mathrm{nc}} \), the
only Casimir operator of \( \mathcal{A}_{\mathrm{nc}} \).

With the tools gathered so far, we are now ready to treat the topic
gauge transformation and gauge invariance.

\subsection{Gauge Transformations and Gauge Invariance}

As we have seen in Subsection \ref{se: q-integra} the \( q- \)integral
is not cyclic anymore. Therefore the usual gauge transformation \( B\rp UBU^{-1} \)
does not lead to a gauge invariance principle for \( q\neq 1. \)
Thus, we have to modify gauge transformations so that for \( q=1 \)
we get the usual, commutative gauge transformation on the one hand
and a reasonable gauge invariance principle for \( q\neq 1 \) on
the other hand.

\subsubsection{Gauge Transformations}

In analogy to the procedure in \cite{Grosse:2000Fuzzy1} we introduce
the following gauge transformation of a one-form \( B \): \begin{equation}
\label{de: gauge transformations of one-forms}
B\rp \alpha (S(\gamma _{(1)}))B\alpha (\gamma _{(2)})\, \, \, ,
\end{equation}
 with \( \gamma \in \mathcal{G} \), where \begin{equation}
\label{de: gauge group G}
\mathcal{G}:=\{\gamma \in U_{q}(e(2)):\, \varepsilon (\gamma )=1,\, \overline{\gamma }=S(\gamma )\}
\end{equation}
 and \( \alpha :U_{q}(e(2))\rp \mathcal{A}_{\mathrm{nc}} \) is the
algebra homomorphism found in Subsection \ref{se: alg hom Uq(e(2)) to Alg_nonc}.
\emph{\( \mathcal{G} \)} is closed under multiplication and to draw
analogy to the commutative case we write \[
\mathcal{G}=e^{\mathcal{H}}\]
 where \[
\mathcal{H}:=\{\gamma \in U_{q}(e(2)):\, \varepsilon (\gamma )=0,\, \overline{\gamma }=S(\gamma )\}\, \, \, .\]
 The image \( \alpha (\mathcal{H}) \) can be interpreted as a sub-algebra
of the algebra of functions \( \mathcal{A}_{\mathrm{nc}} \). \( \overline{\gamma }=S(\gamma ) \)
reflects the unitarity condition of a commutative unitary gauge parameter
and an arbitrary gauge parameter can be written as\[
\alpha (\gamma )(z,\overline{z})=e^{i\alpha (h)(z,\overline{z})}\, \, \, \textrm{with}\, \, \, \overline{h}=-S(h)\, \, \, .\]
 On account of the property (\ref{eq: property of alpha}) and since
the \( \theta ^{i} \) commute with arbitrary functions (\ref{eq: theta commutes with all functions f}),
the gauge transformation for the components reads\[
B_{i}\rp \alpha (S(\gamma _{(1)}))B_{i}\alpha (\gamma _{(2)})=B_{i}\lt \gamma \, \, \, .\]
 With \( B=\Theta +A \) follows then for the gauge transformation
of \( A \):\[
\Theta +A\rp \alpha (S(\gamma _{(1)}))(\Theta +A)\alpha (\gamma _{(2)})\]
and \begin{eqnarray*}
\alpha (S(\gamma _{(1)}))(\Theta +A)\alpha (\gamma _{(2)}) & = & \alpha (S(\gamma _{(1)}))\alpha (\gamma _{(2)})\Theta +\alpha (S(\gamma _{(1)}))[\Theta ,\alpha (\gamma _{(2)})]\\
 &  & +\alpha (S(\gamma _{(1)}))A\alpha (\gamma _{(2)})\\
 & = & \alpha (\varepsilon (\gamma ))\Theta +\alpha (S(\gamma _{(1)}))A\alpha (\gamma _{(2)})+\alpha (S(\gamma _{(1)}))d\alpha (\gamma _{(2)})\\
 & = & \Theta +\alpha (S(\gamma _{(1)}))A\alpha (\gamma _{(2)})+\alpha (S(\gamma _{(1)}))d\alpha (\gamma _{(2)})\\
 & =: & \Theta +A'
\end{eqnarray*}
 where we have used the fact that \( \Theta  \) generates the exterior
differential \( d \) (Lemma \ref{la: generator of exterior differentil: Ttheta}),
that \( \alpha  \) is an algebra homomorphism (Lemma \ref{la: the homomorphism alpha})
and that \( S(\gamma _{(1)})\gamma _{(2)}=\varepsilon (\gamma )=1 \)
for \( \gamma \in \mathcal{G} \). Hence, we conclude:\begin{equation}
\label{eq: q-guage transformation for A}
A\rp A'=\alpha (S(\gamma _{(1)}))A\alpha (\gamma _{(2)})+\alpha (S(\gamma _{(1)}))d(\alpha (\gamma _{(2)}))\, \, \, .
\end{equation}

\subsubsection{Gauge Invariance}

The field strength defined in (\ref{de: Field strength F equals B to the two})
as \( F:=B^{2} \) transforms \emph{{}``gauge-covariantly}'':\begin{eqnarray*}
F & \rp  & \alpha (S(\gamma _{(1)_{(1)}}))B\alpha (\gamma _{(1)_{(2)}})\alpha (S(\gamma _{(2)_{(1)}}))B\alpha (\gamma _{(2)_{(2)}})\\
 & = & \alpha (S(\gamma _{(1)}))B\varepsilon (\gamma )B\alpha (\gamma _{(2)})\\
 & = & \alpha (S(\gamma _{(1)}))F\alpha (\gamma _{(2)})\, \, \, ,
\end{eqnarray*}
 since \( \varepsilon (\gamma )=1 \) for \( \gamma \in \mathcal{G} \).
If we write \( F=f\theta \overline{\theta } \) this yields, since
\( \theta  \) and \( \overline{\theta } \) commute with all functions,
\begin{equation}
\label{eq: gauge trafo of f}
f\rp \alpha (S(\gamma _{(1)}))f\alpha (\gamma _{(2)})\, \, \, .
\end{equation}
We already saw in (\ref{eq: limit q to 1 for F equals B to the two})
that \( F \) has the right semi-classical limit such that we really
gave a reasonable definition for \( F \).

Moreover, the action \( S:=\int ^{q}F(*_{H}F) \) is \emph{gauge invariant:}
As defined in (\ref{de: Hodge dual of functions}), \( *_{H}F=\frac{1}{2}f \).
Therefore, we get with (\ref{eq: gauge trafo of f}) that \[
*_{H}F\rp \alpha (S(\gamma _{(1)}))(*_{H}F)\alpha (\gamma _{(2)})\]
 for gauge transformations. Thus, \emph{\begin{eqnarray*}
S & \rp  & \int ^{q}\alpha (S(\gamma _{(1)}))F(*_{H}F)\alpha (\gamma _{(2)})\\
 & = & \int ^{q}\alpha (S(\gamma _{(1)}))\frac{1}{2}f^{2}\alpha (\gamma _{(2)})\theta \overline{\theta }\, \, \, .
\end{eqnarray*}
} Since the homomorphism \( \alpha  \) has the property \( \alpha (S(u_{(1)}))x\alpha (u_{(2)})=x\lt u \)
for all \( x\in \mathcal{A}_{\mathrm{nc}} \) and \( u\in U_{q}(e(2)) \)
(Lemma \ref{la: the homomorphism alpha}) and the integral is invariant
with respect to the action of \( U_{q}(e(2)) \) (cf. Subsection \ref{se: q-integra}),
we obtain \[
S\rp \int ^{q}(\frac{1}{2}f^{2}\lt \gamma )q^{-2}dzd\overline{z}=\varepsilon (\gamma )\int ^{q}\frac{1}{2}f^{2}\theta \overline{\theta }=\int ^{q}F(*_{H}F)=S\, \, \, .\]
 Therefore, the action is indeed gauge invariant.

\subsubsection{The Semi-classical Limit \protect\( q\rp 1\protect \)}

We consider the first contribution of a gauge transformation of a
one-form \( A \) given in (\ref{eq: q-guage transformation for A}):
\begin{eqnarray*}
\alpha (S(\gamma _{(1)}))A\alpha (\gamma _{(2)}) & = & \alpha (S(\gamma _{(1)}))\alpha (\gamma _{(2)})A+\alpha (S(\gamma _{(1)}))[A,\alpha (\gamma _{(2)})]\\
 & = & \varepsilon (\gamma )A+\alpha (S(\gamma _{(1)}))[A,\alpha (\gamma _{(2)})]=A+\alpha (S(\gamma _{(1)}))[A,\alpha (\gamma _{(2)})]
\end{eqnarray*}
 for \( \gamma \in \mathcal{G} \). In the limit \( q\rp 1 \), the
commutator vanishes and we conclude\[
\alpha (S(\gamma _{(1)}))A\alpha (\gamma _{(2)})\stackrel{q\rp 1}{\longrightarrow }A\, \, \, .\]
 On the other hand, the second term in the transformation of \( A \)
reads with \( df=dz^{i}\partial _{z^{i}}(f) \) (Lemma \ref{la: generator of exterior differentil: Ttheta}):\begin{equation}
\label{eq: trafo of A second term}
\alpha (S(\gamma _{(1)}))d(\alpha (\gamma _{(2)}))=\alpha (S(\gamma _{(1)}))dz^{i}\partial _{z^{i}}(\alpha (\gamma _{(2)}))\, \, \, .
\end{equation}
 Let us calculate an example to get a better understanding of these
deformed gauge transformations. If we take \( \gamma =q^{2J} \),
we first see that \( \varepsilon (\gamma )=1 \) and \( \overline{\gamma }=q^{-2J}=S(\gamma ) \)
(see (\ref{eq: commrel and structure maps for Uq(e(2))}) and (\ref{eq: complex conjugation of J})).
Thus, \( \gamma  \) is an element of \( \mathcal{G} \). Using the
structure maps of \( J \) given in (\ref{eq: commrel and structure maps for Uq(e(2))}),
we get:

\begin{equation}
\label{eq: auf dem Weg zum kl. Limit der Eichtrafos}
(S\te 1)\Delta \gamma =q^{-2J}\te q^{2J}\, \, \, .
\end{equation}
Since \( \alpha (q^{-2iJ})=z\overline{z} \), we obtain with \( q=e^{h} \)
that \( \alpha (-2ihJ)=\mathrm{ln}(z\overline{z}) \). Hence \[
\alpha (q^{2J})=e^{i\mathrm{ln}(z\overline{z})}\]
 and we finally obtain with (\ref{eq: auf dem Weg zum kl. Limit der Eichtrafos}):\[
\alpha (S(\gamma _{(1)}))d(\alpha (\gamma _{(2)}))=e^{-i\mathrm{ln}(z\overline{z})}d(e^{i\mathrm{ln}(z\overline{z})})\, \, \, .\]
 For \( q\rp 1 \) this becomes\[
id(\mathrm{ln}(z\overline{z}))=i\partial _{z^{i}}(\mathrm{ln}(z\overline{z}))dz^{i}\, \, \, .\]
 Altogether we see that, if we write \( \lambda :=\mathrm{ln}(z\overline{z}) \),
the gauge transformation of \( A \) becomes in the semi-classical
 limit \( q\rp 1 \): \[
A\rp A'=A+\partial _{z^{i}}\lambda (z,\overline{z})dz^{i}\]
 and this is indeed a commutative gauge transformation. This gives
us an impression of the semi-classical of the \( q- \)deformed gauge
transformations. 

Nevertheless, this is not fully satisfying. We still have to examine
whether the same result follows for an arbitrary gauge parameter \( \lambda  \).
Unfortunately, it was not possible to give a general answer in the
framework of this thesis. Further considerations at this point are
missing and could be part of ongoing research on this topic.

However, we want to add some general considerations for the interested
reader. The gauge group we considered so far is not the most general
choice. It is possible to make the following generalization for the
{}``gauge group'' \( \mathcal{G} \) \cite{Steinacker:privatecom}:

Let \( \mathcal{G} \) be a subset of a Hopf algebra \( H \) such
that the following requirements are satisfied: 

\begin{enumerate}
\item There exists an algebra homomorphism \[
\alpha :H\rp \mathcal{A}_{\mathrm{nc}}\, \, \, .\]

\item For all \( U\in \mathcal{G} \) we have \begin{equation}
\label{eq: requirement for alpha}
\mathcal{D}(\alpha (U))=\alpha (S^{-2}(U))\, \, \, ,
\end{equation}
 with \( \mathcal{D} \) as defined in (\ref{de: operator D (cyclic property of the q-integral)}).
\item For all \( U\in \mathcal{G} \) we have \( \varepsilon (U)=1 \).
\end{enumerate}
Let us point out that in particular \( H=U_{q}(e(2)) \) and \( \mathcal{G} \)
as defined in (\ref{de: gauge group G}) satisfy those demands. We
just have to use the homomorphism \( \alpha  \) constructed in Subsection
\ref{se: alg hom Uq(e(2)) to Alg_nonc} to check that the second condition
is satisfied: It is easy to verify that \[
S^{2}(U)=q^{2iJ}Uq^{-2iJ}\, \, \, \Leftrightarrow \, \, \, S^{-2}(U)=q^{-2iJ}Uq^{2iJ}\]
 for all \( U\in U_{q}(e(2)) \). Thus, we have to check whether\[
\alpha (S^{-2}(U))=\alpha (q^{-2iJ})\alpha (U)\alpha (q^{2iJ})\stackrel{!}{=}\mathcal{D}(\alpha (U))\, \, \, .\]
 Since \( \alpha (q^{-2iJ})=z\overline{z} \) and \( z\overline{z}f(z,\overline{z})(z\overline{z})^{-1}=f(q^{-2}z,q^{2}\overline{z})=\mathcal{D}(f(z,\overline{z})) \)
for all \\
\( f\in \mathcal{A}_{\mathrm{c}} \), this indeed is true. 

Defining gauge transformations again as the adjoint action of an element
\\
\( U\in \mathcal{G} \), i.e. \[
B\rp \alpha (S(U_{(1)}))B\alpha (U_{(2)})\, \, \, ,\]
we obtain that the action \( S:=\int ^{q}F(*_{H}F) \) is gauge invariant:\emph{\begin{eqnarray*}
S & \rp  & \int ^{q}\alpha (S(\gamma _{(1)}))F(*_{H}F)\alpha (\gamma _{(2)})\\
 &  & =\int ^{q}F(*_{H}F)\alpha (\gamma _{(2)})\mathcal{D}(\alpha (S(\gamma _{(1)})))\\
 &  & =\int ^{q}F(*_{H}F)\alpha (\gamma _{(2)})(\alpha (S^{-1}(\gamma _{(1)})))\\
 &  & =\varepsilon (\gamma )\int ^{q}F(*_{H}F)=S\, \, \, ,
\end{eqnarray*}
} where we used in the second line the cyclic property of the \( q- \)integral
(Lemma \emph{}\ref{la: cyclic property of the q-integral}) and in
the last but one step the second property from above.

The hope is to find by this generalization a gauge group \( \mathcal{G} \)
that is big enough together with an adequate homomorphism \( \alpha  \)
such that we get in the semi-classical limit \( q\rp 1 \) classical
gauge transformations even for arbitrary gauge parameters.

\subsection{Summary and Last Remarks}

In this section we established an \( E_{q}(2)- \)covariant gauge
field theory. First we found a \( U_{q}(e(2))- \)invariant integral.
This integral is uniquely determined by the invariance requirement.
We saw that unfortunately this integral does not admit a trace property
(cf. (\ref{eq: cyclic property of the q-integral})) such that we
already concluded at this point, that defining gauge transformations
of a field strength as usual by conjugation with an unitary element,
will not lead to an invariant action. We further studied the properties
of the \( E_{q}(2)- \)symmetric space developing a covariant differential
calculus in all detail. To make calculations easier, a convenient
basis of one forms, a frame, was introduced as well as a generator
for the \( q- \)deformed exterior differential. This made it possible
to talk about one- and two-forms in a nice language, finally enabling
us to define a gauge field \( A \) and a field strength \( F \)
that becomes the commutative field strength in the semi-classical
 limit \( q\rp 1 \). At this point gauge transformations still have
not been introduced but after having established an algebra homomorphism
\( \alpha :U_{q}(e(2))\rp \mathcal{A}_{\mathrm{nc}} \) with the property
\( f\lt X=\alpha (S(X_{(1)}))f\alpha (X_{(2)}) \) for all \( X\in U_{q}(e(2)) \)
and \( f\in \mathcal{A}_{\mathrm{nc}} \) we defined gauge transformations
as the action of elements of a certain sub-algebra of \( U_{q}(e(2)) \),
that we considered as the gauge group, on functions in \( \mathcal{A}_{\mathrm{nc}} \)
(see \ref{de: gauge transformations of one-forms}). The homomorphism
\( \alpha  \) gave the gauge group an interpretation as functions
in \( \mathcal{A}_{\mathrm{nc}} \) and the way of defining gauge
transformations itself led to the notion of an gauge invariant action
that is \( E_{q}(2)-\textrm{invariant} \) as well. At the end, we
tried to give the \( q-\textrm{deformed } \)gauge transformations
a classical interpretation in the limit \( q\rp 1 \). Although the
explanation is surely not fully satisfying it leads us to the assumption
that we indeed introduced reasonable gauge transformations which become
ordinary gauge transformations for \( q\rp 1 \). Nevertheless, more
detailed considerations are certainly missing at this point and give
incentive to ongoing research on this topic. Moreover, future work
could be to develop some analogons of the Seiberg-Witten maps we considered
in Section \ref{se: Generalization of constant case}. The aim would
be to express noncommutative quantities in terms of the commutative
ones in such a way, that commutative gauge transformations of the
commutative quantities induce the {}``deformed'' gauge transformations
we considered in this section. If such a Seiberg-Witten map exists,
it should be possible, together with the star-product formalism developed
in the previous section, to express all noncommutative gauge fields
in terms of the commutative ones and expand them in the deformation
parameter similarly as we did in Section \ref{se: Generalization of constant case}.
This would allow us to read off explicitly order by order the corrections
this noncommutative theory predicts for the commutative theory, just
as in the previous section but this time for a theory that respects
the background symmetry of the space.

However, before solving this problem we surely have to understand
better the nature of the {}``deformed'' symmetry transformations
we introduced. Since elements in \( U_{q}(e(2)) \) are in general
not group-like, we get that for example the gauge transformation of
a product of functions implies transforming each factor differently
because of \( fg\lt X=(f\lt X_{(1)})(g\lt X_{(2)}) \). This curious
transformation property has to be reconciled with the physical interpretation.

\chapter{Conclusion}

We studied in the first chapter how far a change of a star product
changes the resulting action in the case of canonical deformation
of the space, i.e. in the case of a constant Poisson tensor \( \theta ^{ij} \).
The result is, that in general two different star products will lead
to two different actions but that for a large class of star products
the corresponding actions indeed end up to be equal. The presumption
is that a physical reason implies a restriction of all allowed ordering
prescriptions and that we are obliged to consider only particular
star products that fit in this physical requirement. For those star
products, the actions would all be equal. Thus, the gauge field theory
developed using the Moyal-Weyl star product would actually be independent
of the choice of star product within this set.

In the second part of this thesis two approaches to gauge field theory
on the \( E_{q}(2)- \)symmetric plane were established. The first,
as generalization of the theory developed in the case of a constant
Poisson structure, admits the expression of the noncommutative physical
fields in terms of the commutative ones. An explicit expansion of
the action in orders of the deformation parameter \( h \) leads to
new interactions that do not exist in the commutative theory. However,
a freedom in defining the gauge field and the field strength could
not be avoided. Moreover, the fact that we introduced infinitesimal
gauge transformations of a gauge field by the star commutator with
the gauge parameter, forced us to introduce an integral with trace
property which in turn is not \( E_{q}(2)- \)invariant. 

In a second approach we set up a theory that is \( E_{q}(2)- \)covariant
constructing an \( E_{q}(2)- \)invariant integral and an \( E_{q}(2)- \)differential
calculus. We defined a noncommutative field strength which approaches
the classical field strength in the semi-classical  limit \( q\rp 1 \).
It is a result of the demand for \( E_{q}(2)-\textrm{covariance} \)
of the theory that the integral is not cyclic. That is why we were
forced to introduce \( q- \)deformed gauge transformations. We found
a gauge invariant action and discussed the semi-classical limit of
gauge transformations.

The conclusion of the work in the second chapter is the following:
If we establish gauge field theories on quantum spaces we encounter
the following dilemma: Either we introduce gauge transformations of
gauge fields as conjugation with an unitary element. In this case
we have to introduce a cyclic integral, which will in general not
be invariant under the quantum group symmetry, to get a gauge invariant
action. Or we consider the quantum group covariance as the fundamental
concept. We then have to construct an invariant integral which at
the end is in general not cyclic, forcing us to define deformed gauge
transformations that are difficult to interpret.

In this thesis we exemplified this dilemma discussing two approaches
to establish gauge field theory on the \( E_{q}(2)-\textrm{covariant } \)plane.
We pointed out the conceptual problems we meet in each case and presented
possible solutions.

\appendix

\chapter{Change of basis \protect\( z,\overline{z}\leftrightarrow x,y\protect \)\label{appendix: change of basis z,zbar to x,y}}

Let us define a change of basis by \begin{equation}
\label{app; definition of transition phi}
\phi (z,\overline{z})=\left( \begin{array}{c}
\frac{1}{2}(z+\overline{z})\\
\frac{1}{2}(z-\overline{z})
\end{array}\right) =:\left( \begin{array}{c}
x\\
y
\end{array}\right) \, \, \, .
\end{equation}
 Then one-forms transform as follows:\begin{eqnarray*}
A & := & \tilde{a}_{i}(x,y)dx^{i}\\
 & = & \tilde{a}_{i}(\phi (z,\overline{z}))\frac{\partial \phi ^{i}}{\partial z^{j}}dz^{j}\\
 & = & \frac{1}{2}(\tilde{a}_{1}(\phi (z,\overline{z}))-i\tilde{a}_{2}(\phi (z,\overline{z}))dz+\frac{1}{2}(\tilde{a}_{1}(\phi (z,\overline{z}))+i\tilde{a}_{2}(\phi (z,\overline{z}))d\overline{z}\\
 & =: & a_{z}dz+a_{\overline{z}}d\overline{z}
\end{eqnarray*}
 such that we conclude:\begin{equation}
\label{app: transformation of one-forms for x,y to z,zbar}
\begin{array}{ccc}
a_{z}(z,\overline{z}) & = & \frac{1}{2}\left\{ \tilde{a}_{1}(\phi (z,\overline{z}))-i\tilde{a}_{2}(\phi (z,\overline{z}))\right\} \\
a_{\overline{z}}(z,\overline{z}) & = & \frac{1}{2}\left\{ \tilde{a}_{1}(\phi (z,\overline{z}))+i\tilde{a}_{2}(\phi (z,\overline{z}))\right\} \, \, \, .
\end{array}
\end{equation}
Two-forms transform in the following way:\begin{eqnarray*}
F & := & \tilde{f}_{ij}(x,y)dx^{i}\wedge dx^{j}\\
 & = & \tilde{f}_{ij}(\phi (z,\overline{z}))\frac{\partial \phi ^{i}}{\partial z^{k}}dz^{k}\wedge \frac{\partial \phi ^{j}}{\partial z^{l}}dz^{l}\\
 & = & (\frac{1}{4}i\tilde{f}_{12}(\phi (z,\overline{z}))-\frac{1}{4}i\tilde{f}_{21}(\phi (z,\overline{z})))dz\wedge d\overline{z}+(-\frac{1}{4}i\tilde{f}_{12}(\phi (z,\overline{z}))+\frac{1}{4}i\tilde{f}_{21}(\phi (z,\overline{z})))\\
 & =: & f_{z\overline{z}}dz\wedge d\overline{z}+f_{\overline{z}z}d\overline{z}\wedge dz\, \, \, .
\end{eqnarray*}
 Therefore we get\[
\begin{array}{ccc}
f_{z\overline{z}} & = & \frac{1}{4}i\tilde{f}_{12}(\phi (z,\overline{z}))-\frac{1}{4}i\tilde{f}_{21}(\phi (z,\overline{z}))\\
f_{\overline{z}z} & = & -\frac{1}{4}i\tilde{f}_{12}(\phi (z,\overline{z}))+\frac{1}{4}i\tilde{f}_{21}(\phi (z,\overline{z}))\, \, \, .
\end{array}\]
This becomes, if \( \tilde{f}_{ij} \) is antisymmetric\begin{equation}
\label{app: trafo of antisymm twoform for x,y to z,zbar}
\begin{array}{ccc}
f_{z\overline{z}} & = & \frac{1}{2}i\tilde{f}_{12}(\phi (z,\overline{z}))\\
f_{\overline{z}z} & = & -\frac{1}{2}i\tilde{f}_{12}(\phi (z,\overline{z}))\, \, \, .
\end{array}
\end{equation}
In Chapter 2, we need to know \( f_{z\overline{z}} \) for the case
where \( \tilde{f} \) is the commutative, abelian field strength,
i.e. \( \tilde{f}_{12}(x,y)=\partial _{x}\tilde{a}_{2}-\partial _{y}\tilde{a}_{1} \).
To calculate \( f_{z\overline{z}} \) we first want to express the
partial derivatives \( \partial _{x} \) and \( \partial _{y} \)
in terms of partial derivatives with respect to \( z=x+iy \) and
\( \overline{z}=x-iy \): \[
\begin{array}{c}
\partial _{x}=\frac{\partial z}{\partial x}\partial _{z}+\frac{\partial \overline{z}}{\partial x}\partial _{\overline{z}}=\partial _{z}+\partial _{\overline{z}}\\
\partial _{y}=\frac{\partial z}{\partial y}\partial _{z}+\frac{\partial \overline{z}}{\partial y}\partial _{\overline{z}}=i\partial _{z}-i\partial _{\overline{z}}\, \, \, .
\end{array}\]
 Using this we can calculate\begin{equation}
\label{app: f_zzbar in dependence of a}
\begin{array}{rcl}
f_{z\overline{z}} & = & \frac{1}{2}i\tilde{f}_{12}(\phi (z,\overline{z}))\\
 & = & \frac{1}{2}i((\partial _{z}+\partial _{\overline{z}})\tilde{a}_{2}(\phi (z,\overline{z}))-(i\partial _{z}-i\partial _{\overline{z}})\tilde{a}_{1}(\phi (z,\overline{z})))\\
 & = & i((\partial _{z}+\partial _{\overline{z}})\tilde{a}_{2}(\phi (z,\overline{z}))+i(\partial _{z}-\partial _{\overline{z}})\tilde{a}_{1}(\phi (z,\overline{z})))\\
 & = & \frac{1}{2}\partial _{z}(\tilde{a}_{1}(\phi (z,\overline{z}))+i\tilde{a}_{2}(\phi (z,\overline{z})))+\frac{1}{2}\partial _{\overline{z}}(\tilde{a}_{1}(\phi (z,\overline{z}))-i\tilde{a}_{2}(\phi (z,\overline{z})))\\
 & = & \partial _{z}a_{\overline{z}}-\partial _{\overline{z}}a_{z}\, \, \, .
\end{array}
\end{equation}

\chapter{The Right-Action of \protect\( U_{q}(e(2))\protect \) on \protect\( \mathcal{A}_{\mathrm{nc}}\protect \)\label{appendix: Action of Uq on A_nc}}

Knowing the structure maps for \( J,T,\overline{T}\in U_{q}(e(2)) \)
(see (\ref{eq: commrel and structure maps for Uq(e(2))})) and their
action on \( z,\overline{z} \) (see (\ref{eq: action of Uq on generators z, zbar}))
we can determine the action of \( J,T,\overline{T} \) on arbitrary
functions using \( (xy)\lt A=(x\lt A_{(1)})(y\lt A_{(2)}) \) for
arbitrary \( x,y\in \mathcal{A}_{\mathrm{nc}},\, A\in U_{q}(e(2)) \)
(cf. Definition \ref{de: def of action on an algebra}). Since an
arbitrary function \( f(z,\overline{z})\in \mathcal{A}_{\mathrm{nc}} \)
can be decomposed in \( f(z,\overline{z})=\sum _{k\in \mathbb {Z}}z^{k}f_{k}(z\overline{z}) \)
(see (\ref{eq: decomposition of arbitrary f})) it is sufficient to
know the action on a summand \[
z^{k}f(z\overline{z})\, \, \, ,\]
 where \( f \) is a formal power series in \( z\overline{z} \).
We will derive the formulas even for negative powers of \( z\overline{z} \),
i.e. \( f(z\overline{z})=\sum _{l\in \mathbb {Z}}a_{l}(z\overline{z})^{l} \).
We start with the action on \( z^{k} \):

\begin{Cl}

For \( k\in \mathbb {Z} \) we have\begin{eqnarray}
z^{k}\lt T & = & \frac{1-q^{-2k}}{1-q^{-2}}z^{k-1}\nonumber \\
z^{k}\lt \overline{T} & = & 0\label{app: action of Uq on z^k} \\
z^{k}\lt J & = & i^{k}z^{k}\, \, \, .\nonumber 
\end{eqnarray}

\end{Cl}

\begin{proof}

The first equation can be shown by induction. Let us start to prove
it for \( k>0 \):

\begin{itemize}
\item We have \( z\lt T=1 \) such that the claim is true for \( k=1 \).
\item Supposing the claim to be true for \( k \) we find for \( k+1 \),
using \( \Delta (T)=T\te q^{2iJ}+1\te T \) as well as the actions
of \( T \) and \( J \) on \( z \) given in (\ref{eq: action of Uq on generators z, zbar}),
we get\begin{eqnarray*}
z^{k+1}\lt T & = & (z^{k}\lt T)(z\lt q^{2iJ})+z^{k}(z\lt T)\\
 & = & \frac{1-q^{-2k}}{1-q^{-2}}z^{k-1}q^{-2}z+z^{k}\\
 & = & \frac{1-q^{-2k+1}}{1-q^{-2}}z^{k}\, \, \, .
\end{eqnarray*}
Thus, the claim is proved for \( k>0. \) But for \( k=0 \) we get
\( 1\lt T=0=\frac{1-q^{0}}{1-q^{-2}} \) and the claim is true in
this case, too. For \( k<0 \) we first of all need to know how \( T \)
acts on \( z^{-1} \). We have \[
0=1\lt T=(z^{-1}z)\lt T=(z^{-1}\lt T)(z\lt q^{2iJ})+z^{-1}(z\lt T)=(z^{-1}\lt T)q^{-2}z+z^{-1}\]
 such that we conclude\[
z^{-1}\lt T=-q^{2}z^{-2}\, \, \, .\]
 This is consistent with (\ref{app: action of Uq on z^k}) since in
fact \( -q^{2}=\frac{1-q^{2}}{1-q^{_{-2}}} \). An induction as done
for the case \( k>0 \) finally proves the claim for \( k<0 \), too
and with that follows the claim for \( k\in \mathbb {Z} \). 
\end{itemize}
The last two equations finally follow immediately with \( z\lt \overline{T}=0 \),
\( z\lt J=iz \) and \( \Delta (\overline{T})=\overline{T}\te q^{2iJ}+1\te \overline{T} \)
first for \( k>0 \) and then with \( 1\lt \overline{T}=0=z^{-1}\lt \overline{T} \)
also for \( k\leq 0 \). 

\end{proof}

To determine the action on \( f(z\overline{z})=\sum _{l\in \mathbb {Z}}a_{l}(z\overline{z})^{l} \)
we start considering the action on a summand \( (z\overline{z})^{l} \):

\begin{Cl}

For \( l\in \mathbb {Z} \) we have\begin{eqnarray}
(z\overline{z})^{l}\lt T & = & q^{2}\frac{1-q^{-2l}}{1-q^{-2}}(z\overline{z})^{l-1}\overline{z}\nonumber \\
(z\overline{z})^{l}\lt \overline{T} & = & -q^{2}\frac{1-q^{2l}}{1-q^{2}}(z\overline{z})^{l-1}z\label{app: action on (zzbar)^l} \\
(z\overline{z})^{l}\lt J & = & (z\overline{z})^{l}\, \, \, .\nonumber 
\end{eqnarray}

\end{Cl}

\begin{proof}

The last equation follows immediately with \( z\lt J=iz \), \( \overline{z}\lt J=-i\overline{z} \)
and \( \Delta (J)=J\te 1+1\te J \). The first equation follows again
by induction. We start treating the case \( l>0 \):

\begin{itemize}
\item We have \( (z\overline{z})\lt T=(z\lt T)(\overline{z}\lt q^{2iJ})+z(\overline{z}\lt T)=q^{2}\overline{z} \)
and the claim is proved for \( l=1 \).
\item Supposing the claim to be true for \( l \) we get, using as in the
previous proof that \( \Delta (T)=T\te q^{2iJ}+1\te T \) as well
as the actions of \( T \) and \( J \) on \( z \) given in (\ref{eq: action of Uq on generators z, zbar}),\begin{eqnarray*}
(z\overline{z})^{l+1}\lt T & = & ((z\overline{z})^{l}\lt T)((z\overline{z})\lt q^{2iJ})+(z\overline{z})^{l}((z\overline{z})\lt T)\\
 & = & q^{2}\frac{1-q^{-2l}}{1-q^{-2}}(z\overline{z})^{l-1}\overline{z}(z\overline{z})+(z\overline{z})^{l}q^{2}\overline{z}\\
 & = & q^{2}\frac{1-q^{-2l-2}}{1-q^{-2}}(z\overline{z})^{l}\overline{z}
\end{eqnarray*}
 such that the claim follows for \( l+1. \)
\end{itemize}
If \( l=0 \), then \( 1\lt T=0 \), which is consistent with the
claim. To derive the action of \( T \) on \( (z\overline{z})^{-1} \)
we calculate\[
0=((z\overline{z})^{-1}(z\overline{z}))\lt T=((z\overline{z})^{-1}\lt T)z\overline{z}+(z\overline{z})^{-1}((z\overline{z})\lt T)=((z\overline{z})^{-1}\lt T)z\overline{z}+(z\overline{z})^{-1}q^{2}\overline{z}\]
such that we can conclude\[
(z\overline{z})^{-1}\lt T=-(z\overline{z})^{-2}\overline{z}\]
which is consistent with (\ref{app: action on (zzbar)^l}), too. For
\( l<0 \) the claim then follows by induction similar as for the
case \( l>0. \) 

Finally, the second equation follows by a similar induction as done
for the first equation.

\end{proof}

Putting these results together and using \( f(z\overline{z})=\sum _{l\in \mathbb {Z}}a_{l}(z\overline{z})^{l} \)
we obtain\begin{eqnarray*}
z^{k}f(z\overline{z})\lt T & = & (z^{k}\lt T)(f(z\overline{z})\lt q^{2iJ})+z^{k}(f(z\overline{z})\lt T)\\
 & = & \frac{1-q^{-2k}}{1-q^{-2}}z^{k-1}f(z\overline{z})+z^{k}\sum _{l\in \mathbb {Z}}a_{l}q^{2}\frac{1-q^{-2l}}{1-q^{-2}}(z\overline{z})^{l-1}\overline{z}\\
 & = & \frac{1-q^{-2k}}{1-q^{-2}}z^{k-1}f(z\overline{z})+z^{k-1}\sum _{l\in \mathbb {Z}}a_{l}q^{2}\frac{1-q^{-2l}}{1-q^{-2}}q^{2(l-1)}(z\overline{z})^{l}\\
 & = & \frac{z^{k-1}}{1-q^{-2}}((1-q^{-2k})f(z\overline{z})+\sum _{l\in \mathbb {Z}}a_{l}(q^{2l}-1)(z\overline{z})^{l})\\
 & = & \frac{z^{k-1}}{1-q^{-2}}(f(q^{2}z\overline{z})-q^{-2k}f(z\overline{z}))\, \, \, ,
\end{eqnarray*}
 where we used the commutation relation \( z\overline{z}=q^{2}\overline{z}z \),
too. A similar calculation finally leads to the following results:\begin{eqnarray}
z^{k}f(z\overline{z})\lt T & = & \frac{z^{k-1}}{1-q^{-2}}(f(q^{2}z\overline{z})-q^{-2k}f(z\overline{z}))\nonumber \\
z^{k}f(z\overline{z})\lt \overline{T} & = & \frac{q^{4}}{1-q^{2}}z^{k+1}\frac{f(z\overline{z})-f(q^{-2}z\overline{z})}{z\overline{z}}\label{app: action of J,T,Tbar on z^kf(zzbar))} \\
z^{k}f(z\overline{z})\lt J & = & i^{k}z^{k}f(z\overline{z})\, \, \, .\nonumber 
\end{eqnarray}

\bibliographystyle{diss}
\bibliography{literature}

\begin{thebibliography}{10}

\bibitem{Snyder:1947}
H.~S. Snyder,
\newblock {\em Quantized space-time},
\newblock Phys. Rev. {\bf 71}, 38 (1947).

\bibitem{Seiberg:1999vs}
N.~Seiberg and E.~Witten,
\newblock {\em String theory and noncommutative geometry},
\newblock JHEP {\bf 09}, 032 (1999), hep-th/9908142.

\bibitem{Madore:Buch}
J.~Madore,
\newblock {\em An Introduction to Noncommutative Differential Geometry and its
  Physical Applications},
\newblock Cambridge University Press, secon ed. (1999) 200 p. (London
  Mathematical Society Lecture Note Series 257).

\bibitem{Madore:2000en}
J.~Madore, S.~Schraml, P.~Schupp, and J.~Wess,
\newblock {\em Gauge theory on noncommutative spaces},
\newblock Eur. Phys. J. {\bf C16}, 161 (2000), hep-th/0001203.

\bibitem{Jurco:2001rq}
B.~Jurco, L.~{M\"oller}, S.~Schraml, P.~Schupp, and J.~Wess,
\newblock {\em Construction of non-Abelian gauge theories on noncommutative
  spaces},
\newblock Eur. Phys. J. {\bf C21}, 383 (2001), hep-th/0104153.

\bibitem{Jurco:2000ja}
B.~Jurco, S.~Schraml, P.~Schupp, and J.~Wess,
\newblock {\em Enveloping algebra valued gauge transformations for non- Abelian
  gauge groups on non-commutative spaces},
\newblock Eur. Phys. J. {\bf C17}, 521 (2000), hep-th/0006246.

\bibitem{Jurco:2000dx}
B.~Jurco, P.~Schupp, and J.~Wess,
\newblock {\em Nonabelian noncommutative gauge fields and Seiberg-Witten map},
\newblock Mod. Phys. Lett. {\bf A16}, 343 (2001), hep-th/0012225.

\bibitem{Schupp:2001we}
P.~Schupp,
\newblock {\em Non-Abelian gauge theory on noncommutative spaces},
\newblock (2001), hep-th/0111038.

\bibitem{Jurco:2001my}
B.~Jurco, P.~Schupp, and J.~Wess,
\newblock {\em Nonabelian noncommutative gauge theory via noncommutative extra
  dimensions},
\newblock Nucl. Phys. {\bf B604}, 148 (2001), hep-th/0102129.

\bibitem{Waldmann}
S.~Waldmann,
\newblock {\em Zur Deformationsquantisierung in der klassischen Mechanik:
  Observablen, {Zust\"ande} und Darstellungen},
\newblock Thesis, ALU-Freiburg, Chair Prof. Dr. H. {R\"omer}  (1999),
  http://idefix.physik.uni-freiburg.de/~stefan/.

\bibitem{Bayen:1978ha}
F.~Bayen, M.~Flato, C.~Fronsdal, A.~Lichnerowicz, and D.~Sternheimer,
\newblock {\em Deformation Theory and Quantization},
\newblock Ann. Phys. {\bf 111}, 61 (1978).

\bibitem{Gerstenhaber}
M.~Gerstenhaber and S.~Schack,
\newblock {\em Algebraic Cohomology and Deformation Theory},
\newblock M. Hazewinkel and M.Gerstenhaber: Deformation Theory of Algebras and
  Structures and Applications ,
\newblock Kluwer Academic Publishers Group, Dordrecht (1988), 11-264.

\bibitem{Kontsevich:1997vb}
M.~Kontsevich,
\newblock {\em Deformation quantization of Poisson manifolds, I},
\newblock (1997), q-alg/9709040.

\bibitem{Gutt}
M.~Bertelson, M.~Cahen, and S.~Gutt,
\newblock {\em Equivalence of Star Products},
\newblock Class. Quantum Grav. {\bf 14}, A93 (1997).

\bibitem{Moyal:1949sk}
J.~E. Moyal,
\newblock {\em Quantum mechanics as a statistical theory},
\newblock Proc. Cambridge Phil. Soc. {\bf 45}, 99 (1949).

\bibitem{Behr}
W.~Behr,
\newblock {\em The Standard Model on Noncommutative Spacetime},
\newblock Diploma-Thesis, {LMU-M\"unchen}, Chair Prof. J. Wess  (2002).

\bibitem{Syko}
A.~Sykora,
\newblock {\em Elektrostatik im Quantenraum},
\newblock Diploma-Thesis, {LMU-M\"unchen}, Chair Prof. J. Wess  (2001).

\bibitem{FaddeevResTak:1987}
L.~Faddeev, N.~Reshetikhin, and L.~Takhtajan,
\newblock {\em Quantization of Lie Groups and Lie Algebras},
\newblock Algebra Anal. {\bf 1}, 187 (1987).

\bibitem{Schupp:1992ex}
P.~Schupp, P.~Watts, and B.~Zumino,
\newblock {\em The Two-dimensional quantum Euclidean algebra},
\newblock Lett. Math. Phys. {\bf 24}, 141 (1992), hep-th/9206024.

\bibitem{Klimyk:1997eb}
A.~Klimyk and K.~{Schm\"udgen},
\newblock {\em Quantum groups and their representations},
\newblock Berlin, Germany: Springer (1997) 552 p.

\bibitem{Schraml}
S.~Schraml,
\newblock {\em Untersuchung nichtkommutativer {R\"aume} als Grundlage
  physikalischer Probleme},
\newblock Thesis, {LMU-M\"unchen}, Chair Prof. J. Wess  (2000).

\bibitem{Felder}
G.~Felder and B.~Shoiket,
\newblock {\em Deformation quantization with traces},
\newblock math.QA/0002057.

\bibitem{Grosse:2000Fuzzy1}
H.~Grosse, J.~Madore, and H.~Steinacker,
\newblock {\em Field theory on the q-deformed fuzzy sphere. I},
\newblock J. Geom. Phys. {\bf 38}, 308 (2001), hep-th/0005273.

\bibitem{Grosse:2001pr}
H.~Grosse, J.~Madore, and H.~Steinacker,
\newblock {\em Field theory on the q-deformed fuzzy sphere. II: Quantization},
\newblock J. Geom. Phys. {\bf 43}, 205 (2002), hep-th/0103164.

\bibitem{Steinacker:2001fj}
H.~Steinacker,
\newblock {\em Quantum field theory on the q-deformed fuzzy sphere},
\newblock (2001), hep-th/0105126.

\bibitem{Kassel:1995xr}
C.~Kassel,
\newblock {\em Quantum groups},
\newblock New York, USA: Springer (1995) 531 p. (Graduate text in mathematics,
  155).

\bibitem{Chaichian:1996ah}
M.~Chaichian and A.~P. Demichev,
\newblock {\em Introduction to quantum groups},
\newblock Singapore: World Scientific (1996) 343 p.

\bibitem{Koe}
H.~T. Koelink,
\newblock {\em {The quantum group of plane motions and the Hahn-Exton
  $q$-Bessel function}},
\newblock Duke Math. J. {\bf 76}, 483 (1994).

\bibitem{Chaichian:1999wy}
M.~Chaichian, A.~Demichev, and P.~Presnajder,
\newblock {\em Quantum field theory on the noncommutative plane with {$E_q(2)$}
  symmetry},
\newblock J. Math. Phys. {\bf 41}, 1647 (2000), hep-th/9904132.

\bibitem{Steinacker:1995jh}
H.~Steinacker,
\newblock {\em Integration on quantum Euclidean space and sphere in N-
  dimensions},
\newblock J. Math. Phys. {\bf 37 Nr.9} (1996), q-alg/9506020.

\bibitem{Woronowicz:1989rt}
S.~L. Woronowicz,
\newblock {\em Differential calculus on compact matrix pseudogroups (quantum
  groups)},
\newblock Commun. Math. Phys. {\bf 122}, 125 (1989).

\bibitem{Zumino:1991}
B.~Zumino,
\newblock {\em Introduction to the Differential Geometry of Quantum Groups},
\newblock In: K. Sch{\"u}dgen (Ed.), Math. Phys. X, Proc. X-th IAMP Conf.
  Leibzig  (1991),
\newblock Springer (1992).

\bibitem{Jurco:1991}
B.~Jurco,
\newblock {\em Differential calculus on quantized simple Lie groups},
\newblock Lett. Math. Phys. {\bf 22}, 177 (1991).

\bibitem{Wess:1991vh}
J.~Wess and B.~Zumino,
\newblock {\em Covariant Differential Calculus on the Quantum Hyperplane},
\newblock Nucl. Phys. Proc. Suppl. {\bf 18B}, 302 (1991).

\bibitem{Steinacker:privatecom}
H.~Steinacker,
\newblock {\em private communication}.

\end{thebibliography}

\chapter*{Acknowledgements}

I would like to thank all who made this diploma-thesis possible.

Especially, I would like to thank Prof. Dr. Julius Wess who kindly
excepted me in his group and provided me with this interesting topic
for my diploma thesis.

I am very grateful to Prof. Dr. Peter Schupp and Dr. Harold Steinacker
for supervising me. They gave me guidance and helped me with many
useful explanations and suggestions.

Furthermore, I would like to thank everyone in the group for the friendly,
pleasant and productive ambiance. A lot of interesting and helpful
discussions helped me to acquire a deeper understanding of the thesis'
subject matter as well as let me have an enjoyable time. In this context
special thanks go to Fabian Bachmaier (who additionally helped me
with {}``Mathematica''), Wolfgang Behr, Dr. Christian Blohmann,
Dr. Branislav Jur\v co, Dr. John Madore, D\v zo Mikulovi\' c and Andreas
Sykora. 

Of course my parents Elisabeth and Johann Meyer merit my particular
gratitude. They made it possible for me to study physics supporting
me in every regard. I appreciate a lot their faith and confidence
towards me.

Last but surely not least I want to say thank you to my girl-friend
Walkiria for all her patience and love.

\chapter*{Erkl"arung}

Hiermit erkl"are ich, dass ich die vorliegende Arbeit selbstst"andig
und unter aus-schlie"slicher Verwendung der angegebenen Quellen und
Hilfsmittel verfasst habe.
\end{document}